\documentclass[12pt,runningheads]{article}
\usepackage[utf8]{inputenc}
\usepackage{array,multicol,mathpazo}
\usepackage[normalem]{ulem}
\usepackage{fullpage}
\usepackage{soul}
\usepackage{floatrow}
\usepackage[top=1in, bottom=1in, left=1in, right=1in]{geometry}
\usepackage{setspace}
   
\fontsize{12}{0}
\usepackage{amsmath}
\usepackage{amssymb}
\usepackage{bm}
\usepackage{amsthm}
\usepackage{mathtools} 
\usepackage{exscale}
\usepackage[mathscr]{eucal}
\usepackage{bm}
\usepackage{eqlist} 
\usepackage[final]{graphicx}
\usepackage[dvipsnames]{color}
\usepackage{verbatim}
\usepackage[square,sort,comma]{natbib}
\usepackage{natbib}
\usepackage{caption}
\usepackage{enumerate}
\usepackage{subcaption}
\usepackage{algpseudocode}
\usepackage{rotating}
\usepackage{etoolbox}
\usepackage{color}
\usepackage[linesnumbered,ruled]{algorithm2e}
\usepackage[thinc]{esdiff}
\usepackage[shortlabels]{enumitem}
\PassOptionsToPackage{hyphens}{url}\usepackage{hyperref}

\makeatletter
\patchcmd{\env@cases}{1.2}{0.6}{}{}
\makeatother

\DeclareGraphicsExtensions{.pdf, .jpg}

\theoremstyle{definition}

\newtheorem{condition}{Condition}
\newtheorem{proposition}{Proposition}
\newtheorem{proposition.a}{Proposition A\ignorespaces}
\newtheorem{proposition.s}{Proposition S\ignorespaces}

\newtheorem{remark.s}{Remark S\ignorespaces}
\newtheorem{thm}{Theorem}
\newtheorem{thm.s}{Theorem S\ignorespaces}
\newtheorem{cor}{Corollary}
\newtheorem{cor.s}{Corollary S\ignorespaces}

\newtheorem{lem.s}{Lemma S\ignorespaces}
\newcommand{\A}{\bm{A}}

\newcommand{\x}{\bm{x}}

\newcommand{\z}{\bm{z}}

\newcommand{\w}{\bm{w}}

\newcommand{\y}{\bm{y}}
\newcommand{\X}{\bm{X}}

\newcommand{\h}{{h}}
\newcommand{\Z}{\bm{Z}}
\newcommand{\I}{\bm{I}}
\newcommand{\W}{\bm{W}}
\newcommand{\R}{\bm{R}}
\newcommand{\Q}{\bm{Q}}
\newcommand{\K}{\bm{K}}
\newcommand{\U}{\bm{U}}
\newcommand{\V}{\bm{V}}
\newcommand{\bmP}{\bm{P}}
\newcommand{\bmS}{\bm{S}}

\newcommand{\bmeps}{\bm{\epsilon}}
\newcommand{\bmbeta}{\bm{\beta}}

\newcommand{\bmeta}{\bm{\eta}}

\newcommand{\tE}{\mbox{E}}

\newcommand{\diagg}{\mbox{Diag}}

\newcommand{\tr}{\mbox{tr}}
\newcommand{\bbR}{\mathbb{R}}

\newcommand{\bmI}{\mbox{I}}
\newcommand{\bmSigma}{\bm{\Sigma}}

\makeatletter
\newcommand*\rel@kern[1]{\kern#1\dimexpr\macc@kerna}
\newcommand*\widebar[1]{%
  \begingroup
  \def\mathaccent##1##2{%
    \rel@kern{0.8}%
    \overline{\rel@kern{-0.8}\macc@nucleus\rel@kern{0.2}}%
    \rel@kern{-0.2}%
  }%
  \macc@depth\@ne
  \let\math@bgroup\@empty \let\math@egroup\macc@set@skewchar
  \mathsurround\z@ \frozen@everymath{\mathgroup\macc@group\relax}%
  \macc@set@skewchar\relax
  \let\mathaccentV\macc@nested@a
  \macc@nested@a\relax111{#1}%
  \endgroup
}
\makeatother
\usepackage{newfloat}
\usepackage[labelfont=bf]{caption}
\usepackage{tabularx}
\usepackage{lipsum}
\def\bxz{\color{black}}
\title{\Large
On block-wise and reference panel-based estimators for genetic data prediction in high dimensions} 
\author{Bingxin Zhao\footnote{Department of Statistics, Purdue University. Email: bingxin@purdue.edu}, Shurong Zheng\footnote{School of Mathematics and Statistics, Northeast Normal University. Emai: zhengsr@nenu.edu.cn}, and Hongtu Zhu\footnote{ Department of Biostatistics, University of North Carolina at Chapel Hill. Email: htzhu@email.unc.edu}\\~\\
}
\begin{document}
\maketitle
\date{\vspace{-3ex}}
\date{}
\abstract{

Genetic prediction of complex traits and diseases has attracted enormous attention in precision medicine, mainly because it has the potential to translate discoveries from genome-wide association studies (GWAS) into medical advances.  
As the high dimensional covariance matrix (or the linkage disequilibrium (LD) pattern) of genetic variants has a block-diagonal structure, many existing methods attempt to account for the dependence among variants in predetermined local LD blocks/regions. 
Moreover, due to privacy restrictions and data protection concerns, genetic variant dependence in each LD block is typically estimated from external reference panels rather than the original training dataset. 
This paper presents a unified analysis of block-wise and reference panel-based estimators in a high-dimensional prediction framework without sparsity restrictions. 
We find that, surprisingly, even when the covariance matrix has a block-diagonal structure with well-defined boundaries, block-wise estimation methods adjusting for local dependence can be substantially less accurate than methods controlling for the whole covariance matrix.
Further, estimation methods built on the original training dataset and external reference panels are likely to have varying performance in high dimensions, which may reflect the cost of having only access to summary level data from the training dataset. 
This analysis is based on our novel results in random matrix theory for  block-diagonal covariance matrix.
We numerically evaluate our results using extensive simulations and the large-scale UK Biobank real data analysis of $36$ complex traits.\\

}

\section{Introduction}\label{sec1}
Genome-wide association studies (GWAS) have been widely used to examine the relationship between thousands of complex traits and millions of genetic variants in the human genome \citep{visscher201710}. Recently,  genetic risk prediction is used to translate the knowledge learned from massive GWAS studies into clinical advances in precision medicine \citep{torkamani2018personal}.
Numerous statistical methods have been developed   to improve genetic risk prediction accuracy  \citep{pain2021evaluation}. Now, 
genetic risk prediction has been widely evaluated across diverse clinical fields, resulting in thousands of publications every year \citep{zhao2021polygenic}. For example, the genetic risk scores have been successfully applied to quantify the susceptibility and progression of important clinical outcomes, such as glaucoma \citep{craig2020multitrait}, Parkinson’s disease \citep{liu2021genome},  and breast cancer \citep{fritsche2020cancer}.  

\begin{figure}[t]
\includegraphics[page=2,width=1\linewidth]{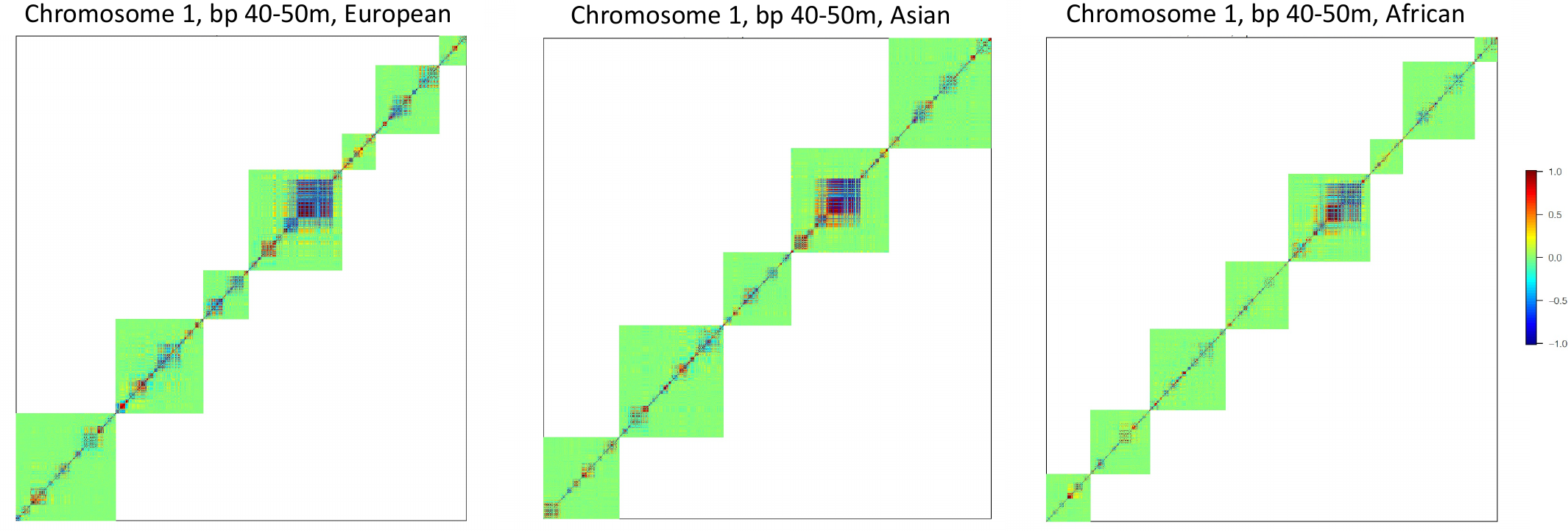}
  \caption{Illustration of the LD blocks in the human genome. 
  We display one genomic region (bp 40-50m) on chromosome 1 and show multiple LD blocks within this region in European (EUR, left), Asian (ASN, middle), and African (AFR, right) ancestries, respectively. The block boundaries are identified by \cite{berisa2016approximately} and the correlation patterns are estimated using real genotype data from the UK Biobank study \citep{bycroft2018uk}. 
}
\label{fig1}
\end{figure}

In genetic risk prediction, a major challenge is how to take the linkage disequilibrium (LD) pattern into account.
The LD pattern $\bmSigma$ is a population-based parameter that describes the covariance structure among the $p$ genetic variants  within a given population \citep{pasaniuc2017dissecting}. 
Based on empirical evidence, such LD pattern exhibits a block-diagonal structure (Figure~\ref{fig1}). 
Then we may assume 
$\bmSigma=\diagg(\bmSigma_{1}, \ldots, \bmSigma_{l},\ldots,\bmSigma_K)$ for $K$ local LD blocks.
Therefore, many existing methods use block-wise estimators and they explicitly adjust for the local dependence among variants within each LD block 
\citep{marquez2020ldpred,ge2019polygenic,vilhjalmsson2015modeling,mak2017polygenic,hu2017leveraging,yang2020accurate,lloyd2019improved,pattee2020penalized,song2020leveraging}. 
For example, let $\X=[\X_1,\cdots,\X_l\cdots,\X_K]$ be genetic variant data in the $K$ LD blocks and $\y$ be a continuous trait, a block-wise ridge estimator $\widetilde{\bmbeta}_{B}(\lambda)$ can be given by $\widetilde{\bmbeta}_{B}(\lambda)=\{\widetilde{\bmbeta}_{B_1}(\lambda)^T,\ldots,\widetilde{\bmbeta}_{B_l}(\lambda)^T,\ldots,\widetilde{\bmbeta}_{B_K}(\lambda)^T\}^T$, where 
$\widetilde{\bmbeta}_{B_l}(\lambda)=(\X_l^T\X_l+\lambda \I_{p_l})^{-1}\X_l^T\y$ with $\lambda \in (0,\infty)$ \citep{vilhjalmsson2015modeling,ge2019polygenic}. The $\widetilde{\bmbeta}_{B}(\lambda)$ is different from the traditional ridge estimator $\widetilde{\bmbeta}_{R}(\lambda)=(\X^T\X+\lambda \I_p)^{-1}\X^T\y$, which ignores the block-diagonal structure of $\bmSigma$ and adjusts for the whole LD pattern using a genome-wise sample covariance matrix. 

Furthermore, most of the existing methods are built on the public available GWAS marginal summary-level data $\widetilde{\bmbeta}_{S} \propto \X^T\y$. 
Due to privacy restrictions and data protection concerns about sharing individual-level genetic data $\X$, genetic variant correlations in each LD block are typically estimated from an external reference panel dataset $\W=[\W_1,\cdots,\W_l\cdots,\W_K]$, which are usually independent from $\X$  and $\y$. For the block-wise ridge estimator,  a reference panel-based version can be 
$\widetilde{\bmbeta}_{BW}(\lambda)=\{\widetilde{\bmbeta}_{BW_1}(\lambda)^T,$ $\ldots,\widetilde{\bmbeta}_{BW_l}(\lambda)^T,\ldots,\widetilde{\bmbeta}_{BW_K}(\lambda)^T\}^T$, where
$\widetilde{\bmbeta}_{BW_l}(\lambda)=(\W_l^T\W_l+\lambda \I_{p_l})^{-1}\X_l^T\y$. 
The 1000 Genomes reference panel \citep{10002015global} is used to estimate the LD pattern 
in many methods (e.g., \cite{ge2019polygenic,vilhjalmsson2015modeling,mak2017polygenic,hu2017leveraging,wang2021testing}).
Other frequently used reference panels include the UK10K  \citep{uk10k2015uk10k}, the TOPMed \citep{taliun2021sequencing}, and the in-house testing datasets \citep{lloyd2019improved,yang2020accurate}.
Although block-wise estimators and reference panels are extremely popular in genetic risk prediction, we know little about their statistical properties. 

The aim of this paper is to provide a comprehensive and unified analysis for the prediction accuracy of various block-wise ridge-type estimators in a high-dimensional non-sparsity framework \citep{zhao2019cross}. 
Specifically, we investigate two fundamental questions. \begin{itemize}
\item The first question is for the block-wise approach: 
When the boundaries in $\bmSigma$ are known, will $\widetilde{\bmbeta}_{B}(\lambda)$ have a better prediction accuracy than $\widetilde{\bmbeta}_{R}(\lambda)$?  
\end{itemize}
In practice, $\widetilde{\bmbeta}_{B}(\lambda)$ can be constructed  simultaneously from the $K$ LD blocks, which is computationally more efficient than $\widetilde{\bmbeta}_{R}(\lambda)$. 
Therefore, we are likely to sacrifice prediction accuracy for computational efficiency. 
Alternatively, $\widetilde{\bmbeta}_{B}(\lambda)$ may have better prediction accuracy and higher computational efficiency by making better use of the block-diagonal data structure.  
\begin{itemize} 
\item
The second question is related to the reference panel: Do 
$\widetilde{\bmbeta}_{BW}(\lambda)$ and $\widetilde{\bmbeta}_{B}(\lambda)$ have the same prediction performance? \end{itemize} Comparing $\widetilde{\bmbeta}_{B}(\lambda)$ and $\widetilde{\bmbeta}_{BW}(\lambda)$ will enable us to quantify the influence of the reference panel on genetic risk prediction.  
To answer these questions, we use recent advances in random matrix theory \citep{dobriban2018high} and develop our novel results for  block-diagonal covariance matrix. 
As high-dimensional data with block-diagonal covariance structures are widely available in many fields, our results may provide  insights for a variety of prediction problems.

There are two major theoretical findings. First, we reveal that, even when $\bmSigma$ has a block-diagonal structure with known block boundaries, block-wise ridge estimator $\widetilde{\bmbeta}_{B}(\lambda)$ adjusting for local dependence may have lower accuracy than the traditional ridge estimator $\widetilde{\bmbeta}_{R}(\lambda)$ adjusting for global covariance. 
The reduction is primarily determined by the high dimensionality of predictors (i.e., genetic variations), as well as by the signal-to-noise ratio (otherwise known as heritability) and the training data sample size. Given the rapid growth of global biobank-scale GWAS samples \citep{zhou2021global}, such reductions are likely to become more prevalent for many heritable complex traits.
Second, the performance of $\widetilde{\bmbeta}_{BW}(\lambda)$ that use external reference panels is likely to vary compared to the $\widetilde{\bmbeta}_{B}(\lambda)$ built directly on the training datasets, which may reflect the cost of having only access to GWAS summary-level data from the training dataset. 
In summary, our results suggest that we may achieve higher prediction accuracy if we simultaneously adjusting for the whole LD across the entire genome using the original training dataset. 

Methodologically, we develop a novel framework for addressing  the question of how to better use the block-wise data structure. One solution might be block-wise low-rank models. 
Although low-rank estimators have been widely applied to high-dimensional data problems, they have been less popular in genetic risk prediction since most existing methods directly utilize genetic variants for prediction. 
Within LD blocks, low-rank structure often occurs as nearby variants are highly correlated (Figure~\ref{fig1}). 
To understand whether low-rank estimators can achieve better genetic prediction accuracy, we further extend our analysis to evaluate a set of block-wise local principal components (BLPC) based methods, which use BLPCs instead of the original genetic variants to perform genetic prediction. 
Intuitively, BLPCs are useful for reducing the dimension of genetic predictors and aggregating small contributions of causal variants. 
Particularly, local correlations among genetic variants are removed because the local PCs within each block are orthogonal to each other.
We numerically demonstrate that BLPC-based methods can produce better prediction accuracy than conventional genetic variant-based methods for many complex traits in the UK Biobank study \citep{bycroft2018uk}.

The rest of the paper proceeds as follows.
In Section~\ref{sec2}, we introduce the model setups and estimators. 
In Section~\ref{sec3}, we provide the results for block-wise estimators.  
Section~\ref{sec4} extends the results to block-wise reference panel-based estimators. 
Section~\ref{sec5} studies the BLPC-based methods. 
Sections~\ref{sec6} performs simulation studies and real data analysis to numerically verify our asymptotic results in finite samples and illustrate the performance in the UK Biobank study. 
We discuss a few future topics in Section~\ref{sec7}. 
Most of the technical details are provided in the supplementary file.
\section{LD blocks and block-wise ridge  estimators}\label{sec2}
\subsection{Model setups}\label{sec2.1}
Consider two independent GWAS that are conducted for the same continuous complex trait with the same $p$ genetic variants:
\begin{itemize}
\item Training GWAS: $(\X,\y)$, where $\X=(\x_{1},\ldots,\x_{p}) \in \bbR^{n\times p}$ and $\y \in \bbR^{n \times 1}$;  
\item Testing GWAS: $(\Z,\y_{z})$, where $\Z=(\z_{1},\ldots,\z_{p}) \in \bbR^{n_{z}\times p}$ and $\y_{z} \in \bbR^{n_{z} \times 1}$. 
\end{itemize}
Here $\y$ and $\y_{z}$ are the complex traits measured in two independent cohorts with sample sizes $n$ and $n_z$, respectively. Without loss of generality, 
it is assumed that there are $m\le p$ causal variants with nonzero effects and  $\X_{(1)} \in \bbR^{n \times m}$ and $\Z_{(1)}  \in \bbR^{n_{z} \times m}$ denote 
the corresponding data matrices among the $p$ genetic variants in $\X$ and $\Z$, respectively. 
The linear additive polygenic models \citep{jiang2016high} assume
\begin{flalign}
\y= \X\bmbeta+\bmeps=\X_{(1)}\bmbeta_{(1)}+\bmeps  \quad \text{and}  \quad
\y_{z}=\Z\bmbeta+\bmeps_{z}= \Z_{(1)}\bmbeta_{(1)}+\bmeps_{z},  
\label{equ2.1}
\end{flalign}
where $\bmbeta_{(1)}$ is a $m \times 1$ vector of nonzero causal genetic effects, $\bmbeta$ is a $p \times 1$ vector consisting of $\bmbeta_{(1)}$ and $p-m$ zeros, 
and $\bmeps$ and $\bmeps_{z}$ represent independent random error vectors. 
In addition, we define an external  genotype reference panel dataset for the same $p$ genetic variants as
\begin{itemize}
\item Genotype reference panel: $\W=(\w_{1},\ldots,\w_{p}) \in \bbR^{n_{w}\times p}$.
\end{itemize}
The assumptions on genetic variant data $\X$, $\Z$, and $\W$ are summarized in Condition~\ref{con1}.
\begin{condition}
\label{con1}
\begin{enumerate}[(a).]
\item Let $\X={\X_0}\bmSigma^{1/2}$, $\Z={\Z_0}\bmSigma^{1/2}$, and $\W={\W_0}\bmSigma^{1/2}$, where the entries of $\X_0$, $\Z_0$, and $\W_0$ are real-value i.i.d. random variables with mean zero, variance one, and a finite  $4$th order moment.
The $\bmSigma$ is a $p\times p$ population level deterministic positive definite matrix of $p$ genetic variants.  
We have uniformly bounded eigenvalues in $\bmSigma$ in the sense that $0<c\le \lambda_{min}(\bmSigma) \le \lambda_{max}(\bmSigma)\le C$ for all $p$ and some constants $c,C$, where 
$\lambda_{min}(\A)$ and $\lambda_{max}(\A)$ are the smallest and largest eigenvalues of a generic matrix $\A$, respectively. The $\bmSigma^{1/2}$ denotes any nonnegative square root of $\bmSigma$. 
For simplicity, we also assume $\bmSigma_{ii}=1$, $i=1,\ldots,p$, or equivalently, $\X$, $\Z$, and $\W$ have been column-standardized.
\item The $\bmSigma$ has a block-diagonal structure with $K$  blocks, whose boundaries are known, denoted by 
 $\bmSigma=\diagg(\bmSigma_{1}, \ldots, \bmSigma_{l},\ldots,\bmSigma_K),$ 
where $\bmSigma_{l}$ is a  $p_l\times p_l$ matrix 
for $l \in 1, \ldots,  K$  such that $\sum_{l=1}^{K} p_l=p$. 
Let
$\X_l=(\x_{\tilde{p}_{_{l-1}}+1},\ldots,\x_{\tilde{p}_l})$, $\W_l=(\w_{\tilde{p}_{_{l-1}}+1},\ldots,\w_{\tilde{p}_l})$, and 
$\Z_l=(\z_{\tilde{p}_{_{l-1}}+1},\ldots,\z_{\tilde{p}_l})$ be the submatrices of $\X$, $\W$, and $\Z$ corresponding to $\bmSigma_l$, respectively, where $\tilde{p}_l=p_1+\cdots+p_l$. 
\item For $l=1,\ldots,K$, we define the empirical spectral distribution (ESD) of  $\bmSigma_l$ as $F^{\bmSigma_l}_{p_l}(x)=p_l^{-1}\cdot\sum^{p_l}_{i=1}\bmI(\lambda_i(\bmSigma_l)\le x )$, $x \in \bbR$, where $\lambda_i(\A)$ is the $i-$th eigenvalue of a generic matrix $\A$. 
As $p_l \to \infty$, let $\{\bmSigma_{{l}}\}_{p_l>1}$ be a sequence of matrices, we assume that the sequence of corresponding ESDs $\{F^{\bmSigma_l}_{p_l}(x)\}_{p_l>1}$
converges weakly to a limit probability distribution $H_l(x)$ for $x \in \bbR$, called the limiting spectral distribution (LSD), for each $l$. 
Similarly, we assume the $p\times p$ population level correlation matrix $\bmSigma$ has a LSD, denoted as $H(x)$ for $x \in \bbR$.
\item 
As $\mbox{min}(n,n_z,n_w,p) \to \infty$, we assume
$\omega_n=p/n \to \omega\in (0,\infty)$, $\omega_{w_n}=p/n_w \to \omega_{w} \in (0,\infty)$, and  $\omega_{z_n}=p/n_z \to \omega_{z}\in (0,\infty)$. In addition, for $l=1,\ldots,K$, we assume 
$\omega_{l_n}=p_l/n \to \omega_l \in (0,\infty)$, $\omega_{w_{l_n}}=p_l/n_w \to \omega_{w_{l}} \in (0,\infty)$, and $\omega_{z_{l_n}}=p_l/n_z \to \omega_{z_{l}} \in (0,\infty)$.
\end{enumerate}
\end{condition}
Conditions~\ref{con1}~(a)~and~\ref{con1}~(c) are frequently used in random matrix theory \citep{bai2004clt,bai2008large,bai2010spectral,yao2015sample}. 
In Condition~\ref{con1}~(b),   $\bmSigma$ is assumed to be  a block-diagonal structure with $K$ local LD blocks of genetic variants in GWAS. For each global population, the human genome can be divided into thousands of largely independent LD blocks, while 
individuals from the same population have similar LD boundaries. 
The LD boundaries can be determined in two ways in practice. First, the approximately independent LD blocks can be estimated from reference panel for each population.  
For example, the genome can be divided into $1,701$ independent LD blocks for European ancestry, $2,582$ blocks for African American ancestry, and $1,445$ blocks for Asian ancestry \citep{berisa2016approximately}. Second, it is possible to estimate LD blocks with certain equal window size (for instance, $1$ centimorgan) and to vary the window sizes to test the robustness of estimation \citep{bulik2015ld}. 
In Condition~\ref{con1}~(d), we assume that the sample size and the number of genetic variants are proportional to each other. Similar conditions are typically used for studying non-sparse problems in high dimensions \citep{dobriban2018high,jiang2016high,dicker2011dense}.

\begin{condition}\label{con2}
As $p\to \infty$, we assume $\gamma_p=m/p \to \gamma \in(0, \infty)$. 
\end{condition}

  Condition~\ref{con2}  assumes that the number of causal variants $m$ is proportional to   $p$  \citep{jiang2016high}. 
It has been observed that a large number of genetic variants together contribute to many complex traits,   referred to be the polygenic or omnigenic genetic architecture \citep{timpson2018genetic}. As a result, we use a high-dimensional framework for GWAS without sparsity restrictions.

Next, we introduce the condition on nonzero genetic effects $\bmbeta_{(1)}$ and random error vectors $\bmeps$ and $\bmeps_{z}$. 
\begin{condition}\label{con3}
Let $F(0,V)$ represent a generic distribution with mean zero, (co)variance $V$, and finite fourth order moments. 
We assume that the distribution of $\bmbeta_{(1)}$ is independent of the genetic variant data $\X$, $\Z$, and  $\W$ and 
satisfies
$
\begin{matrix} 
\bmbeta_{(1)}
\end{matrix}
\sim F 
  (
\begin{matrix} 
\bm{0}
\end{matrix},
\begin{matrix} 
p^{-1} \cdot\bmSigma_{\beta} 
\end{matrix}
  ),
$ 
where 
$\bmSigma_{\beta}=\sigma_{\beta}^2 \cdot \I_m$.
In addition, $\bmeps$ and $\bmeps_z$ are independent random variables satisfying
\begin{flalign*}
\begin{pmatrix} 
\bmeps\\
\bmeps_z 
\end{pmatrix}
\sim F 
\left \lbrack
\begin{pmatrix} 
\bm{0}\\
\bm{0}
\end{pmatrix},
\begin{pmatrix} 
\bmSigma_{\epsilon\epsilon} & \bm{0} \\
\bm{0} & \bmSigma_{\epsilon_z\epsilon_z}
\end{pmatrix}
\right \rbrack,
\end{flalign*} 
where 
$\bmSigma_{\epsilon\epsilon}=\sigma^2_\epsilon \cdot \I_n$ and
$\bmSigma_{\epsilon_z\epsilon_z}=\sigma^2_{\epsilon_z} \cdot \I_{n_z}$.
\end{condition}
In Condition~\ref{con3}, we model $\bmbeta_{(1)}$ as i.i.d. random variables with finite moments \citep{dobriban2018high,jiang2016high}, reflecting the fact that most of genetic variants have a small contribution to complex traits in GWAS. 
In practice, it is likely that a subgroup of genetic variants has a greater impact on a complex trait than other variants  \citep{finucane2018heritability}. We will  discuss the robustness of our i.i.d. random effect assumption in the Discussion section. 
The genetic heritability $h^2_{\beta}$ of $\y$ can be defined as
$\h^2_{\beta}=(\bmbeta^T\X^T\X\bmbeta)/(\bmbeta^T\X^T\X\bmbeta+\bmeps^T\bmeps)\in (0,1)$,
which describes how much variation in a trait can be attributed to genetic factors. The $\h^2_{\beta}$ is closely related to the signal-to-noise (STN) ratio commonly used in statistical literature. For example, the STN ratio defined in \cite{dobriban2018high} can be denoted as $\h^2_{\beta}/(1-\h^2_{\beta})$. 

\subsection{Block-wise ridge-type estimators}\label{sec2.2}
Based on the predetermined LD boundaries, {\bxz many existing genetic risk prediction methods explicitly account for local correlations within each block  \citep{marquez2020ldpred,ge2019polygenic,vilhjalmsson2015modeling,mak2017polygenic,hu2017leveraging,yang2020accurate,lloyd2019improved,pattee2020penalized,song2020leveraging}}.
We study  three block-wise ridge-type estimators of $\bmbeta$ depending on how we estimate $\bmSigma_l$s as follows. 
\paragraph{i) Block-wise ridge estimator with $\X$}
The $\bmSigma_l$s' can be approximated by 
$\widehat{\bmSigma}_{B}=\diagg(\widehat{\bmSigma}_{B_1},\ldots,\widehat{\bmSigma}_{B_l},\ldots,\widehat{\bmSigma}_{B_K})$ based on the training data  $\X$, 
where 
$\widehat{\bmSigma}_{B_l}=n^{-1}\X_l^T\X_l$ 
for $l= 1,\ldots,K$. The corresponding black-wise ridge estimator is 
\begin{flalign*}
\widehat{\bmbeta}_{B}(\lambda)=\big\{\widehat{\bmbeta}_{B_1}(\lambda)^T,\ldots,\widehat{\bmbeta}_{B_l}(\lambda)^T,\ldots,\widehat{\bmbeta}_{B_K}(\lambda)^T\big\}^T, \quad \lambda \in (0,\infty),
\end{flalign*}
where $\widehat{\bmbeta}_{B_l}(\lambda)=n^{-1}\big(\widehat{\bmSigma}_{B_l}+\lambda \I_{p_l} \big)^{-1}\X_l^T\y$ {\bxz for}  $l \in 1,\ldots,K$.

\paragraph{ii) Block-wise ridge estimator with $\W$}
The $\bmSigma_l$s' can be approximated by  $\widehat{\bmSigma}_{BW}=\diagg(\widehat{\bmSigma}_{BW_1},\ldots,\widehat{\bmSigma}_{BW_l},\ldots,\widehat{\bmSigma}_{BW_K})$  based on the external reference panel $\W$, where $\widehat{\bmSigma}_{BW_l}=n_w^{-1}\W_l^T\W_l$ for $l = 1,\ldots,K$. 
The corresponding  block-wise ridge estimator is 
\begin{flalign*}
\widehat{\bmbeta}_{BW}(\lambda)=\big\{\widehat{\bmbeta}_{BW_1}(\lambda)^T,\ldots,\widehat{\bmbeta}_{BW_l}(\lambda)^T,\ldots,\widehat{\bmbeta}_{BW_K}(\lambda)^T\big\}^T, \quad \lambda \in (0,\infty),
\end{flalign*}
where $\widehat{\bmbeta}_{BW_l}(\lambda)=n_w^{-1}\big(\widehat{\bmSigma}_{BW_l}+\lambda \I_{p_l} \big)^{-1}\X_l^T\y$ for $l = 1,\ldots,K$. 
\paragraph{iii) Block-wise ridge estimator with $\Z$}
The $\bmSigma_l$s' can also be approximated by 
$\widehat{\bmSigma}_{BZ}=\diagg(\widehat{\bmSigma}_{BZ_1},\ldots,\widehat{\bmSigma}_{BZ_l},\ldots,\widehat{\bmSigma}_{BZ_K})$ based on 
  the testing data $\Z$,  where  $\widehat{\bmSigma}_{BZ_l}=n_z^{-1}\Z_l^T\Z_l$ for $l=1,\ldots,K$.
The corresponding  block-wise ridge estimator is 
\begin{flalign*}
\widehat{\bmbeta}_{BZ}(\lambda)=\big\{\widehat{\bmbeta}_{BZ_1}(\lambda)^T,\ldots,\widehat{\bmbeta}_{BZ_l}(\lambda)^T,\ldots,\widehat{\bmbeta}_{BZ_K}(\lambda)^T\big\}^T, \quad \lambda \in (0,\infty),
\end{flalign*}
where $\widehat{\bmbeta}_{BZ_l}(\lambda)=n_z^{-1}\big(\widehat{\bmSigma}_{BZ_l}+\lambda \I_{p_l} \big)^{-1}\X_l^T\y$ for $l= 1,\ldots,K$.

Because  $\widehat{\bmbeta}_{B}(\lambda)$, $\widehat{\bmbeta}_{BW}(\lambda)$, and  $\widehat{\bmbeta}_{BZ}(\lambda)$ only need to estimate the correlations within local blocks, they are computationally efficient when both the dimension and sample size are large. 
The $\widehat{\bmbeta}_{B}(\lambda)$ is related to the traditional ridge estimator
$\widehat{\bmbeta}_{R}(\lambda)=n^{-1}\big(\widehat{\bmSigma}+\lambda \I_{p} \big)^{-1}\X^T\y$ for  $\lambda \in (0,\infty),$ 
where $\widehat{\bmSigma}=n^{-1}\X^T\X$.
The $\widehat{\bmSigma}_{B}$ provides a better estimate of $\bmSigma$ than $\widehat{\bmSigma}$ when $K>1$, since  $\widehat{\bmSigma}$ ignores the block-diagonal structure of $\bmSigma$. However,  we will demonstrate that $\widehat{\bmbeta}_{B}(\lambda)$ generally has lower prediction accuracy than $\widehat{\bmbeta}_{R}(\lambda)$. {\bxz 
A scalable algorithm of $\widehat{\bmbeta}_{R}(\lambda)$ has been developed recently for GWAS data  \citep{qian2020fast}.}
Moreover, 
  $\widehat{\bmbeta}_{BW}(\lambda)$ and $\widehat{\bmbeta}_{BZ}(\lambda)$ are popular since   they   can be easily obtained by assembling 
marginal GWAS summary statistics $\widehat{\bmbeta}_{S}= n^{-1}\X^T\y$  \citep{pasaniuc2017dissecting}  with correlations estimated from the publicly available reference panel $\W$ \citep{ge2019polygenic} or in-house testing GWAS $(\Z,\y_z)$ \citep{yang2020accurate}. 


We will quantify the high-dimensional prediction accuracy  of $\widehat{\bmbeta}_{B}(\lambda)$, $\widehat{\bmbeta}_{BW}(\lambda)$, and  $\widehat{\bmbeta}_{BZ}(\lambda)$  based on  their  asymptotic limits. 
Due to the specified block-diagonal structure and the use of reference panels, such asymptotic limits   are quite complicated so that  we have to resort to some  novel techniques of random matrix theory  \citep{bai2004clt,bai2008large}. 
Specifically, we use the out-of-sample prediction $R^2$ to compare the finite sample performance of all three   estimators, {\bxz which is a common measure in  genetic risk prediction \citep{qian2020fast}.}  
Let $\widehat{\bmbeta}$ be a generic $p\times 1$ estimator of $\bmbeta$, the out-of-sample predictor is given by $\widehat{\bmS}_{\Z}=\Z\widehat{\bmbeta}$.
Then, the out-of-sample $R^2$ is given by  $A^2=(\y_{z}^T\widehat{\bmS}_{\Z})^2/(\big\Vert\y_{z}\big\Vert\cdot\big\Vert\widehat{\bmS}_{\Z}\big\Vert)^2$, where  $\Vert\y_{z}\Vert^2=\y_{z}^T\y_{z}$ and $\Vert\widehat{\bmS}_{\Z}\big\Vert^2=\widehat{\bmS}_{\Z}^T\widehat{\bmS}_{\Z}$.
\section{Block-wise ridge estimator  }\label{sec3}
In this section, we quantify the out-of-sample performance of $\widehat{\bmbeta}_{B}(\lambda)$ in explicit form. Its major technical challenge is to obtain the limit of trace functions involving the population  covariance  matrix $\bmSigma$, the sample covariance matrix $\widehat{\bmSigma}$, and the block-wise  covariance matrix $\widehat{\bmSigma}_{B}$, since 
the standard random matrix theory does not provide simple expressions for such limit.
We start from the definition of Stieltjes transform.
Specifically, for a given distribution $F(t)$ with support on $[0,\infty)$, the Stieltjes transform is defined as $m_F(z)=\int_{0}^{\infty} (t-z)^{-1} dF(t)$ \citep{dobriban2018high}. See Section~B.2 in \cite{bai2010spectral} for more details.  
The ESD of $\widehat{\bmSigma}_{B_l}$ is denoted by
$F^{\widehat{\bmSigma}_{B_l}}_{p_l}(x)=p_l^{-1}\cdot\sum^{p_l}_{i=1}\bmI(\lambda_i(\widehat{\bmSigma}_{B_l})\le x )$ for $x \in \bbR$. 
As $p_l \to \infty$, it is well-known \citep{yao2015sample} that $F^{\widehat{\bmSigma}_{B_l}}_{p_l}(x)$ converges weakly to the LSD of $\widehat{\bmSigma}_{B_l}$, denoted as $M_l(x)$,  with probability one.
The Stieltjes transform of $M_l(x)$, denoted as $m_l(z)$, can be implicitly deﬁned by the Marchenko-Pastur equation given in   equation~(\ref{equ2}) below \citep{marchenko1967distribution,silverstein1995strong,bai2008large}.

We first establish the following results for the trace functions involved in $\widehat{\bmbeta}_{B}(\lambda)$ and state them in the following theorem, whose proof is provided in the supplementary file. 

\begin{thm}\label{thm1}
Under Condition~\ref{con1}, as $\mbox{min}(n, \
p_1, \ldots, p_K)\rightarrow\infty$, the limits of 
\begin{flalign}
p^{-1}\tr\{(\widehat{\bmSigma}-\widehat{\bmSigma}_{B})(\widehat{\bmSigma}_{B}+\lambda\I_p)^{-1}\bmSigma\} \quad\mbox{and}\quad p^{-1}\tr\{(\widehat{\bmSigma}-\widehat{\bmSigma}_{B})(\widehat{\bmSigma}_{B}+\lambda\I_p)^{-2}\bmSigma\}
\label{equ0}
\end{flalign}
are zeros
and the limit of 
\begin{flalign}
p^{-1}\tr\{(\widehat{\bmSigma}-\widehat{\bmSigma}_{B})(\widehat{\bmSigma}_{B}+\lambda\I_p)^{-1}\bmSigma(\widehat{\bmSigma}-\widehat{\bmSigma}_{B})(\widehat{\bmSigma}_{B}+\lambda\I_p)^{-1}\}
\label{equ1}
\end{flalign}
is 
$
{\omega^{-1}}
\sum_{l,h=1,\ldots,K;~{l\neq h}}{\omega_l \omega_h } \cdot\{1-\lambda m_l(-\lambda)\}\{1-\lambda m_h(-\lambda)\} +o_p(1),
$
where
\begin{flalign}
m_l(z)=\int \frac{1}{t\{1-\omega_l-\omega_l{ z} m_l(z)\}-z} d H_l(t),
\label{equ2}
\end{flalign}
in which  $H_l(t)$ is the LSD of $\bmSigma_{B_l}$ and $m_l(z)$ is the Stieltjes transform of $M_l(x)$ for  $l=1, \ldots, K$.
\end{thm}

The trace functions in equations~(\ref{equ0}) and~(\ref{equ1}) are related to the influence of using block-wise estimator $\widehat{\bmSigma}_B$ instead of the simple covariance matrix $\widehat{\bmSigma}$ in high-dimensional ridge regression. Let $\widehat{\bmSigma}_\Delta=\widehat{\bmSigma}-\widehat{\bmSigma}_{B}$, Theorem~\ref{thm1} indicates that the trace functions with one $\widehat{\bmSigma}_\Delta$ have zero limits if the the block boundaries of $\bmSigma$ are correctly specified.
However, the trace functions with more than one $\widehat{\bmSigma}_\Delta$ can be complicated and generally have nonzero limits.  
In our analysis, the trace function in~(\ref{equ1}) has two $\widehat{\bmSigma}_\Delta$s and its limit is determined by the LSD of all block-wise covariance metrics $\widehat{\bmSigma}_{B_l}$s in $\widehat{\bmSigma}_{B}$.
The information of these LSDs is contained in their Stieltjes transforms, which are linked to the LSDs of population level covariance metrics $\bmSigma_{B_l}$s based on  the Marchenko-Pastur equation. 
Specifically, the limit is related to the 
sum of products of the Stieltjes transform of the LSD of each pair of block-wise covariance metrics.

To obtain the asymptotic prediction accuracy  of $\widehat{\bmbeta}_{B}(\lambda)$, we need the following additional assumptions. 
\begin{condition}\label{con4}
Let $\A$ be a generic $p\times p$ matrix and $\A_m$ be an $m\times m$ sub-matrix of $\A$ corresponding to the $m$ causal variants with nonzero effects. As $\mbox{min}(n,p)\to \infty$,  we assume
$\tr(\A_m)/\tr(\A)=\gamma +o_p(1)$ for $\A=\bmSigma(\widehat{\bmSigma}_B+\lambda \I_p)^{-1}\widehat{\bmSigma}$ and $\widehat{\bmSigma}(\widehat{\bmSigma}_B+\lambda \I_p)^{-1}\bmSigma(\widehat{\bmSigma}_B+\lambda \I_p)^{-1}\widehat{\bmSigma}$.
\end{condition}
Condition~\ref{con4} describes some mild conditions on the homogeneity between the $m$ causal variants and the $p-m$ null variants, reflecting by the assumption that the ratio between $\tr(\A_m)$ and $\tr(\A)$ is close to $m/p$.

The next theorem summarizes the asymptotic prediction accuracy  of $\widehat{\bmbeta}_{B}(\lambda)$, denoted as $A^2_{B}(\lambda)$.
\begin{thm}\label{thm2}
Under polygenic model~(\ref{equ2.1}) and Conditions~\ref{con1}~-~\ref{con4},  as $\mbox{min}(n$, $n_z$, $p_l)\rightarrow\infty$, the limit of $A^2_{B}(\lambda)$ is
{\fontsize{10.2}{20}
\begin{flalign*}
\frac{\big[1-\lambda \R_1(\lambda)  \big]^2\cdot \h_{\beta}^4}
{\Big[1+\lambda^2\R_2(\lambda)-2\lambda\R_1(\lambda)+\R_3(\lambda)\Big]\cdot\h_{\beta}^2
+ \Big[\R_1(\lambda)-\lambda  \R_2(\lambda)\Big] \cdot\omega(1-\h_{\beta}^2)}+o_p(1),
\end{flalign*}
}%
where 
\begin{flalign*}
&\R_1(\lambda)=p^{-1}\sum_{l=1}^{K} \tr\{(a_l\bmSigma_l+\lambda\I_{p_l})^{-1}\bmSigma_l\}, \quad \\
&\R_2(\lambda)=p^{-1} \sum_{l=1}^{K}\tr\{(a_l\bmSigma_l+\lambda\I_{p_l})^{-2}(\I_{p_l}-\dot{a}_l\bmSigma_l)\bmSigma_l\}, \quad \mbox{and}\\
&
\R_3(\lambda)=\omega^{-1}\sum_{l,h=1,\ldots,K;~{l\neq h}}  \omega_l \omega_h \{1-\lambda m_l(-\lambda)\}\{1-\lambda m_h(-\lambda)\}.
\end{flalign*}
Here  $a_l$ is the unique positive solution of 
$$1-a_l
=\omega_l \cdot \Big\{1- \lambda \int_{o}^{\infty} (a_lt+\lambda)^{-1} dH_l(t)  \Big\}
=\omega_l \cdot \Big\{1-  \tE_{H_l(t)}\big(\frac{\lambda}{a_lt+\lambda}\big)\Big\} $$
and $\dot{a}_l$ is  the first order derivative of  $a_l$ and given by 
$$\dot{a}_l
=\frac{\omega_l \cdot \tE_{H_l(t)}\big\{\frac{a_lt}{(a_lt+\lambda)^2}\big\} }{
-1-\omega_l\lambda\cdot \tE_{H_l(t)}\big\{\frac{t}{(a_lt+\lambda)^2}\big\}}. $$
When $\lambda=\lambda^{*}\equiv \omega\cdot(1-\h_{\beta}^2)/\h_{\beta}^2$, $A^2_{B}(\lambda)$ is maximized and the optimal prediction accuracy is  given by 
\begin{flalign}
\label{equ3.1}
&A^2_{B}(\lambda^{*})=\h_{\beta}^2\cdot
\frac{\big\{1-\lambda^{*} \R_1(\lambda^{*})  \big\}^2}
{1-\lambda^{*}\R_1(\lambda^{*})+\R_3(\lambda^{*})}
+o_p(1). 
\end{flalign}
\end{thm}
\begin{figure}
\includegraphics[page=1,width=0.8\linewidth]{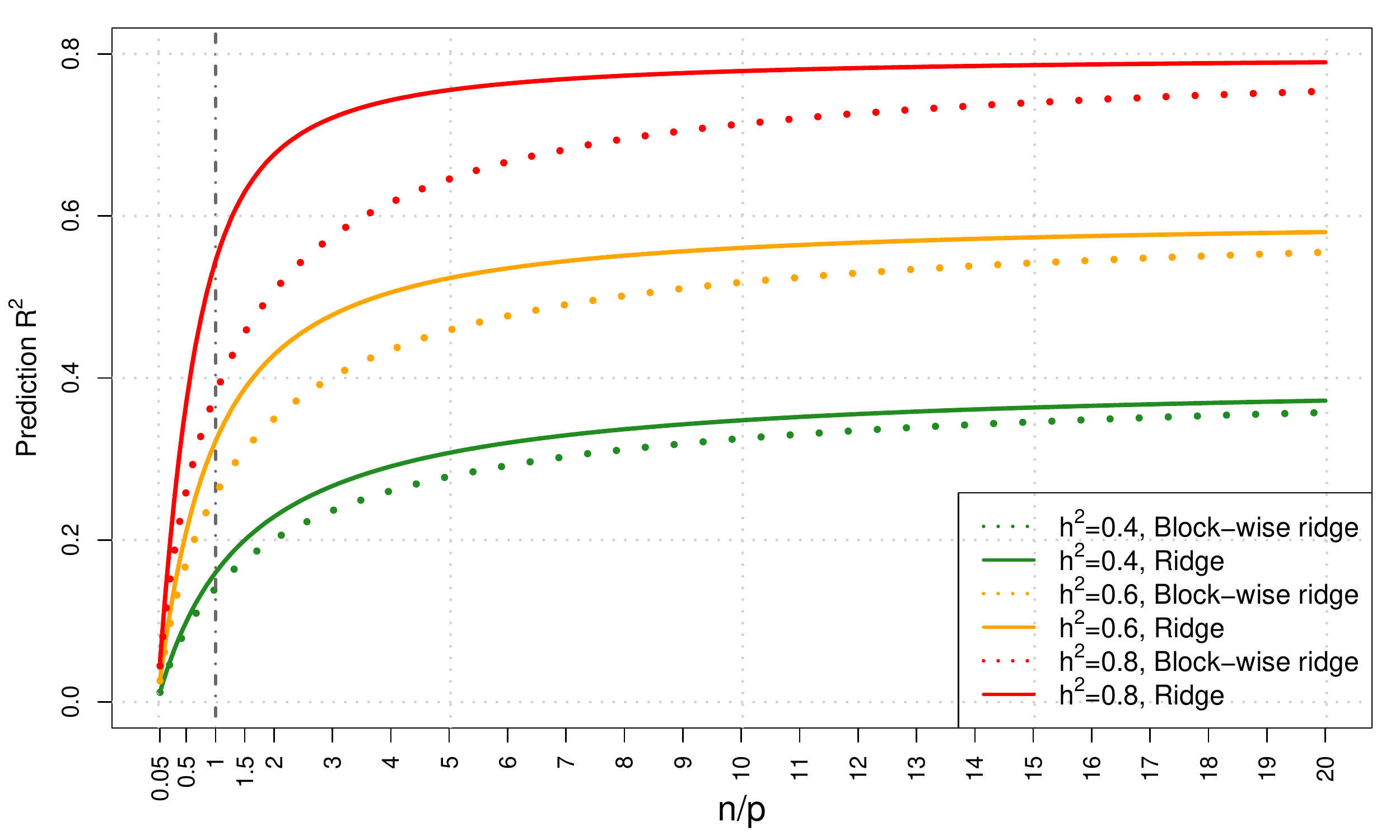}
  \caption{Comparing the prediction accuracy of block-wise ridge estimator $\widehat{\bmbeta}_{B}(\lambda^{*})$ and  traditional  ridge estimator $\widehat{\bmbeta}_{R}(\lambda^{*})$ at different heritability ($\h^2$) levels and $n/p$ ratios. 
We simulate the data with $20$ independent blocks, each of which has a block size $p_l=50$ ($p=1,000$). An auto-correlation structure is given within each block and the auto-correlation coefficient is $\rho_b=0.5$. The vertical line represents $n/p=1$.
}
\label{fig2}
\end{figure}
Theorem~\ref{thm2} shows that  besides the heritability $\h_{\beta}^2$, 
the prediction accuracy of $\widehat{\bmbeta}_{B}(\lambda^{*})$ is also determined by the training GWAS sample size (represented by {\bxz $\omega$}) and 
the two traces $\R_1(\lambda^{*})$ and $\R_3(\lambda^{*})$. 
The $\R_1(\lambda^{*})$ as the limit of $p^{-1}\tr\{(\widehat{\bmSigma}_B+\lambda^{*}\I_{p})^{-1}\bmSigma\}$  is related to the LSD of $\bmSigma_l$s, describing the influence of the LD pattern on the prediction accuracy. For general $\bmSigma$, $\R_1(\lambda)$ does not have a closed-form expression. In Theorem~\ref{thm2}, we use the notation of  deterministic equivalents  \citep{sheng2020one,serdobolskii2007multiparametric}. 
It is also possible to express $\R_1(\lambda)$ using classical Stieltjes transforms, say 
 $\R_1(\lambda)=\omega^{-1}\sum_{l=1}^{K} \big\{(\lambda v_l(-\lambda))^{-1}-1\big\},$ 
where $v_l(-\lambda)=\omega_l\{m_l(-\lambda)-\lambda^{-1}\}+\lambda^{-1}$ \citep{dobriban2018high,zhao2019cross}. 
The $\R_2(\lambda)$ is the limit of $p^{-1}\tr\{(\widehat{\bmSigma}_B+\lambda\I_{p})^{-2}\bmSigma\}$, which can be obtained by taking the first order derivative of $v_l(-\lambda)$. 
When $\lambda=\lambda^*$, the $\R_2(\lambda)$-related terms cancel out in $A^2_{B}(\lambda^{*})$ and $\widehat{\bmbeta}_{B}(\lambda)$ has the optimal prediction accuracy. Thus, $\R_2(\lambda)$ represents the decay in the performance of the ridge estimator when a sub-optimal tuning parameter $\lambda$ is used.
We also discuss $\R_3(\lambda^{*})$ as the limit of $p^{-1}\tr\{(\widehat{\bmSigma}-\widehat{\bmSigma}_{B})(\widehat{\bmSigma}_{B}+\lambda\I_p)^{-1}\bmSigma(\widehat{\bmSigma}-\widehat{\bmSigma}_{B})(\widehat{\bmSigma}_{B}+\lambda\I_p)^{-1}\}$,  describing  the impact of  $\widehat{\bmSigma}_\Delta=\widehat{\bmSigma}-\widehat{\bmSigma}_{B}$.  The $\widehat{\bmSigma}_\Delta$
 is the off-diagonal portion of the sample covariance that is lost if we only adjust for local dependency using $\widehat{\bmSigma}_{B}$. 
Although each entry of $\widehat{\bmSigma}_\Delta$ has mean zero, its variance is nonzero and  the limit of $\R_3(\lambda^{*})$ is nonzero. 
Therefore, $\R_3(\lambda^{*})$ can explain most of the reduction of prediction accuracy due to only locally accounting for the LD pattern in genetic risk prediction. 

We   compare 
the prediction accuracy of $\widehat{\bmbeta}_{B}(\lambda^{*})$ and $\widehat{\bmbeta}_{R}(\lambda^{*})$. 
Similar to Theorem ~\ref{thm2},   the prediction accuracy of  $\widehat{\bmbeta}_{R}(\lambda^{*})$ \citep{dobriban2018high}  is given by  
\begin{flalign}
\label{equ3.2}
A^2_{R}(\lambda^{*})=
\h_{\beta}^2\cdot\big\{ 1-
\lambda^{*}\R_{1_R}(\lambda^{*})\big\}
+o_p(1),
\end{flalign}
where $\R_{1_R}(\lambda^{*})=p^{-1}\tr\{(a_r\bmSigma+\lambda^{*}\I_p)^{-1}\bmSigma\} 
=p^{-1}\sum_{l=1}^{K} \tr\{(a_r\bmSigma_l+\lambda^{*}\I_{p_l})^{-1}\bmSigma_l\}$ is the limit of $p^{-1}\tr\{(\widehat{\bmSigma}+\lambda^{*}\I_p)^{-1}\bmSigma\}$  and $a_r$ is the unique positive solution of 
$1-a_r
=\omega \cdot \{1- \lambda^{*} \int_{o}^{\infty} (a_rt+\lambda^{*})^{-1} dH(t) \}$.
Suppose   $\R_3(\lambda^{*})$ becomes zero,   $A^2_{B}(\lambda^{*})$ in~(\ref{equ3.1}) increases and we have $A^2_{B}(\lambda^{*})=
\h_{\beta}^2\cdot\big\{ 1-
\lambda^{*}\R_{1}(\lambda^{*})\big\}
+o_p(1)$, which has the same functional form as the $A^2_{R}(\lambda^{*})$ in~(\ref{equ3.2}). 
Figure~\ref{fig2} provides a numerical comparison between $A^2_{R}(\lambda^{*})$ and $A^2_{B}(\lambda^{*})$ across different $\omega$ and $\h^2_\beta$. 
When either $n$ is much larger than $p$ (or  very small $\omega$) or $p$ is much larger than $n$ (or very big $\omega$), 
$A^2_{R}(\lambda^{*})$ and $A^2_{B}(\lambda^{*})$ are close to each other. 
However, when $n$ and $p$ are comparable, $\widehat{\bmbeta}_{R}(\lambda^{*})$ consistently  outperforms $\widehat{\bmbeta}_{B}(\lambda^{*})$, especially for highly heritable traits. 
Figure~\ref{fig2} also indicates that their difference is maximized when $n$ is a few times larger than $p$. 
For most complex traits, the current GWAS sample size $n$ is smaller than the number of genetic variants $p$. Thus, our analysis suggests that the difference between  $A^2_{R}(\lambda^{*})$ and $A^2_{B}(\lambda^{*})$   becomes larger as the GWAS sample size increases. In practice, local LD adjustments with $\widehat{\bmSigma}_{B}$ is convenient and computationally efficient, but adjusting the whole LD with $\widehat{\bmSigma}$   leads to better prediction for highly heritable traits studied in large-scale GWAS training  data.  

\section{Reference panel-based estimators}\label{sec4}
In this section, we quantify the out-of-sample performance of  the reference panel-based estimators $\widehat{\bmbeta}_{BW}(\lambda)$ and $\widehat{\bmbeta}_{BZ}(\lambda)$, in which  $\bmSigma_l$s are estimated from $\W$ and $\Z$, respectively.  Let   $v_{w_l}(-\lambda)=\omega_{w_l}\{m_{w_l}(-\lambda)-\lambda^{-1}\}+\lambda^{-1}$, 
and
$
m_{w_l}(z)=\int [t\{1-\omega_{w_l}-\omega_{w_l} z m_{w_l}(z)\}-z]^{-1} d H_l(t)
$
denotes the  Stieltjes  transform  of the LSD of $\widehat{\bmSigma}_{BW_l}$, denoted as $M_{w_l}(x)$. 
We obtain the limits of 
the trace functions involved in $\widehat{\bmbeta}_{BW}(\lambda)$ and $\widehat{\bmbeta}_{BZ}(\lambda)$ and state them below.  
\begin{thm}\label{thm3}
Under Condition~\ref{con1}, as $\mbox{min}(n_w$, $p_l)\rightarrow\infty$, the limit of 
$
p^{-1} \sum_{l=1}^{K} \tr\{(\widehat{\bmSigma}_{BW_l}+\lambda \I_{p_l})^{-1}\bmSigma_l(\widehat{\bmSigma}_{BW_l}+\lambda\I_{p_l})^{-1}\bmSigma_l\}
$ is approximated by 
\begin{equation*}
 p^{-1} \sum_{l=1}^{K}
\Big[1-\frac{n^{-1}\tr\{(\K_l+\lambda \I_{p_l})^{-1}\bmSigma_l(\K_l+\lambda \I_{p_l})^{-1}\bmSigma_l \}}{[1+n^{-1}\tr\{(\K_l+\lambda \I_{p_l})^{-1}\bmSigma_l \}]^2}\Big]^{-1}  \tr\{(\K_l+\lambda \I_{p_l})^{-1}\bmSigma_l(\K_l+\lambda \I_{p_l})^{-1}\bmSigma_l \},  
\end{equation*}
where $\K_l=[1+n^{-1}\tE\tr\{(\widehat{\bmSigma}_{BW_l}+\lambda \I_{p_l})^{-1}\bmSigma_{l}\}]^{-1} \bmSigma_{l}   =\lambda v_{w_l}(-\lambda)\bmSigma_{l} $.  
In addition, the limit of 
$
p^{-1} \sum_{l=1}^{K} \tr\{(\widehat{\bmSigma}_{BW_l}+\lambda \I_{p_l})^{-1}\bmSigma_l(\widehat{\bmSigma}_{BW_l}+\lambda\I_{p_l})^{-1}\bmSigma_l^2\}
$ 
is approximated by  
\begin{flalign*}
p^{-1} \sum_{l=1}^{K}
\frac{[1+n^{-1}\tr\{(\K_l+\lambda \I_{p_l})^{-1}\bmSigma_l\}]^2\cdot  \tr\{(\K_l+\lambda \I_{p_l})^{-1}\bmSigma_l(\K_l+\lambda \I_{p_l})^{-1}\bmSigma_l^2 \} }
{[1+n^{-1}\tr\{(\K_l+\lambda \I_{p_l})^{-1}\bmSigma_l\}]^2 - n^{-1}
\tr\{(\K_l+\lambda \I_{p_l})^{-1}\bmSigma_l(\K_l+\lambda \I_{p_l})^{-1}\bmSigma_l \}}.
\end{flalign*}
\end{thm}
Theorem~\ref{thm3} shows how   $ \tr\{(\widehat{\bmSigma}_{BW_l}+\lambda \I_{p_l})^{-1}\bmSigma_l(\widehat{\bmSigma}_{BW_l}+\lambda\I_{p_l})^{-1}\bmSigma_l\}$ and 
$ \tr\{(\widehat{\bmSigma}_{BW_l}+\lambda \I_{p_l})^{-1}\bmSigma_l(\widehat{\bmSigma}_{BW_l}+\lambda\I_{p_l})^{-1}\bmSigma_l^2\}$ relate to $H_l(t)$, the LSD of $\bmSigma_l$. For example, the latter is a function of $\tr\{(\K_l+\lambda \I_{p_l})^{-1}\bmSigma_l(\K_l+\lambda \I_{p_l})^{-1}\bmSigma_l^2 \}$,  $\tr\{(\K_l+\lambda \I_{p_l})^{-1}\bmSigma_l(\K_l+\lambda \I_{p_l})^{-1}\bmSigma_l \}$, and $\tr\{(\K_l+\lambda \I_{p_l})^{-1}\bmSigma_l\}$. All of them are functions of $m_{w_l}(z)$, the Stieltjes transform of $M_{w_l}(x)$, which is closely related to $H_l(t)$. 
These results may be helpful for a wide range of ridge-type problems in high dimensions. To obtain the asymptotic prediction accuracy  of $\widehat{\bmbeta}_{BW}(\lambda)$ and $\widehat{\bmbeta}_{BZ}(\lambda)$, we need the following additional assumptions. 
\begin{condition}\label{con5}
As $\mbox{min}(n,n_w,n_z,p)\to \infty$,  we assume 
$\tr(\A_m)/\tr(\A)=\gamma +o_p(1)$ for $\A=\widehat{\bmSigma}(\widehat{\bmSigma}_{BZ}+\lambda \I_p)^{-1}\widehat{\bmSigma}_Z(\widehat{\bmSigma}_{BZ}+\lambda \I_p)^{-1}\widehat{\bmSigma}$, $\widehat{\bmSigma}(\widehat{\bmSigma}_{BZ}+\lambda \I_p)^{-1}\bmSigma$, 
$\widehat{\bmSigma}(\widehat{\bmSigma}_{BW}+\lambda \I_p)^{-1}\bmSigma(\widehat{\bmSigma}_{BW}+\lambda \I_p)^{-1}\widehat{\bmSigma}$,
and 
$\bmSigma(\widehat{\bmSigma}_{BW}+\lambda \I_p)^{-1}\bmSigma$,
where $\widehat{\bmSigma}_Z=n_z^{-1}\Z^T\Z$.
\end{condition}

Similar to Condition~\ref{con4}, Condition~\ref{con5} is a set of   mild assumptions on the overall homogeneity between the $m$ causal variants and the $p-m$ null variants. Based on these assumptions, the prediction accuracy of  $\widehat{\bmbeta}_{BW}(\lambda)$ and that of  $\widehat{\bmbeta}_{BZ}(\lambda)$, donated separately as $A^2_{BW}(\lambda)$ and $A^2_{BZ}(\lambda)$, respectively,  are given in the next theorem. 

\begin{thm}\label{thm4}
Under polygenic model~(\ref{equ2.1}) and Conditions~\ref{con1}, \ref{con2}, \ref{con3}, and \ref{con5}, as $\mbox{min}(n$, $n_w$, $n_z$, $p_l)\rightarrow\infty$, we have 
\begin{flalign*}
A^2_{BW}(\lambda)=
\frac{\Q_{1}^2(\lambda)\cdot \h_{\beta}^4}
{\Q_{2}(\lambda) \cdot \h_{\beta}^2
+ 
\Q_{3}(\lambda)\cdot \omega }+o_p(1),
\end{flalign*}
where  $\Q_{1}(\lambda)$, $\Q_{2}(\lambda)$, and 
$ \Q_{3}(\lambda)$ are, respectively, given by 
\begin{flalign*}
&
 p^{-1} \sum_{l=1}^K \tr\{(a_{w_l}\bmSigma_l+\lambda\I_{p_l})^{-1}\bmSigma_l^2 \},
\\
&p^{-1} {\sum_{l=1}^{K}}
\frac{[1+n^{-1}\tr\{(\K_l+\lambda \I_{p_l})^{-1}\bmSigma_l\}]^2\cdot  \tr\{(\K_l+\lambda \I_{p_l})^{-1}\bmSigma_l(\K_l+\lambda \I_{p_l})^{-1}\bmSigma_l^2 \} }
{[1+n^{-1}\tr\{(\K_l+\lambda \I_{p_l})^{-1}\bmSigma_l\}]^2 - n^{-1}
\tr\{(\K_l+\lambda \I_{p_l})^{-1}\bmSigma_l(\K_l+\lambda \I_{p_l})^{-1}\bmSigma_l \}},
\end{flalign*}
and 
\begin{flalign*}
&p^{-1} {\sum_{l=1}^{K}}
\Big[1-\frac{n^{-1}\tr\{(\K_l+\lambda \I_{p_l})^{-1}\bmSigma_l(\K_l+\lambda \I_{p_l})^{-1}\bmSigma_l \}}{[1+n^{-1}\tr\{(\K_l+\lambda \I_{p_l})^{-1}\bmSigma_l \}]^2}\Big]^{-1}  \tr\{(\K_l+\lambda \I_{p_l})^{-1}\bmSigma_l(\K_l+\lambda \I_{p_l})^{-1}\bmSigma_l\}.
\end{flalign*}
Here $a_{w_l}$ is the unique positive solution of 
$1-a_{w_l}
=\omega_{w_l} \cdot \{1- \lambda \int_{o}^{\infty} (a_{w_l}t+\lambda)^{-1} dH_l(t) \}$. In addition, we have 
\begin{flalign*}
&A^2_{BZ}(\lambda)=
\frac{\big\{1-\lambda \Q_{4}(\lambda) \big\}^2\cdot \h_{\beta}^4}
{\{\Q_{5}(\lambda)-\lambda\Q_{6}(\lambda)\}\cdot\h_{\beta}^2
+ \{\Q_{4}(\lambda)-\lambda\Q_{7}(\lambda)\}\cdot \omega }
+o_p(1),
\end{flalign*}
where 
\begin{flalign*}
&\Q_{4}(\lambda)
=p^{-1}\sum_{l=1}^{K}\tr\{(a_{z_l}\bmSigma_l+\lambda\I_{p_l})^{-1}\bmSigma_l\},\quad
\Q_{5}(\lambda)
=p^{-1} \sum_{l=1}^{K} \tr\{(a_{z_l}\bmSigma_l+\lambda\I_{p_l})^{-1}\bmSigma_l^2\},\\
&\Q_{6}(\lambda)=p^{-1} \sum_{l=1}^{K} \tr\{(a_{z_l}\bmSigma_l+\lambda\I_{p_l})^{-2}(\I_{p_l}-\dot{a}_{z_l}\bmSigma_l)\bmSigma_l^2\}, \quad \mbox{and} \\
&\Q_{7}(\lambda)=p^{-1}\sum_{l=1}^{K} \tr\{(a_{z_l}\bmSigma_l+\lambda\I_{p_l})^{-2}(\I_{p_l}-\dot{a}_{z_l}\bmSigma_l)\bmSigma_l\}. 
\end{flalign*}
Moreover, $a_{z_l}$ is the unique positive solution of $1-a_{z_l}
=\omega_{z_l} \cdot \big\{1- \lambda \int_{o}^{\infty} (a_{z_l}t+\lambda)^{-1} dH_l(t)  \big\}$
and $\dot{a}_{z_l}$ is the first order derivative of $a_{z_l}$, given by 
$$\dot{a}_{z_l}
=\frac{\omega_{z_l} \cdot \tE_{H_l(t)}\big\{\frac{a_{z_l}t}{(a_{z_l}t+\lambda)^2}\big\} }{
-1-\omega_{z_l} \cdot\lambda \tE_{H_l(t)}\big\{\frac{t}{(a_{z_l}t+\lambda)^2}\big\}}. $$
\end{thm}
\begin{figure}[t]
\centering
\includegraphics[page=1,width=0.9\linewidth]{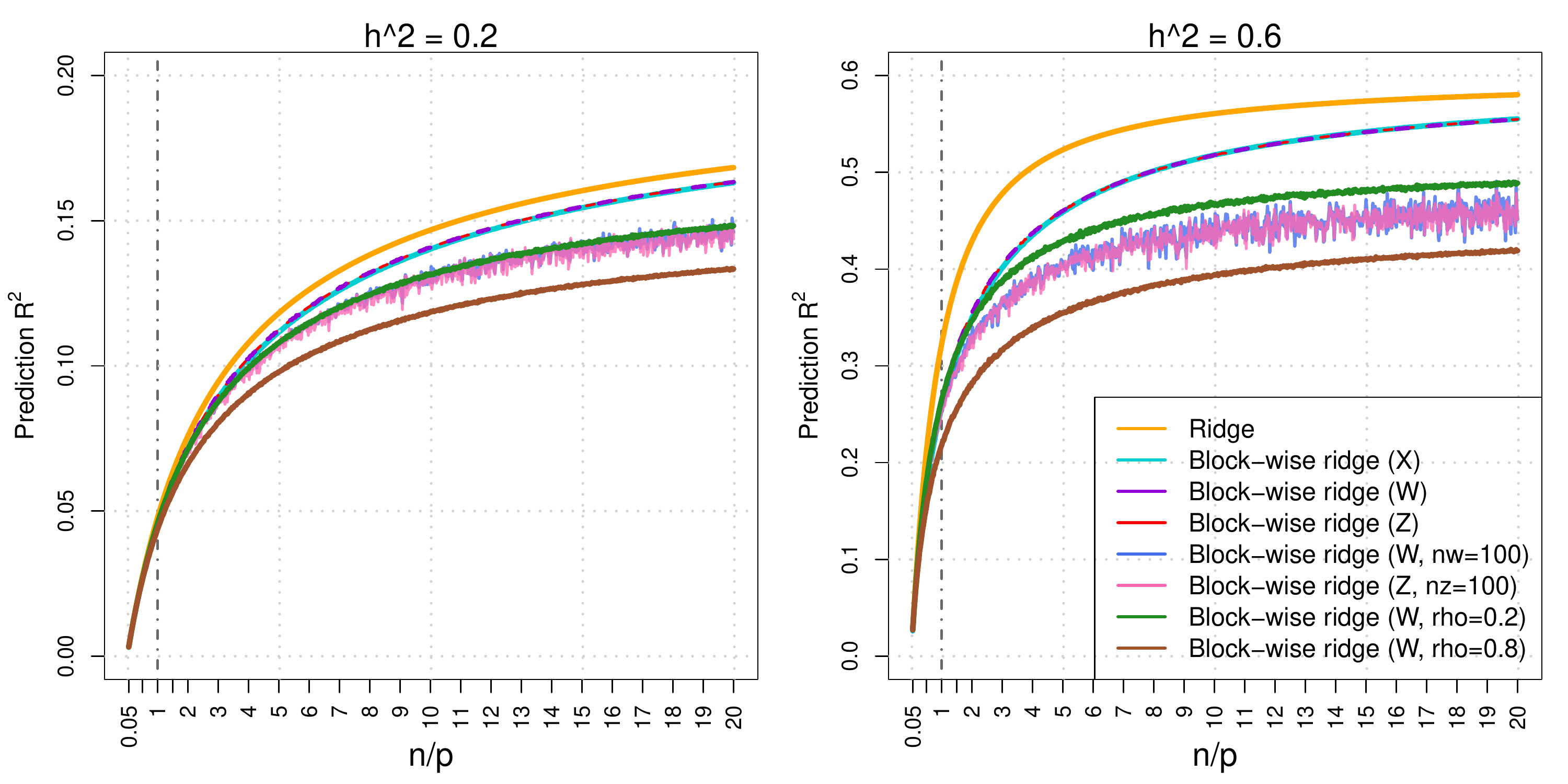}
  \caption{Comparing the prediction accuracy of block-wise ridge estimators $\widehat{\bmbeta}_{B}(\lambda^{*})$, $\widehat{\bmbeta}_{BW}(\lambda^{*})$, $\widehat{\bmbeta}_{BZ}(\lambda^{*})$,
  and traditional  ridge estimator $\widehat{\bmbeta}_{R}(\lambda^{*})$ at different $n/p$ ratios with $\h_\beta^2=0.2$ (left panel) or $0.6$ (right panel).
We simulate the data with $20$ independent blocks, each of which has a block size $p_l=50$ ($p=1,000$). An auto-correlation structure is given within each block and the auto-correlation coefficient is $\rho_b=0.5$. We also illustrate the 
performance of $\widehat{\bmbeta}_{BW}(\lambda^{*})$ and $\widehat{\bmbeta}_{BZ}(\lambda^{*})$ when $n_w=n_z=100$, and  the
performance of $\widehat{\bmbeta}_{BW}(\lambda^{*})$ when the $\rho_b$ in $\W$ is $0.8$ or $0.2$. The vertical line represents $n/p=1$. 
}
\label{fig3}
\end{figure}
Theorem~\ref{thm4} shows that $A^2_{BW}(\lambda)$ is related to $\Q_{1}(\lambda)$, $\Q_{2}(\lambda)$, and $\Q_{3}(\lambda)$ and $A^2_{BZ}(\lambda)$ is related to $\Q_{4}(\lambda)$, $\Q_{5}(\lambda)$, $\Q_{6}(\lambda)$, and $\Q_{7}(\lambda)$. Therefore, the asymptotic limits of $A^2_{B}(\lambda)$,   $A^2_{BW}(\lambda)$, and $A^2_{BZ}(\lambda)$ are different due to  different data sets used to estimate   $\bmSigma_l$s. 
Although the three estimators have different analytical forms, we investigate whether they are numerically close to each other   when the sample sizes of $\X$, $\W$, and $\Z$ are the same (that is, $n=n_w=n_z$). 
 Figure~\ref{fig3} and Supplementary Figure~1 show that 
  $A^2_{B}(\lambda)$,   $A^2_{BW}(\lambda)$, and $A^2_{BZ}(\lambda)$ are truly close to each other  across a wide range of $n/p$ ratios. 
It is also clear that 
all of the three block-wise estimators have lower prediction accuracy than the ridge estimator $\widehat{\bmbeta}_{R}(\lambda)$. 
In practice, the sample size  of reference panel and that of  testing GWAS are typically much smaller than that of training GWAS. 
For example, the popular 1000 Genomes reference panel \citep{10002015global} only has hundreds of subjects for each population. 
In Figure~\ref{fig3}, we also examine the impact of sample sizes $n_w$ and $n_z$ on the reference panel-based methods. 
Smaller sample size of reference panels can lead to a substantial reduction and larger variation in prediction accuracy. Additionally, we evaluate the influence of the mismatch between the LD in training GWAS dataset $\X$ and in the reference panel dataset $\W$, denoted as $\bmSigma_X$ and $\bmSigma_W$, respectively. 
Figure~\ref{fig3} suggests that the mismatch between $\bmSigma_X$ and $\bmSigma_W$ can dramatically reduce the performance of $\widehat{\bmbeta}_{BW}(\lambda)$, especially when the training GWAS sample size $n$ is large. Overall, these findings suggest that large reference panels  and  using a reference panel with its LD matching the original GWAS are required to ensure reliable  genetic risk prediction. 

Although $\widehat{\bmbeta}_{BW}(\lambda)$ and $\widehat{\bmbeta}_{BZ}(\lambda)$ have similar numerical performance to $\widehat{\bmbeta}_{B}(\lambda)$ when $n=n_w=n_z$, it is worth mentioning that this is not always true for reference panel-based estimators that lack a block-wise structure. 
Generally, reference panel-based estimators can behave very differently from the original ridge estimator in high dimensions. 
To better understand the reference panel-based estimators, we consider the prediction accuracy of two non-block ridge estimators 
\begin{flalign*}
\widehat{\bmbeta}_{RW}(\lambda)=n_w^{-1}\big(\widehat{\bmSigma}_W+\lambda \I_{p} \big)^{-1}\X^T\y\quad \mbox{and}\quad  
\widehat{\bmbeta}_{RZ}(\lambda)=n_z^{-1}\big(\widehat{\bmSigma}_Z+\lambda \I_{p} \big)^{-1}\X^T\y
\end{flalign*}
for $\lambda \in (0,\infty),$  
where $\widehat{\bmSigma}_W=n_w^{-1}\W^T\W$.
Supplementary Figure~2 illustrates that  $\widehat{\bmbeta}_{RW}(\lambda)$ and $\widehat{\bmbeta}_{RZ}(\lambda)$ can have much worse performance than $\widehat{\bmbeta}_{R}(\lambda)$, even when there is no LD mismatch between training GWAS and the reference panel. 
Let $A^2_{RW}(\lambda)$ and $A^2_{RZ}(\lambda)$ be the asymptotic prediction accuracy of $\widehat{\bmbeta}_{RW}(\lambda)$ and that of  $\widehat{\bmbeta}_{RZ}(\lambda)$, respectively. 
The following corollary further demonstrates this point in a special case $\bmSigma=\I_p$.
\begin{cor}
\label{cor1}
Under polygenic model~(\ref{equ2.1}) and Conditions~\ref{con1},~\ref{con2},~\ref{con3}, and ~\ref{con5}, with  $\widehat{\bmSigma}_{BW}$ and $\widehat{\bmSigma}_{BZ}$ in Condition~\ref{con5} being replaced by $\widehat{\bmSigma}_{W}$ and $\widehat{\bmSigma}_{Z}$, respectively, as $\mbox{min}(n$, $n_w$, $n_z$, $p)\rightarrow\infty$, when $\bmSigma=\I_p$, we have 
\begin{flalign*}
&A^2_{RW}(\lambda)=\frac{\h_{\beta}^4}
{\h_{\beta}^2+\omega}\cdot\frac{1}{1-\dot{b}_w}
+o_p(1) \quad \mbox{and} \quad
A^2_{RZ}(\lambda)=\frac{\h_{\beta}^4}
{\h_{\beta}^2+\omega}\cdot \frac{b_z^2+\lambda}{b_z+\lambda}
+o_p(1),
\end{flalign*}
where $b_w=(1/2)\cdot\{\sqrt{(\lambda+\omega_w-1)^2+4\lambda}-(\lambda+\omega_w-1)\}$,
$\dot{b}_w=-(\omega_wb_w)/\{\omega_w\lambda+(b_w+\lambda)^2\}$, and $b_z=(1/2)\cdot\{\sqrt{(\lambda+\omega_z-1)^2+4\lambda}-(\lambda+\omega_z-1)\}$.
\end{cor}

In this special case, we have closed-form expressions for Stieltjes transforms and asymptotic limits corresponding to $\widehat{\bmbeta}_{RW}(\lambda)$ and  $\widehat{\bmbeta}_{RZ}(\lambda)$. 
Let $\widehat{\bmbeta}_S=n^{-1}\X^T\y$ be the marginal estimator and  $A^2_{S}$ denotes the corresponding prediction accuracy.  
The prediction accuracy of $\widehat{\bmbeta}_S$  and that of  $\widehat{\bmbeta}_R(\lambda)$ are, respectively, given by  
\begin{flalign*}
&A^2_{S}=\frac{\h_{\beta}^4}
{\h_{\beta}^2+\omega}
+o_p(1) \quad \mbox{and} \quad
A^2_{R}(\lambda)=\frac{\h_{\beta}^4\cdot b_r^2}
{(b_r^2-\lambda^2\dot{b}_r)\h_{\beta}^2+(b_r+\lambda\dot{b}_r)\omega(1-\h_{\beta}^2)}
+o_p(1),
\end{flalign*}
where   $b_r=(1/2)\cdot\{\sqrt{(\lambda+\omega-1)^2+4\lambda} - (\lambda+\omega-1)\}$ and $\dot{b}_r=-(\omega b_r)/\{\omega\lambda+(b_r+\lambda)^2\}$.
For $\lambda,\omega_w,\omega_z \in (0,\infty)$,  we have 
$\max\{A^2_{RW}(\lambda),A^2_{RZ}(\lambda)\} < A^2_{S} < A^2_{R}(\lambda)$ due to  $\dot{b}_w < 0$ and $b_z^2 < b_z$. 
Therefore, as illustrated in Supplementary Figure 3, $\widehat{\bmbeta}_R(\lambda)$ can outperform $\widehat{\bmbeta}_S$ when $\bmSigma=\I_p$, but $\widehat{\bmbeta}_{RW}(\lambda)$ and $\widehat{\bmbeta}_{RZ}(\lambda)$  may have worse prediction performance. 
These results highlight the difference between the  reference panel-based estimators and those estimators directly built on the training data set in high dimensions. 
In summary, it is important to use the reference panel-based estimators with caution and awareness of potential issues especially when the structure of $\bmSigma$ is unknown. 

\section{Block-wise principal component analysis}\label{sec5}

In this section, we 
 extend our analysis to study the   performance of block-wise local principal components (BLPCs) in genetic risk prediction. 
In Figure~\ref{fig1},  evidence of low-rank structures can be seen within these blocks, so applying PCA to each of these blocks may improve prediction accuracy.  
In   training GWAS, suppose singular value decomposition (SVD) is separately performed on each of the LD blocks $\X_l$s. 
For the $l-$th block,  we have $\X_l=\widetilde{\U}_l\widetilde{\bm{D}}_l\widetilde{\V}_l^T$ for $l \in 1,\ldots,K$,
where 
$\widetilde{\bm{D}}_l=\mbox{Diag}(d_{l_1},\ldots,d_{l_{r_l}})$ are the $r_l$ positive singular values, and 
$\widetilde{\U}_l=[\bm{u}_{l_1},\ldots,\bm{u}_{l_{r_l}}] \in \bbR^{n\times r_l}$ and $\widetilde{\V}_l=[\bm{v}_{l_1},\ldots,\bm{v}_{l_{r_l}}] \in \bbR^{p_l\times r_l}$ are the left and right singular vectors, respectively.
Among the $r_l$ left singular vectors, $q_l$ ($1\le q_l\le r_l$)  of them are selected for prediction, denoted by $\U_{l}=\X_l\V_{l} \in \bbR^{n\times q_l}$, with $\V_{l} \in \bbR^{p_l\times q_l}$ being the set of corresponding right singular vectors. 
Then the marginal and block-wise ridge BLPC estimators are, respectively,  defined as 
\begin{flalign*}
\widehat{\bmeta}_{S}=n^{-1}\V^T\X^T\y \quad \mbox{and} \quad
\widehat{\bmeta}_{B}(\lambda)=n^{-1}\R(\lambda)\V^T\X^T\y, \quad \lambda \in (0,\infty),
\end{flalign*}
  where $\V=\mbox{Diag}(\V_{1},\ldots,\V_{l},\ldots,\V_{K})$ is a $p\times q$ projection matrix and 
$$\R(\lambda)=\mbox{Diag}\{(\V_{1}^T\X_1^T\X_1\V_{1}+\lambda\I_{q_1})^{-1},\ldots,(\V_{K}^T\X_K^T\X_K\V_{K}+\lambda\I_{q_K})^{-1}\}$$
is a block-wise ridge-type sample covariance estimator for the BLPCs with $q=\sum_{l=1}^K q_l$.
Let $\Z\V$ be the set of projected genetic variants in the testing GWAS dataset, 
the prediction accuracy of $\widehat{\bmeta}_{S}$ and that of $\widehat{\bmeta}_{B}(\lambda)$, denoted as $A^2_{PC_{S}}$ and $A^2_{PC_{R}}(\lambda)$, respectively,  are given in the following proposition with  additional assumptions listed in Condition~\ref{con6}.

\begin{condition}\label{con6}
As $\min(n,p)\to \infty$, we assume 
$\tr(\A_m)/\tr(\A)=\gamma +o_p(1)$ for 
$\A=\bmSigma\V\V^T\widehat{\bmSigma}$, 
$\widehat{\bmSigma}\V\V^T\bmSigma\V\V^T\widehat{\bmSigma}$, 
$\bmSigma\V\R(\lambda)\V^T\widehat{\bmSigma}$, and 
$\widehat{\bmSigma}\V\R(\lambda)\V^T\bmSigma\V\R(\lambda)\V^T\widehat{\bmSigma}$.
\end{condition}
\begin{proposition}\label{pop1}
Under polygenic model~(\ref{equ2.1}) and Conditions~\ref{con1}, \ref{con2}, \ref{con3}, and \ref{con6}, as $\mbox{min}(n$, $n_z$, $p_l)\rightarrow\infty$, we have 
\begin{flalign*}
&A^2_{PC_{S}}=
\frac{\bmP_1^2\cdot \h_{\beta}^4}
{\bmP_2\cdot\h_{\beta}^2
+ \bmP_3\cdot\omega(1-\h_{\beta}^2)}
+o_p(1) \quad \mbox{and} \\
&A^2_{PC_{R}}(\lambda)=
\frac{\bmP_4(\lambda)^2\cdot \h_{\beta}^4}
{\bmP_5(\lambda)\cdot\h_{\beta}^2
+ \bmP_6(\lambda)\cdot\omega(1-\h_{\beta}^2)}
+o_p(1),
\end{flalign*}
where 
$\bmP_1=p^{-1}\tr(\bmSigma\V\V^T\widehat{\bmSigma})$, 
$\bmP_2=p^{-1}\tr(\V\V^T\bmSigma\V\V^T\widehat{\bmSigma}^2)$,
$\bmP_3=p^{-1}\tr(\V\V^T\bmSigma\V\V^T\widehat{\bmSigma})$,
$\bmP_4(\lambda)=p^{-1}\tr(\bmSigma\V\R(\lambda)\V^T\widehat{\bmSigma})$, 
$\bmP_5(\lambda)=p^{-1}\tr(\V\R(\lambda)\V^T\bmSigma\V\R(\lambda)\V^T\widehat{\bmSigma}^2 )$, and 
$\bmP_6(\lambda)=p^{-1}\tr(\V\R(\lambda)\V^T\bmSigma\V\R(\lambda)\V^T\widehat{\bmSigma} )$.
\end{proposition}

Proposition~\ref{pop1} suggests that BLPC-based estimators differ from genetic variants-based ones primarily due to the $p\times q$ block-wise projection matrix $\V$. The $\V_l$ maps the $p_l$ genetic variants  within the $l-$th LD block onto $q_l$ BLPCs. By taking only the top-ranked BLPCs that can explain a large proportion of genetic variation,   $q_l$ is typically much smaller   than $p_l$, and thus BLPCs can reduce the dimension of predictors, while   aggregating small genetic effects.  
 The projection matrix $\V$ influences prediction accuracy primarily by adding  $\V\V^T$ (or $\V\R(\lambda)\V^T$) into the trace functions on both the numerator and denominator of $A^2_{PC_{S}}$ (or $A^2_{PC_{R}}$). 
As a result, whether BLPC can improve prediction performance depends on whether adding $\V\V^T$ 
(or $\V\R(\lambda)\V^T$) can increase the ratio between the numerator and denominator in $A^2_{PC_{S}}$ (or $A^2_{PC_{R}}$). 
For example, the prediction accuracy of $\widehat{\bmbeta}_S$ can be given by 
\begin{flalign*}
&A^2_{S}=
\frac{\big\{p^{-1}\tr(\bmSigma\widehat{\bmSigma}) \big\}^2\cdot \h_{\beta}^4}
{p^{-1}\tr(\bmSigma\widehat{\bmSigma}^2)\cdot\h_{\beta}^2
+ p^{-1}\tr(\bmSigma\widehat{\bmSigma})\cdot\omega(1-\h^2_{\beta})}
+o_p(1), 
\end{flalign*}
which is the same as $A^2_{PC_{S}}$ if we set $\V\V^T=\I_p$. {It would be interesting to further express the functions of $\V$ in terms of the LSD of $\bmSigma_l$s, but that would be complicated in our setups and beyond the scope of this paper. } In later sections, we will numerically evaluate the performance of BLPC-based estimators. 

\section{Numerical experiments}\label{sec6}
\subsection{Simulated genotype data}\label{sec6.1}
To illustrate the finite sample performance of the estimators proposed above, 
we simulate $p=10,000$ genetic variants for $10,000$ independent individual samples in  training data $\X$, external reference panel $\W$, and testing data $\Z$, respectively. 
To mimic the block-diagonal LD patterns, we construct $\bmSigma$ with $5$ independent blocks (block size $=2,000$). 
Correlations among the genetic variants within each block are estimated from one genomics region on chromosome $1$ using the European subjects in the 1000 Genomes reference panel \citep{10002015global}, and variants belonging to different blocks are independent.  
The minor allele frequency (MAF) $f$ of each genetic variant  is independently sampled from Uniform $[0.05, 0.45]$. Each entry of $\X_0$, $\W_0$, and $\Z_0$ is then independently generated from $\{0,1,2\}$ with probabilities $\{(1-f)^2,2f(1-f),f^2\}$, respectively.
The casual genetic effects are simulated from  $\bmbeta_{(1)}\sim MVN(\bm{0},\I_m/p)$. We set $\h_\beta^2=0.2$, $0.4$, or $0.8$, and vary the sparsity from $0.05$ to $0.6$ (i.e, $m/p=0.05,0.1,0.2,0.4$, or $0.6$). 
The linear polygenic model~(\ref{equ2.1}) is used to generate $\y$ and $\y_z$. 
We illustrate the numerical results of the following estimators:
1) marginal estimator $\widehat{\bmbeta}_{S}$ (no LD adjustment, Marginal); 
2) traditional ridge estimator $\widehat{\bmbeta}_{R}(\lambda^*)$, where $\lambda^*$ is the optimal regularizer  (Ridge);
3) block-wise ridge estimator based on $\X$, $\widehat{\bmbeta}_{B}(\lambda^*)$ (Block-wise-ridge);
4) block-wise ridge estimator based on $\W$, $\widehat{\bmbeta}_{BW}(\lambda^*)$ 
(Block-wise-ridge-W);  
5) block-wise ridge estimator based on $\Z$, $\widehat{\bmbeta}_{BZ}(\lambda^*)$  (Block-wise-ridge-Z);
6) ridge estimator based on $\W$, $\widehat{\bmbeta}_{RW}(\lambda^*)$  (Ridge-W); and 
7) ridge estimator based on $\Z$, $\widehat{\bmbeta}_{RZ}(\lambda^*)$  (Ridge-Z).
In addition, we reduce the sample size of  $\W$ to check whether the performance of  $\widehat{\bmbeta}_{BW}(\lambda^*)$ is sensitive to the size of reference panel. Thus, the following 
two estimators are added: 
8) block-wise ridge estimator based on $n_w=100$ subjects of $\W$
(Block-wise-ridge-W-small);  and 
9) ridge estimator based on $n_w=100$ subjects of $\W$ (Ridge-W-small). 
A total of $200$ replications are conducted for each simulation condition, and we compare these estimators using  prediction $R$-squared. 
\begin{figure}[t]
\includegraphics[page=2,width=1\linewidth]{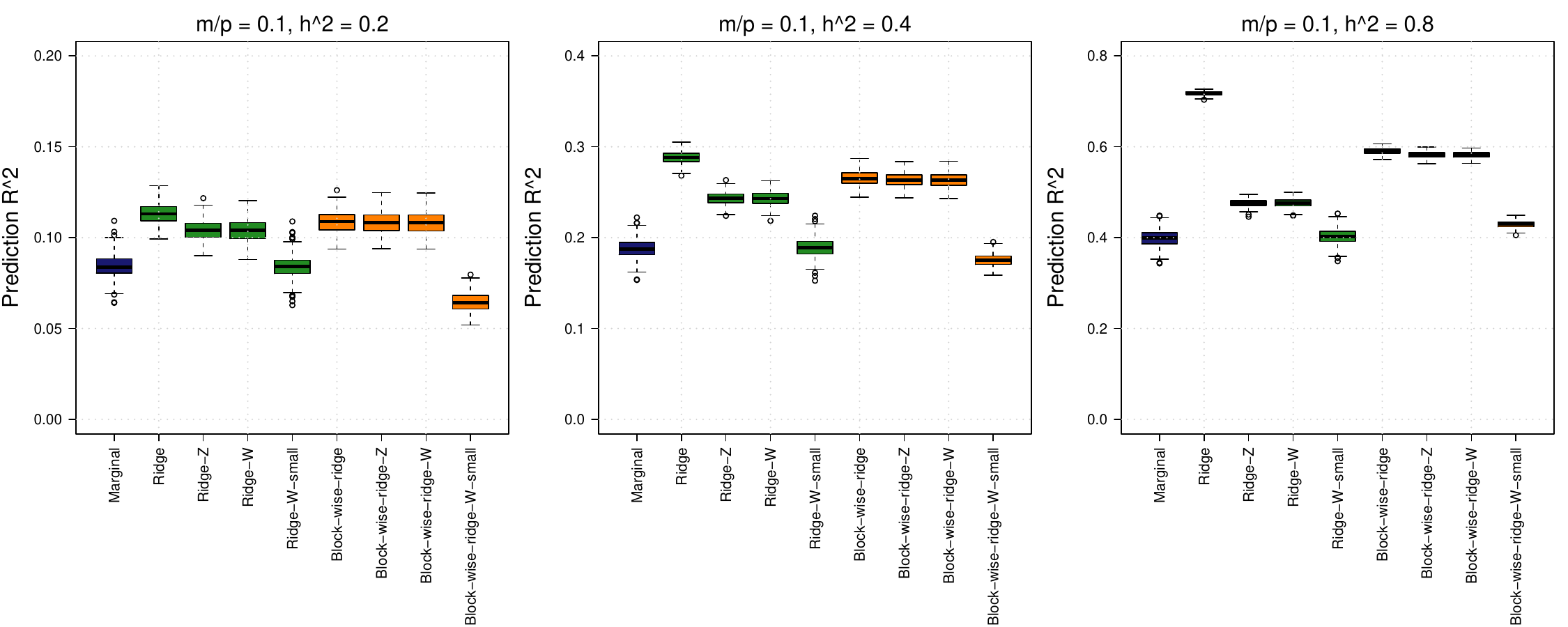}
  \caption{Prediction accuracy of different estimators. 
    We set $p=n$ $=n_z$ $=n_w$ $=10,000$, $m/p=0.1$, and vary $\h_\beta^2$ from $0.2$ to $0.8$.
     From left to right: Marginal, $\widehat{\bmbeta}_S$; 
  Ridge, $\widehat{\bmbeta}_R(\lambda^*)$; 
   Ridge-Z, $\widehat{\bmbeta}_{RZ}(\lambda^*)$; 
  Ridge-W, $\widehat{\bmbeta}_{RW}(\lambda^*)$; 
Ridge-W-small, $\widehat{\bmbeta}_{RW}(\lambda^*)$ with $n_w=100$;
Block-wise-ridge, $\widehat{\bmbeta}_{B}(\lambda^*)$; 
  Block-wise-ridge-Z, $\widehat{\bmbeta}_{BZ}(\lambda^*)$;
  Block-wise-ridge-W, $\widehat{\bmbeta}_{BW}(\lambda^*)$; and 
    Block-wise-ridge-W-small, $\widehat{\bmbeta}_{BW}(\lambda^*)$ with $n_w=100$.
}
\label{fig4}
\end{figure}

We have the following observations. 
First, as expected, the finite-sample performance of all the nine estimators matches with our theoretical results discussed above. See Figure~\ref{fig4} and Supplementary Figures~4-5 for details. Specifically, the traditional ridge estimator $\widehat{\bmbeta}_{R}(\lambda^*)$ performs the best among all estimators in all settings. 
When $n=n_w=n_z$, the three block-wise ridge estimators $\widehat{\bmbeta}_{B}(\lambda^*)$, $\widehat{\bmbeta}_{BW}(\lambda^*)$, and $\widehat{\bmbeta}_{BZ}(\lambda^*)$ have very similar performance, but they are worse than  $\widehat{\bmbeta}_{R}(\lambda^*)$, especially when the heritability is high. 
The performance of $\widehat{\bmbeta}_{RW}(\lambda^*)$ and $\widehat{\bmbeta}_{RZ}(\lambda^*)$ can be even lower than that of the three block-wise estimators. 
These results indicate that the class of block-wise and reference panel-based estimators, although efficient, can be sub-optimal compared to   $\widehat{\bmbeta}_{R}(\lambda^*)$, which   accounts for  the whole LD pattern estimated from the training dataset. All estimators perform consistently across sparsity levels. We also find that smaller reference panel may lead to lower performance for $\widehat{\bmbeta}_{BW}(\lambda^*)$ and $\widehat{\bmbeta}_{rW}(\lambda^*)$. 
Overall, our simulation results suggest that 
even when   $\bmSigma$ has a block-diagonal structure and the boundaries of the blocks are correctly assigned, the estimators based on  $\widehat{\bmSigma}_{B}$ have a lower prediction accuracy than the traditional ridge estimator based on the sample covariance estimator $\widehat{\bmSigma}$.

Next, we examine the effect of heterogeneity in $\W$ on the prediction performance of the reference panel-based estimators. Specifically, we construct $\bmSigma$ with $100$ independent blocks (block size $=50$). Genetic variants within the same block have pair-wise correlation $\rho_b=0.5$ in the training data $\X$.
We change   $\rho_b$ in $\W$ to $0.2$ and $0.8$, respectively, resulting in four new estimators:
10) block-wise ridge estimator based on $\W$ with $\rho_b=0.2$ 
(Block-wise-ridge-W-rho02);  
11) block-wise ridge estimator based on $\W$ with $\rho_b=0.8$ 
(Block-wise-ridge-W-rho08);  
12) ridge estimator based on $\W$ with $\rho_b=0.2$ 
(Ridge-W-rho02); and 
13) ridge estimator based on $\W$ with $\rho_b=0.8$ 
(Ridge-W-rho08). 
We set $m/p=0.6$ and $\h_\beta^2=0.4$ or $\h_\beta^2=0.8$. The other settings are exactly the same as in the previous cases. 
Supplementary Figure~6 shows that the LD heterogeneity may negatively impact the performance of  $\widehat{\bmbeta}_{BW}(\lambda^*)$ and $\widehat{\bmbeta}_{EW}(\lambda^*)$. For example, the prediction accuracy of $\widehat{\bmbeta}_{BW}(\lambda^*)$ can reduce from $51.2\%$ to $37.3\%$ when the $\rho_b$ in $\W$ changes from $0.5$ to $0.8$. These findings indicate that it is crucial to use  good reference panels that match  the training dataset for good prediction accuracy.

\subsection{BLPC in UK biobank data simulations}\label{sec6.2}
To evaluate the performance of BLPC-based estimators compared with traditional variant-based estimators, we perform simulations using genotype data from the UK Biobank (UKB) study \citep{bycroft2018uk}.
After downloading the imputed genotype data, we apply the following quality control (QC) procedures: excluding subjects with more than $10\%$ missing genotypes, only including variants with MAF $>0.01$, genotyping rate $>90\%$, and passing Hardy-Weinberg test ($p$-value $> 1\times 10^{-7}$).
For the simulation, we randomly select $110,000$ unrelated individuals of British ancestry from the QC'ed data set, which contains $8,932,279$ variants over $488,371$ subjects. 
Among the $110,000$ selected samples,
$100,000$ are randomly picked and used as training samples, and the prediction performance is evaluated with the remaining $10,000$ testing individuals.
Furthermore, we limit our analysis to $653,122$ variants that overlap with the HapMap3 reference panel \citep{international2010integrating}, which is a popular choice to balance accuracy and computational burden in reference panel-based approaches \citep{ge2019polygenic}. 
We set  $\h_{\beta}^2$ to $0.6$ and randomly select $200,000$ genetic variants to be causal variants, whose effects are independently generated from $N(0,1/p)$ using GCTA \citep{yang2011gcta}.

We first examine the following genetic variant-based methods: 
1) genetic variant-based marginal estimator (Variant-marginal); 
2) genetic variant-based marginal estimator after LD-based pruning to remove highly correlated variants (pruning window size $250$kb, step size $50$, and $r^2=0.3$) (Variant-marginal-prune); and
3) genetic variant block-wise reference panel-based ridge estimator (Variant-block-wise-ridge-W). 
The 1000 Genomes Phase 3 subjects of European ancestry \citep{10002015global} serve as our reference panel.
Next, we generate BLPCs for each of the $1,701$ European independent genomics regions defined in \cite{berisa2016approximately}. 
We consider the following BLPC-based estimators: 
1) BLPC-based marginal estimator (BLPC-marginal); 
2) BLPC block-wise ridge estimator with $\lambda=c\times \lambda_0$, $\lambda_0=q/n$ and  $c=10,1,0.1,0.01$, and $0$, respectively (BLPC-block-ridge); and 
3) BLPC block-wise reference panel-based ridge estimator with $\lambda=c\times \lambda_0$, where the sample covariance metrics of BCPCs are estimated from the testing data instead of the training data (BLPC-block-wise-ridge-Z). We perform $100$ simulation replications. 

The results are displayed in Supplementary Figure~7. 
The marginal screening estimator using top-ranked BLPCs that account for $50\%$ genetic variations outperforms all genetic variant-based estimators, including the variant-based marginal screening and block-wise ridge estimators.
The marginal estimator with $35\%$ top-ranked BLPCs performs similar to that with $50\%$ top-ranked BLPCs, while the marginal estimator with $80\%$ top-ranked BLPCs has much worse performance and is similar to the variant-based marginal estimator. These results indicate that BLPCs may be able to better control the dependence of genetic effects and  aggregate small contributions of causal variants.    
Additionally, BLPC block-wise ridge estimators with proper tuning parameters can further improve the prediction accuracy over the BLPC marginal estimator. In our simulations, the $q/n$ tuning parameter or a slightly smaller  one (for example $0.1\times q/n$) works well. A larger tuning parameter (for example $10\times q/n$) may over-regularize the coefficients and 
result in worse performance than the BLPC marginal estimator.  
In summary, estimators built on block-wise local principal  components may outperform the traditional variant-based estimators in genetic data prediction. 
\subsection{UK Biobank real data analysis}\label{sec6.3}
We predict $36$ complex traits from different trait domains in the UK Biobank study, such as anthropometric traits, blood traits, cardiovascular traits, mental health traits, and cardiorespiratory traits (Supplementary Table~1).
Specifically, we select $350,000$ unrelated White British subjects as training samples. 
The average sample size for these complex traits is about $300,000$.  
The performance is mainly tested on $19,000$ unrelated White but non-British subjects. 
Based on the results of our simulation study, we keep the top-ranked BLPCs for each block that can explain more than $50\%$ of the genetic variance.  
The same set of covariates is adjusted in the training and testing data, including age, age-squared, sex, and the top 40 genetic principal components provided by the UK Biobank study \citep{bycroft2018uk}. 
The prediction accuracy on the testing samples is measured by the partial $R$-squared in  linear models, while adjusting for the covariates listed above. 
Similar to our simulation analysis, 
we evaluate and compare the following estimators
1) genetic variant-based  marginal  estimator  (Variant-marginal); 
2) genetic variant-based  marginal  estimator with pruned variants only (Variant-marginal-prune); 
3) genetic variant block-wise  reference  panel-based  ridge  estimator (Variant-block-wise-ridge-W);
4) BLPC-based  marginal  estimator (BLPC-marginal); and 
5) BLPC block-wise ridge estimator with $\lambda_0$ (BLPC-block-wise-ridge).
To explore whether increasing the block size can increase the prediction accuracy, we also evaluate a BLPC chromosome-wise ridge estimator (BLPC-chr-wise-ridge). 
Specifically, we treat each chromosome as a large block, group all the BLPCs within each chromosome, and estimate the within-chromosome covariance matrix from the training dataset. We then construct ridge estimators with $\lambda=\lambda_0$. 
\begin{figure}[t]
\includegraphics[page=2,width=0.8\linewidth]{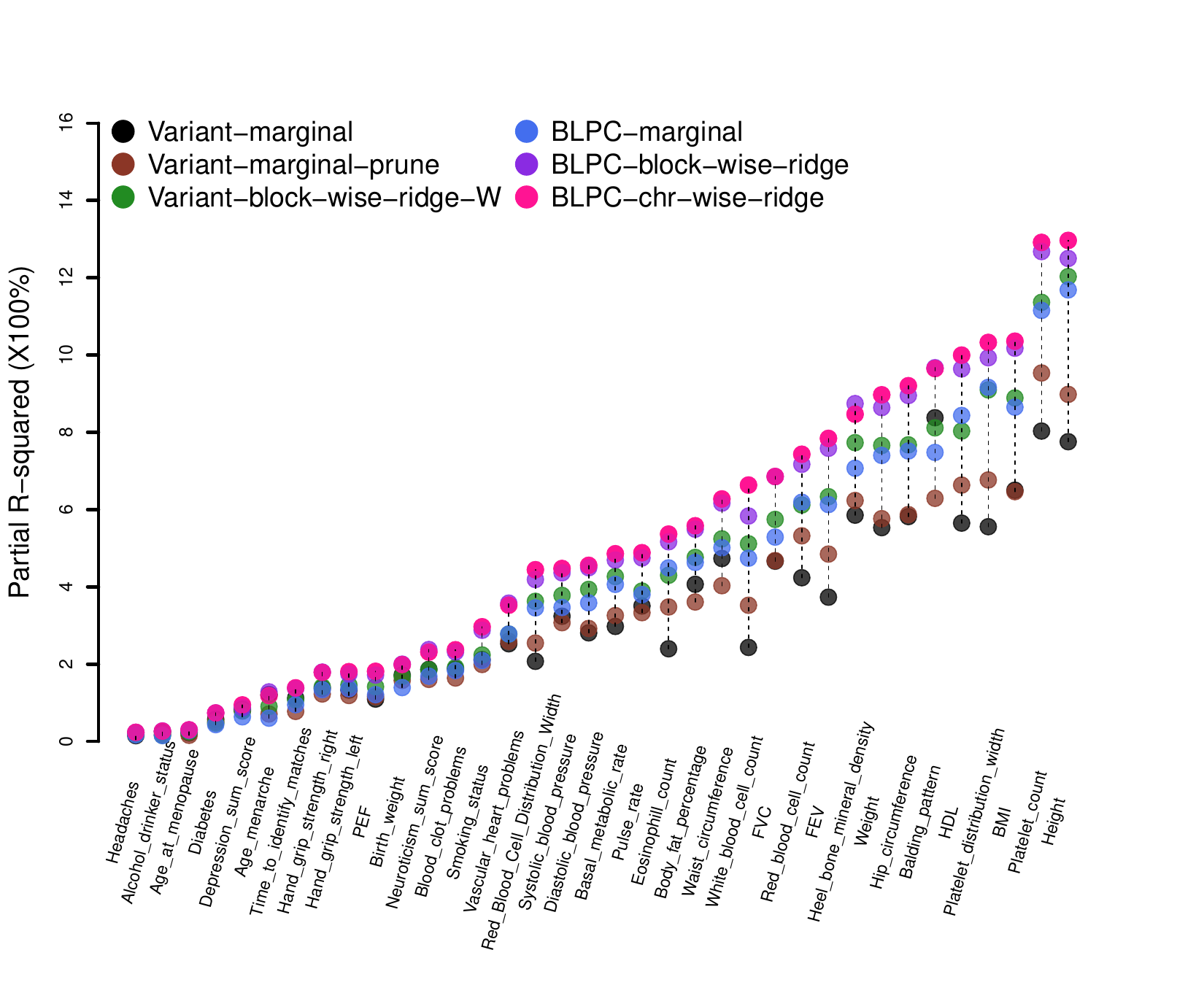}
  \caption{Prediction accuracy of different estimators in the UK Biobank data analysis.
  Variant-marginal: marginal screening with genetic variants; 
  Variant-marginal-prune: marginal screening with pruned genetic variants; 
  Variant-block-wise-ridge-W: block-wise ridge estimator with LDs being estimated from the 1000 Genomes reference panel; 
  BLPC-marginal: marginal screening with BLPCs;
  BLPC-block-wise-ridge: block-wise ridge estimator with BLPCs; and 
  BLPC-chr-wise-ridge: chromosome-wise ridge estimator with BLPCs. 
  More information of these complex traits can be found in the Supplementary Table~1. 
}
\label{fig5}
\end{figure}

Figure~\ref{fig5} illustrates the prediction accuracy of different estimators for the $36$ complex traits.
On average, the best of the three BLPC-based estimators can improve the performance of the best of the three variant-based estimators by $19.8\%$ (Supplementary Table~1). 
For example, the prediction accuracy of high-density lipoprotein can be improved from $8.03\%$ to $9.64\%$ by the BLPC block-wise ridge estimator, and from $8.03\%$ to $10.00\%$ by the BLPC chromosome-wise ridge estimator.
Across the three BLPC-based estimators, the block-wise ridge estimator could improve prediction accuracy by $29.9\%$ over the  marginal estimator, and the chromosome-wise ridge estimator could further improve prediction accuracy by $2\%$ over the block-wise ridge estimator. As chromosome-wise estimators are computationally expensive, these results suggest that block-wise estimators may be a good choice for BLPCs, which balance both prediction performance and computational cost.
In order to compare the performance of these estimators in cross-population prediction, we also tested them on $9,188$ subjects of Asian descent, including Indian, Pakistani, Bangladeshi, and Chinese.
The performance on this Asian testing dataset is summarized in Supplementary Figure~8. 
Although all of the methods have lower prediction accuracy in this Asian dataset, we find that BLPC-based estimators improve variant-based methods by  $22.9\%$, indicating that our BLPCs can be applied to trans-ancestry prediction.
Overall, the results of the UK Biobank data analysis are consistent with those of the simulation and many complex traits can be better predicted with BLPC-based methods.

\section{Discussion}\label{sec7}
There is a high demand for accurate genetic prediction of complex traits and diseases. 
Better genetic prediction will facilitate more widespread applications of precision medicine, such as accelerating the development of genetic medicine and identifying subgroups at higher risk of developing a particular clinical outcome
\citep{torkamani2018personal,martin2019clinical}.
Block-wise and reference panel-based ridge-type estimators are among the most popular methods for predicting genetic risk. 
This paper presents a unified framework for studying these estimators using random matrix theory. 
We show the asymptotic performance gap between block-wise and traditional ridge estimators and quantify the performance of reference panel-based estimators. 
We also examine BLPC-based estimators to better utilize block-diagonal structures of LD patterns, which can reduce genetic variant dimensions and aggregate weak genetic effects. Our theoretical results can also be applied to other high-dimensional random matrix problems and scientific fields with data that have a block-diagonal structure.  

In Condition~\ref{con3}, we assume the causal genetic effects are i.i.d random variables
$
\bmbeta_{(1)}
\sim F 
\left (
\begin{matrix} 
\bm{0_m}
\end{matrix},
\begin{matrix} 
p^{-1} \cdot\bmSigma_{\beta} 
\end{matrix}
\right ),$
where $\bmSigma_{\beta}=\sigma_{\beta}^2 \cdot \I_m$.
An alternative condition for $\bmbeta_{(1)}$ that allows variation of genetic effects could be the independent random effect assumption, in which $\bmSigma_{\beta}=\diagg(\sigma_{\beta_1}^2,\ldots,\sigma_{\beta_{m}}^2)$ with $\sigma_{\beta_i}^2$ being some constants, $i=1,\ldots,m$. 
The main results for i.i.d. random effects still hold under the independent random effects assumption when we use slightly different mild conditions in Conditions~\ref{con4},~\ref{con5}, and~\ref{con6}. The main reason is that the concentration of quadratic forms, such as the Lemma B.26 in \cite{bai2010spectral}, only need the random variables to be independent. The new versions of  Conditions~\ref{con4},~\ref{con5}, and~\ref{con6} under the independent random effects assumption can be found in the supplementary file. 

In this study, we examine the class of ridge-type estimators \citep{hoerl1970ridge} that are not restricted by signal sparsity.
Ridge-type estimators are suitable for weak and moderate genetic signals of polygenic complex traits  \citep{liu2020minimax}. It is also interesting to study sparsity-based penalized methods, such as the Lasso \citep{tibshirani1996regression,pattee2020penalized} or SCAD \citep{fan2001variable}, with block-wise data structures and reference panels. The Lassosum \citep{mak2017polygenic}, for example, is a popular block-wise reference panel-based Lasso-type estimator that has been widely applied to many complex traits in GWAS. 
In addition, further extensions can be made based on the BLPC estimators. 
For example, integration of biological pathways and functional information \citep{hu2017leveraging,marquez2020ldpred} may enhance the low-rank representations of genetic variants and performance of BLPC-based estimators. 
Finally, PCA-based methods rely on linear combinations of genetic variants. We might be able to model block-wise genetic variations in a more effective and flexible framework, such as the neural networks \citep{van2021gennet}.
\section*{Acknowledgement}
We would like to thank Ziliang Zhu, Fei Zou, and  Yue Yang for helpful discussions.
This research has been conducted using the UK Biobank resource (application number $22783$), subject to a data transfer agreement.
We thank the individuals represented in the UK Biobank for their participation and the research teams for their work in collecting, processing and disseminating these datasets for analysis. We would like to thank the University of North Carolina at Chapel Hill and Purdue University and their Research Computing groups for providing computational resources and support that have contributed to these research results. 


\bibliographystyle{rss}
\bibliography{sample.bib}

\begin{thebibliography}{48}
\expandafter\ifx\csname natexlab\endcsname\relax\def\natexlab#1{#1}\fi
\expandafter\ifx\csname url\endcsname\relax
  \def\url#1{\texttt{#1}}\fi
\expandafter\ifx\csname urlprefix\endcsname\relax\def\urlprefix{URL: }\fi

\bibitem[{1000-Genomes-Consortium(2015)}]{10002015global}
1000-Genomes-Consortium (2015) A global reference for human genetic variation.
\newblock \textit{Nature}, \textbf{526}, 68--74.

\bibitem[{Bai and Silverstein(2004)}]{bai2004clt}
Bai, Z. and Silverstein, J.~W. (2004) Clt for linear spectral statistics of
  large-dimensional sample covariance matrices.
\newblock \textit{The Annals of Probability}, \textbf{32}, 553--605.

\bibitem[{Bai and Silverstein(2010)}]{bai2010spectral}
--- (2010) \textit{Spectral analysis of large dimensional random matrices},
  vol.~20.
\newblock Springer.

\bibitem[{Bai and Zhou(2008)}]{bai2008large}
Bai, Z. and Zhou, W. (2008) Large sample covariance matrices without
  independence structures in columns.
\newblock \textit{Statistica Sinica}, \textbf{18}, 425--442.

\bibitem[{Berisa and Pickrell(2016)}]{berisa2016approximately}
Berisa, T. and Pickrell, J.~K. (2016) Approximately independent linkage
  disequilibrium blocks in human populations.
\newblock \textit{Bioinformatics}, \textbf{32}, 283--285.

\bibitem[{Bulik-Sullivan et~al.(2015)Bulik-Sullivan, Loh, Finucane, Ripke,
  Yang, Patterson, Daly, Price, Neale, of~the Psychiatric Genomics~Consortium
  et~al.}]{bulik2015ld}
Bulik-Sullivan, B.~K., Loh, P.-R., Finucane, H.~K., Ripke, S., Yang, J.,
  Patterson, N., Daly, M.~J., Price, A.~L., Neale, B.~M., of~the Psychiatric
  Genomics~Consortium, S. W.~G. et~al. (2015) Ld score regression distinguishes
  confounding from polygenicity in genome-wide association studies.
\newblock \textit{Nature Genetics}, \textbf{47}, 291--295.

\bibitem[{Bycroft et~al.(2018)Bycroft, Freeman, Petkova, Band, Elliott, Sharp,
  Motyer, Vukcevic, Delaneau, O'Connell et~al.}]{bycroft2018uk}
Bycroft, C., Freeman, C., Petkova, D., Band, G., Elliott, L., Sharp, K.,
  Motyer, A., Vukcevic, D., Delaneau, O., O'Connell, J. et~al. (2018) The uk
  biobank resource with deep phenotyping and genomic data.
\newblock \textit{Nature}, \textbf{562}, 203--209.

\bibitem[{Craig et~al.(2020)Craig, Han, Qassim, Hassall, Bailey, Kinzy,
  Khawaja, An, Marshall, Gharahkhani et~al.}]{craig2020multitrait}
Craig, J.~E., Han, X., Qassim, A., Hassall, M., Bailey, J. N.~C., Kinzy, T.~G.,
  Khawaja, A.~P., An, J., Marshall, H., Gharahkhani, P. et~al. (2020)
  Multitrait analysis of glaucoma identifies new risk loci and enables
  polygenic prediction of disease susceptibility and progression.
\newblock \textit{Nature Genetics}, \textbf{52}, 160--166.

\bibitem[{Dicker(2011)}]{dicker2011dense}
Dicker, L.~H. (2011) Dense signals, linear estimators, and out-of-sample
  prediction for high-dimensional linear models.
\newblock \textit{arXiv preprint arXiv:1102.2952}.

\bibitem[{Dobriban and Wager(2018)}]{dobriban2018high}
Dobriban, E. and Wager, S. (2018) High-dimensional asymptotics of prediction:
  Ridge regression and classification.
\newblock \textit{The Annals of Statistics}, \textbf{46}, 247--279.

\bibitem[{Fan and Li(2001)}]{fan2001variable}
Fan, J. and Li, R. (2001) Variable selection via nonconcave penalized
  likelihood and its oracle properties.
\newblock \textit{Journal of the American Statistical Association},
  \textbf{96}, 1348--1360.

\bibitem[{Finucane et~al.(2018)Finucane, Reshef, Anttila, Slowikowski, Gusev,
  Byrnes, Gazal, Loh, Lareau, Shoresh et~al.}]{finucane2018heritability}
Finucane, H.~K., Reshef, Y.~A., Anttila, V., Slowikowski, K., Gusev, A.,
  Byrnes, A., Gazal, S., Loh, P.-R., Lareau, C., Shoresh, N. et~al. (2018)
  Heritability enrichment of specifically expressed genes identifies
  disease-relevant tissues and cell types.
\newblock \textit{Nature Genetics}, \textbf{50}, 621--629.

\bibitem[{Fritsche et~al.(2020)Fritsche, Patil, Beesley, VandeHaar, Salvatore,
  Ma, Peng, Taliun, Zhou and Mukherjee}]{fritsche2020cancer}
Fritsche, L.~G., Patil, S., Beesley, L.~J., VandeHaar, P., Salvatore, M., Ma,
  Y., Peng, R.~B., Taliun, D., Zhou, X. and Mukherjee, B. (2020) Cancer prsweb:
  an online repository with polygenic risk scores for major cancer traits and
  their evaluation in two independent biobanks.
\newblock \textit{The American Journal of Human Genetics}, \textbf{107},
  815--836.

\bibitem[{Ge et~al.(2019)Ge, Chen, Ni, Feng and Smoller}]{ge2019polygenic}
Ge, T., Chen, C.-Y., Ni, Y., Feng, Y.-C.~A. and Smoller, J.~W. (2019) Polygenic
  prediction via bayesian regression and continuous shrinkage priors.
\newblock \textit{Nature Communications}, \textbf{10}, 1--10.

\bibitem[{HapMap3-Consortium(2010)}]{international2010integrating}
HapMap3-Consortium (2010) Integrating common and rare genetic variation in
  diverse human populations.
\newblock \textit{Nature}, \textbf{467}, 52--58.

\bibitem[{van Hilten et~al.(2021)van Hilten, Kushner, Kayser, Arfan~Ikram,
  Adams, Klaver, Niessen and Roshchupkin}]{van2021gennet}
van Hilten, A., Kushner, S.~A., Kayser, M., Arfan~Ikram, M., Adams, H.~H.,
  Klaver, C.~C., Niessen, W.~J. and Roshchupkin, G.~V. (2021) Gennet framework:
  interpretable deep learning for predicting phenotypes from genetic data.
\newblock \textit{Communications Biology}, \textbf{4}, 1--9.

\bibitem[{Hoerl and Kennard(1970)}]{hoerl1970ridge}
Hoerl, A.~E. and Kennard, R.~W. (1970) Ridge regression: Biased estimation for
  nonorthogonal problems.
\newblock \textit{Technometrics}, \textbf{12}, 55--67.

\bibitem[{Hu et~al.(2017)Hu, Lu, Powles, Yao, Yang, Fang, Xu and
  Zhao}]{hu2017leveraging}
Hu, Y., Lu, Q., Powles, R., Yao, X., Yang, C., Fang, F., Xu, X. and Zhao, H.
  (2017) Leveraging functional annotations in genetic risk prediction for human
  complex diseases.
\newblock \textit{PLoS Computational Biology}, \textbf{13}, e1005589.

\bibitem[{Jiang et~al.(2016)Jiang, Li, Paul, Yang and Zhao}]{jiang2016high}
Jiang, J., Li, C., Paul, D., Yang, C. and Zhao, H. (2016) On high-dimensional
  misspecified mixed model analysis in genome-wide association study.
\newblock \textit{The Annals of Statistics}, \textbf{44}, 2127--2160.

\bibitem[{Liu et~al.(2021)Liu, Peng, Liao, Locascio, Corvol, Zhu, Dong,
  Maple-Gr{\o}dem, Campbell, Elbaz et~al.}]{liu2021genome}
Liu, G., Peng, J., Liao, Z., Locascio, J.~J., Corvol, J.-C., Zhu, F., Dong, X.,
  Maple-Gr{\o}dem, J., Campbell, M.~C., Elbaz, A. et~al. (2021) Genome-wide
  survival study identifies a novel synaptic locus and polygenic score for
  cognitive progression in parkinson’s disease.
\newblock \textit{Nature Genetics}, \textbf{53}, 787--793.

\bibitem[{Liu et~al.(2020)Liu, Li and Lin}]{liu2020minimax}
Liu, Y., Li, Z. and Lin, X. (2020) A minimax optimal ridge-type set test for
  global hypothesis with applications in whole genome sequencing association
  studies.
\newblock \textit{Journal of the American Statistical Association}, in press.

\bibitem[{Lloyd-Jones et~al.(2019)Lloyd-Jones, Zeng, Sidorenko, Yengo, Moser,
  Kemper, Wang, Zheng, Magi, Esko et~al.}]{lloyd2019improved}
Lloyd-Jones, L.~R., Zeng, J., Sidorenko, J., Yengo, L., Moser, G., Kemper,
  K.~E., Wang, H., Zheng, Z., Magi, R., Esko, T. et~al. (2019) Improved
  polygenic prediction by bayesian multiple regression on summary statistics.
\newblock \textit{Nature Communications}, \textbf{10}, 1--11.

\bibitem[{Mak et~al.(2017)Mak, Porsch, Choi, Zhou and Sham}]{mak2017polygenic}
Mak, T. S.~H., Porsch, R.~M., Choi, S.~W., Zhou, X. and Sham, P.~C. (2017)
  Polygenic scores via penalized regression on summary statistics.
\newblock \textit{Genetic Epidemiology}, \textbf{41}, 469--480.

\bibitem[{Marchenko and Pastur(1967)}]{marchenko1967distribution}
Marchenko, V.~A. and Pastur, L.~A. (1967) Distribution of eigenvalues for some
  sets of random matrices.
\newblock \textit{Matematicheskii Sbornik}, \textbf{114}, 507--536.

\bibitem[{Marquez-Luna et~al.(2020)Marquez-Luna, Gazal, Loh, Kim, Furlotte,
  Auton, Price, 23andMe Research~Team et~al.}]{marquez2020ldpred}
Marquez-Luna, C., Gazal, S., Loh, P.-R., Kim, S.~S., Furlotte, N., Auton, A.,
  Price, A.~L., 23andMe Research~Team et~al. (2020) Ldpred-funct: incorporating
  functional priors improves polygenic prediction accuracy in uk biobank and
  23andme data sets.
\newblock \textit{bioRxiv}, 375337.

\bibitem[{Martin et~al.(2019)Martin, Kanai, Kamatani, Okada, Neale and
  Daly}]{martin2019clinical}
Martin, A.~R., Kanai, M., Kamatani, Y., Okada, Y., Neale, B.~M. and Daly, M.~J.
  (2019) Clinical use of current polygenic risk scores may exacerbate health
  disparities.
\newblock \textit{Nature Genetics}, \textbf{51}, 584--591.

\bibitem[{Pain et~al.(2021)Pain, Glanville, Hagenaars, Selzam, F{\"u}rtjes,
  Gaspar, Coleman, Rimfeld, Breen, Plomin et~al.}]{pain2021evaluation}
Pain, O., Glanville, K.~P., Hagenaars, S.~P., Selzam, S., F{\"u}rtjes, A.~E.,
  Gaspar, H.~A., Coleman, J.~R., Rimfeld, K., Breen, G., Plomin, R. et~al.
  (2021) Evaluation of polygenic prediction methodology within a
  reference-standardized framework.
\newblock \textit{PLoS Genetics}, \textbf{17}, e1009021.

\bibitem[{Pasaniuc and Price(2017)}]{pasaniuc2017dissecting}
Pasaniuc, B. and Price, A.~L. (2017) Dissecting the genetics of complex traits
  using summary association statistics.
\newblock \textit{Nature Reviews Genetics}, \textbf{18}, 117--127.

\bibitem[{Pattee and Pan(2020)}]{pattee2020penalized}
Pattee, J. and Pan, W. (2020) Penalized regression and model selection methods
  for polygenic scores on summary statistics.
\newblock \textit{PLoS Computational Biology}, \textbf{16}, e1008271.

\bibitem[{Qian et~al.(2020)Qian, Tanigawa, Du, Aguirre, Chang, Tibshirani,
  Rivas and Hastie}]{qian2020fast}
Qian, J., Tanigawa, Y., Du, W., Aguirre, M., Chang, C., Tibshirani, R., Rivas,
  M.~A. and Hastie, T. (2020) A fast and scalable framework for large-scale and
  ultrahigh-dimensional sparse regression with application to the uk biobank.
\newblock \textit{PLoS Genetics}, \textbf{16}, e1009141.

\bibitem[{Serdobolskii(2007)}]{serdobolskii2007multiparametric}
Serdobolskii, V.~I. (2007) \textit{Multiparametric statistics}.
\newblock Elsevier.

\bibitem[{Sheng and Dobriban(2020)}]{sheng2020one}
Sheng, Y. and Dobriban, E. (2020) One-shot distributed ridge regression in high
  dimensions.
\newblock In \textit{International Conference on Machine Learning}, 8763--8772.
  PMLR.

\bibitem[{Silverstein(1995)}]{silverstein1995strong}
Silverstein, J.~W. (1995) Strong convergence of the empirical distribution of
  eigenvalues of large dimensional random matrices.
\newblock \textit{Journal of Multivariate Analysis}, \textbf{55}, 331--339.

\bibitem[{Song et~al.(2020)Song, Jiang, Hou and Zhao}]{song2020leveraging}
Song, S., Jiang, W., Hou, L. and Zhao, H. (2020) Leveraging effect size
  distributions to improve polygenic risk scores derived from summary
  statistics of genome-wide association studies.
\newblock \textit{PLOS Computational Biology}, \textbf{16}, e1007565.

\bibitem[{Taliun et~al.(2021)Taliun, Harris, Kessler, Carlson, Szpiech, Torres,
  Taliun, Corvelo, Gogarten, Kang et~al.}]{taliun2021sequencing}
Taliun, D., Harris, D.~N., Kessler, M.~D., Carlson, J., Szpiech, Z.~A., Torres,
  R., Taliun, S. A.~G., Corvelo, A., Gogarten, S.~M., Kang, H.~M. et~al. (2021)
  Sequencing of 53,831 diverse genomes from the nhlbi topmed program.
\newblock \textit{Nature}, \textbf{590}, 290--299.

\bibitem[{Tibshirani(1996)}]{tibshirani1996regression}
Tibshirani, R. (1996) Regression shrinkage and selection via the lasso.
\newblock \textit{Journal of the Royal Statistical Society. Series B
  (Methodological)}, \textbf{58}, 267--288.

\bibitem[{Timpson et~al.(2018)Timpson, Greenwood, Soranzo, Lawson and
  Richards}]{timpson2018genetic}
Timpson, N.~J., Greenwood, C.~M., Soranzo, N., Lawson, D.~J. and Richards,
  J.~B. (2018) Genetic architecture: the shape of the genetic contribution to
  human traits and disease.
\newblock \textit{Nature Reviews Genetics}, \textbf{19}, 110--125.

\bibitem[{Torkamani et~al.(2018)Torkamani, Wineinger and
  Topol}]{torkamani2018personal}
Torkamani, A., Wineinger, N.~E. and Topol, E.~J. (2018) The personal and
  clinical utility of polygenic risk scores.
\newblock \textit{Nature Reviews Genetics}, \textbf{19}, 581--590.

\bibitem[{UK10K-Consortium(2015)}]{uk10k2015uk10k}
UK10K-Consortium (2015) The uk10k project identifies rare variants in health
  and disease.
\newblock \textit{Nature}, \textbf{526}, 82.

\bibitem[{Vilhj{\'a}lmsson et~al.(2015)Vilhj{\'a}lmsson, Yang, Finucane, Gusev,
  Lindstr{\"o}m, Ripke, Genovese, Loh, Bhatia, Do
  et~al.}]{vilhjalmsson2015modeling}
Vilhj{\'a}lmsson, B.~J., Yang, J., Finucane, H.~K., Gusev, A., Lindstr{\"o}m,
  S., Ripke, S., Genovese, G., Loh, P.-R., Bhatia, G., Do, R. et~al. (2015)
  Modeling linkage disequilibrium increases accuracy of polygenic risk scores.
\newblock \textit{The American Journal of Human Genetics}, \textbf{97},
  576--592.

\bibitem[{Visscher et~al.(2017)Visscher, Wray, Zhang, Sklar, McCarthy, Brown
  and Yang}]{visscher201710}
Visscher, P.~M., Wray, N.~R., Zhang, Q., Sklar, P., McCarthy, M.~I., Brown,
  M.~A. and Yang, J. (2017) 10 years of gwas discovery: biology, function, and
  translation.
\newblock \textit{The American Journal of Human Genetics}, \textbf{101}, 5--22.

\bibitem[{Wang et~al.(2021)Wang, Wang and Li}]{wang2021testing}
Wang, J., Wang, W. and Li, H. (2021) Sparse block signal detection and
  identification for shared cross-trait association analysis.
\newblock \textit{The Annals of Applied Statistics}, in press.

\bibitem[{Yang et~al.(2011)Yang, Lee, Goddard and Visscher}]{yang2011gcta}
Yang, J., Lee, S.~H., Goddard, M.~E. and Visscher, P.~M. (2011) Gcta: a tool
  for genome-wide complex trait analysis.
\newblock \textit{The American Journal of Human Genetics}, \textbf{88}, 76--82.

\bibitem[{Yang and Zhou(2020)}]{yang2020accurate}
Yang, S. and Zhou, X. (2020) Accurate and scalable construction of polygenic
  scores in large biobank data sets.
\newblock \textit{The American Journal of Human Genetics}, \textbf{106},
  679--693.

\bibitem[{Yao et~al.(2015)Yao, Zheng and Bai}]{yao2015sample}
Yao, J., Zheng, S. and Bai, Z. (2015) \textit{Sample covariance matrices and
  high-dimensional data analysis}, vol.~2.
\newblock Cambridge University Press Cambridge.

\bibitem[{Zhao and Zhu(2019)}]{zhao2019cross}
Zhao, B. and Zhu, H. (2019) Cross-trait prediction accuracy of high-dimensional
  ridge-type estimators in genome-wide association studies.
\newblock \textit{arXiv preprint arXiv:1911.10142}.

\bibitem[{Zhao and Zou(2021)}]{zhao2021polygenic}
Zhao, B. and Zou, F. (2021) On polygenic risk scores for complex traits
  prediction.
\newblock \textit{Biometrics}, in press.

\bibitem[{Zhou et~al.(2021)Zhou, GBM-Initiative et~al.}]{zhou2021global}
Zhou, W., GBM-Initiative et~al. (2021) Global biobank meta-analysis initiative:
  Powering genetic discovery across human diseases.
\newblock \textit{medRxiv}.

\end{thebibliography}

\end{document}


\maketitle
\date{\vspace{-5ex}}
\date{}
\section{Proofs}\label{sec1}
\subsection{Theorems~1~and~2 (block-diagonal matrix)}
Under the conditions listed in this paper, we first define some additional notations
\begin{itemize}
\item $c_{n\ell}=p_{\ell}/n\rightarrow c_{\ell}\in(0, \infty)$, $\ell=1,...,K$;
\item $b_{p\ell}=p_{\ell}/p\rightarrow b_{\ell}\in(0, \infty)$, $\ell=1,...,K$;
\item $\bx_i=\bSig^{1/2}{\bf w}_i$, $i=1,...,n$; and 
\item $\bSig_{\ell\ell}$ is the $l$th diagonal block of $\bSig$, which 
has the limiting spectral distribution $H_{\ell}(t)$, $\ell=1,...,K$.
\end{itemize}
The first goal is to obtain  the limits of 
\begin{flalign*}
T_1(\lambda)=p^{-1}\tr\{(\widehat{\bmSigma}-\widehat{\bmSigma}_{B})(\widehat{\bmSigma}_{B}+\lambda\I_p)^{-1}\bmSigma\}\mbox{,}\quad T_2(\lambda)=p^{-1}\tr\{(\widehat{\bmSigma}-\widehat{\bmSigma}_{B})(\widehat{\bmSigma}_{B}+\lambda\I_p)^{-2}\bmSigma\}
\end{flalign*}
and 
\begin{flalign*}
T_3(\lambda)=p^{-1}\tr\{(\widehat{\bmSigma}-\widehat{\bmSigma}_{B})(\widehat{\bmSigma}_{B}+\lambda\I_p)^{-1}\bmSigma(\widehat{\bmSigma}-\widehat{\bmSigma}_{B})(\widehat{\bmSigma}_{B}+\lambda\I_p)^{-1}\}. 
\end{flalign*}
Let $p_{\ell}$ be the dimension of the $\ell$th block of $\bx_i$ and $\tilde{p}_{\ell}=p_1+...+p_{\ell}$. Let
$\bx_i=(\bx_{i1},...,\bx_{iK})^T$ with $\bx_{i\ell}=(x_{i,\tilde{p}_{\ell-1}+1},...,x_{i\tilde{p}_{\ell}})$,
$\br_{i\ell}=n^{-1/2}\bx_{i\ell}$ and $\bx_{i\ell}=\bSig_{\ell\ell}^{1/2}{\bf w}_{i\ell}$.
Following \cite{bai2004clt}, it can be proved that
$T_1(\lambda)$, $T_2(\lambda)$, and $T_3(\lambda)$ have the same limits for $x_{ij}$ and 
\begin{eqnarray}
\hat{x}_{ij}=\frac{x_{ij}I_{|x_{ij}|\leq\eta_n\sqrt{n}}-\rE x_{ij}I_{|x_{ij}|\leq\eta_n\sqrt{n}}}
{{\rm Var}(x_{ij}I_{|x_{ij}|\leq\eta_n\sqrt{n}})}. \label{E0}
\end{eqnarray}
Without loss of generality, we will replace $x_{ij}$ by {\bxz$\hat{x}_{ij}$} in our proof. For simplicity,
we still use the notation $x_{ij}$. Then let 
\begin{eqnarray*}
\tilde{T}_3(\lambda)&=&n^{-1}\tr\{(\hSig-\hSig_B)(\hSig_B+\lambda\bI_p)^{-1}\bSig(\hSig-\hSig_B)(\hSig_B+\lambda\bI_p)^{-1}\}\\
            &=&n^{-1}\tr\{\bSig[(\hSig-\hSig_B)(\hSig_B+\lambda\bI_p)^{-1}]^2\}.
\end{eqnarray*}
Moreover, we have
$$
\hSig-\hSig_B=\left(\begin{array}{cccc}
0&{\bf A}_{12}&\cdots&{\bf A}_{1K}\\
\vdots&\vdots&\vdots&\vdots\\
{\bf A}_{K2}&{\bf A}_{K1}&\cdots&0
\end{array}\right)
$$
and
\begin{eqnarray*}
(\hSig_B+\lambda\bI_p)^{-1}=\left(
\begin{array}{cccc}
\hSig_{B1i}^{-1}-\frac{\hSig_{B1i}^{-1}\br_{i1}\br_{i1}^T\hSig_{B1i}^{-1}}{1+\br_{i1}^T\hSig_{B1i}^{-1}\br_{i1}}&0&\cdots&0\\
0&\hSig_{B2i}^{-1}-\frac{\hSig_{B2i}^{-1}\br_{i2}\br_{i2}^T\hSig_{B2i}^{-1}}{1+\br_{i2}^T\hSig_{B2i}^{-1}\br_{i2}}&0&0\\
\vdots&\vdots&\vdots&\vdots\\
0&0&\cdots&\hSig_{BKi}^{-1}-\frac{\hSig_{BKi}^{-1}\br_{iK}\br_{iK}^T\hSig_{BKi}^{-1}}{1+\br_{iK}^T\hSig_{BKi}^{-1}\br_{iK}}
\end{array}
\right),
\end{eqnarray*}
where $$\hSig_{B\ell}^{-1}=(\sum_{i=1}^n\br_{i\ell}\br_{i\ell}^T+\lambda\bI_{p_{\ell}})^{-1}
=\hSig_{B\ell i}^{-1}-\frac{\hSig_{B\ell i}^{-1}\br_{i\ell}\br_{i\ell}^T\hSig_{B\ell i}^{-1}}{1+\br_{i\ell}^T\hSig_{B\ell i}^{-1}\br_{i\ell}}$$
with $\hSig_{B\ell i}=\sum_{j\not=i}\br_{j\ell}\br_{j\ell}^T+\lambda\bI_{p_{\ell}}.$
Then we have
\begin{eqnarray*}
&&(\hSig-\hSig_B)(\hSig_B+\lambda\bI_p)^{-1}\\
&=&\left(
\begin{array}{cccc}
0&{\bf A}_{12}\hSig_{B2i}^{-1}-\frac{{\bf A}_{12}\hSig_{B2i}^{-1}\br_{i2}\br_{i2}^T\hSig_{B2i}^{-1}}{1+\br_{i2}^T\hSig_{B2i}^{-1}\br_{i2}}&\cdots&
{\bf A}_{1K}\hSig_{BKi}^{-1}-\frac{{\bf A}_{1K}\hSig_{BKi}^{-1}\br_{iK}\br_{iK}^T\hSig_{BKi}^{-1}}{1+\br_{iK}^T\hSig_{BKi}^{-1}\br_{iK}}\\
\vdots&\vdots&\vdots&\vdots\\
{\bf A}_{K1}\hSig_{B1i}^{-1}-\frac{{\bf A}_{K1}\hSig_{B1i}^{-1}\br_{i1}\br_{i1}^T\hSig_{B1i}^{-1}}{1+\br_{i1}^T\hSig_{B1i}^{-1}\br_{i1}}
&{\bf A}_{K2}\hSig_{B2i}^{-1}-\frac{{\bf A}_{K2}\hSig_{B2i}^{-1}\br_{i2}\br_{i2}^T\hSig_{B2i}^{-1}}{1+\br_{i2}^T\hSig_{B2i}^{-1}\br_{i2}}&\cdots&0
\end{array}
\right)\\
&=&\sum_{i=1}^n\left(
\begin{array}{cccc}
0&\frac{\br_{i1}\br_{i2}^T\hSig_{B2i}^{-1}}{1+\br_{i2}^T\hSig_{B2i}^{-1}\br_{i2}}&\cdots&\frac{\br_{i1}\br_{iK}^T\hSig_{BKi}^{-1}}{1+\br_{iK}^T\hSig_{BKi}^{-1}\br_{iK}}\\
\vdots&\vdots&\vdots&\vdots\\
\frac{\br_{iK}\br_{i1}^T\hSig_{B1i}^{-1}}{1+\br_{i1}^T\hSig_{B1i}^{-1}\br_{i1}}&\frac{\br_{iK}\br_{i2}^T\hSig_{B2i}^{-1}}{1+\br_{i2}^T\hSig_{B2i}^{-1}\br_{i2}}
&\cdots&0
\end{array}
\right).
\end{eqnarray*}
It is easy to see that $T_1(\lambda)=T_2(\lambda)=0$. 
For $\tilde{T}_3(\lambda)$, we have 
\begin{eqnarray*}
&\tilde{T}_3(\lambda)=&n^{-1}\tr\{\bSig[(\hSig-\hSig_B)(\hSig_B+\lambda\bI_p)^{-1}]^2\}\\
&=&n^{-1}\tr\Big\{\sum_{i=1}^n\sum_{j=1}^n\bSig\left(
\begin{array}{cccc}
0&\frac{\br_{i1}\br_{i2}^T\hSig_{B2i}^{-1}}{1+\br_{i2}^T\hSig_{B2i}^{-1}\br_{i2}}&\cdots&\frac{\br_{i1}\br_{iK}^T\hSig_{BKi}^{-1}}{1+\br_{iK}^T\hSig_{BKi}^{-1}\br_{iK}}\\
\vdots&\vdots&\vdots&\vdots\\
\frac{\br_{iK}\br_{i1}^T\hSig_{B1i}^{-1}}{1+\br_{i1}^T\hSig_{B1i}^{-1}\br_{i1}}&\frac{\br_{iK}\br_{i2}^T\hSig_{B2i}^{-1}}{1+\br_{i2}^T\hSig_{B2i}^{-1}\br_{i2}}
&\cdots&0
\end{array}
\right)\\
& &\cdot\left(
\begin{array}{cccc}
0&\frac{\br_{j1}\br_{j2}^T\hSig_{B2j}^{-1}}{1+\br_{j2}^T\hSig_{B2j}^{-1}\br_{j2}}&\cdots
&\frac{\br_{j1}\br_{jK}^T\hSig_{BKj}^{-1}}{1+\br_{jK}^T\hSig_{BKj}^{-1}\br_{jK}}\\
\vdots&\vdots&\vdots&\vdots\\
\frac{\br_{jK}\br_{j1}^T\hSig_{B1j}^{-1}}{1+\br_{j1}^T\hSig_{B1j}^{-1}\br_{j1}}
&\frac{\br_{jK}\br_{j2}^T\hSig_{B2j}^{-1}}{1+\br_{j2}^T\hSig_{B2j}^{-1}\br_{j2}}&\cdots&0\\
\end{array}
\right)\Big\}\\
&=&n^{-1}\tr\Big\{\sum_{i=1}^n
\bSig\left(
\begin{array}{cccc}
0&\frac{\br_{i1}\br_{i2}^T\hSig_{B2i}^{-1}}{1+\br_{i2}^T\hSig_{B2i}^{-1}\br_{i2}}&\cdots&\frac{\br_{i1}\br_{iK}^T\hSig_{BKi}^{-1}}{1+\br_{iK}^T\hSig_{BKi}^{-1}\br_{iK}}\\
\vdots&\vdots&\vdots&\vdots\\
\frac{\br_{iK}\br_{i1}^T\hSig_{B1i}^{-1}}{1+\br_{i1}^T\hSig_{B1i}^{-1}\br_{i1}}&\frac{\br_{iK}\br_{i2}^T\hSig_{B2i}^{-1}}{1+\br_{i2}^T\hSig_{B2i}^{-1}\br_{i2}}
&\cdots&0
\end{array}
\right)\\
& &\cdot\left(
\begin{array}{cccc}
0&\frac{\br_{i1}\br_{i2}^T\hSig_{B2i}^{-1}}{1+\br_{i2}^T\hSig_{B2i}^{-1}\br_{i2}}&\cdots
&\frac{\br_{i1}\br_{iK}^T\hSig_{BKi}^{-1}}{1+\br_{iK}^T\hSig_{BKi}^{-1}\br_{iK}}\\
\vdots&\vdots&\vdots&\vdots\\
\frac{\br_{iK}\br_{i1}^T\hSig_{B1i}^{-1}}{1+\br_{i1}^T\hSig_{B1i}^{-1}\br_{i1}}
&\frac{\br_{iK}\br_{i2}^T\hSig_{B2i}^{-1}}{1+\br_{i2}^T\hSig_{B2i}^{-1}\br_{i2}}&\cdots&0\\
\end{array}
\right)\Big\}\\
& &+n^{-1}\tr\Big\{\sum_{i\not=j}\bSig\left(
\begin{array}{cccc}
0&\frac{\br_{i1}\br_{i2}^T\hSig_{B2i}^{-1}}{1+\br_{i2}^T\hSig_{B2i}^{-1}\br_{i2}}&\cdots&\frac{\br_{i1}\br_{iK}^T\hSig_{BKi}^{-1}}{1+\br_{iK}^T\hSig_{BKi}^{-1}\br_{iK}}\\
\vdots&\vdots&\vdots&\vdots\\
\frac{\br_{iK}\br_{i1}^T\hSig_{B1i}^{-1}}{1+\br_{i1}^T\hSig_{B1i}^{-1}\br_{i1}}&\frac{\br_{iK}\br_{i2}^T\hSig_{B2i}^{-1}}{1+\br_{i2}^T\hSig_{B2i}^{-1}\br_{i2}}
&\cdots&0
\end{array}
\right)\\
& &\cdot\left(
\begin{array}{cccc}
0&\frac{\br_{j1}\br_{j2}^T\hSig_{B2j}^{-1}}{1+\br_{j2}^T\hSig_{B2j}^{-1}\br_{j2}}&\cdots
&\frac{\br_{j1}\br_{jK}^T\hSig_{BKj}^{-1}}{1+\br_{jK}^T\hSig_{BKj}^{-1}\br_{jK}}\\
\vdots&\vdots&\vdots&\vdots\\
\frac{\br_{jK}\br_{j1}^T\hSig_{B1j}^{-1}}{1+\br_{j1}^T\hSig_{B1j}^{-1}\br_{j1}}
&\frac{\br_{jK}\br_{j2}^T\hSig_{B2j}^{-1}}{1+\br_{j2}^T\hSig_{B2j}^{-1}\br_{j2}}&\cdots&0\\
\end{array}
\right)\Big\}\\
&=&n^{-1}\tr\Big\{\sum_{i=1}^n
\left(
\begin{array}{cccc}
0&\frac{\bSig_{11}\br_{i1}\br_{i2}^T\hSig_{B2i}^{-1}}{1+\br_{i2}^T\hSig_{B2i}^{-1}\br_{i2}}&\cdots&\frac{\bSig_{11}\br_{i1}\br_{iK}^T\hSig_{BKi}^{-1}}{1+\br_{iK}^T\hSig_{BKi}^{-1}\br_{iK}}\\
\vdots&\vdots&\vdots&\vdots\\
\frac{\bSig_{KK}\br_{iK}\br_{i1}^T\hSig_{B1i}^{-1}}{1+\br_{i1}^T\hSig_{B1i}^{-1}\br_{i1}}
&\frac{\bSig_{KK}\br_{iK}\br_{i2}^T\hSig_{B2i}^{-1}}{1+\br_{i2}^T\hSig_{B2i}^{-1}\br_{i2}}
&\cdots&0
\end{array}
\right)\\
& &\cdot\left(
\begin{array}{cccc}
0&\frac{\br_{i1}\br_{i2}^T\hSig_{B2i}^{-1}}{1+\br_{i2}^T\hSig_{B2i}^{-1}\br_{i2}}&\cdots
&\frac{\br_{i1}\br_{iK}^T\hSig_{BKi}^{-1}}{1+\br_{iK}^T\hSig_{BKi}^{-1}\br_{iK}}\\
\vdots&\vdots&\vdots&\vdots\\
\frac{\br_{iK}\br_{i1}^T\hSig_{B1i}^{-1}}{1+\br_{i1}^T\hSig_{B1i}^{-1}\br_{i1}}
&\frac{\br_{iK}\br_{i2}^T\hSig_{B2i}^{-1}}{1+\br_{i2}^T\hSig_{B2i}^{-1}\br_{i2}}&\cdots&0\\
\end{array}
\right)\Big\}\\
& &+n^{-1}\tr\Big\{\sum_{i\not=j}\left(
\begin{array}{cccc}
0&\frac{\bSig_{11}\br_{i1}\br_{i2}^T\hSig_{B2i}^{-1}}{1+\br_{i2}^T\hSig_{B2i}^{-1}\br_{i2}}&\cdots&\frac{\bSig_{11}\br_{i1}\br_{iK}^T\hSig_{BKi}^{-1}}{1+\br_{iK}^T\hSig_{BKi}^{-1}\br_{iK}}\\
\vdots&\vdots&\vdots&\vdots\\
\frac{\bSig_{KK}\br_{iK}\br_{i1}^T\hSig_{B1i}^{-1}}{1+\br_{i1}^T\hSig_{B1i}^{-1}\br_{i1}}&\frac{\bSig_{KK}\br_{iK}\br_{i2}^T\hSig_{B2i}^{-1}}{1+\br_{i2}^T\hSig_{B2i}^{-1}\br_{i2}}
&\cdots&0
\end{array}
\right)\\
& &\cdot\left(
\begin{array}{cccc}
0&\frac{\br_{j1}\br_{j2}^T\hSig_{B2j}^{-1}}{1+\br_{j2}^T\hSig_{B2j}^{-1}\br_{j2}}&\cdots
&\frac{\br_{j1}\br_{jK}^T\hSig_{BKj}^{-1}}{1+\br_{jK}^T\hSig_{BKj}^{-1}\br_{jK}}\\
\vdots&\vdots&\vdots&\vdots\\
\frac{\br_{jK}\br_{j1}^T\hSig_{B1j}^{-1}}{1+\br_{j1}^T\hSig_{B1j}^{-1}\br_{j1}}
&\frac{\br_{jK}\br_{j2}^T\hSig_{B2j}^{-1}}{1+\br_{j2}^T\hSig_{B2j}^{-1}\br_{j2}}&\cdots&0\\
\end{array}
\right)\Big\}\\
&=&n^{-1}\sum_{j=1}^n\frac{\br_{j2}^T\hSig_{B2j}^{-1}\br_{j2}\cdot\br_{j1}^T\hSig_{B1j}^{-1}\bSig_{11}\br_{j1}}
{(1+\br_{j2}^T\hSig_{B2j}^{-1}\br_{j2})\cdot(1+\br_{j1}^T\hSig_{B1j}^{-1}\br_{j1})}\\
& &+\cdots+n^{-1}\sum_{{\bxz j=1}}^n\frac{\br_{jK}^T\hSig_{BKj}^{-1}\br_{jK}\cdot\br_{j1}^T\hSig_{B1j}^{-1}\bSig_{11}\br_{j1}}
{(1+\br_{jK}^T\hSig_{BKj}^{-1}\br_{jK})\cdot(1+\br_{j1}^T\hSig_{B1j}^{-1}\br_{j1})}+\cdots\\
& &+n^{-1}\sum_{j=1}^n\frac{\br_{j1}^T\hSig_{B1j}^{-1}\br_{j1}\cdot\br_{jK}^T\hSig_{BKj}^{-1}\bSig_{KK}\br_{jK}}
{(1+\br_{j1}^T\hSig_{B1j}^{-1}\br_{j1})\cdot(1+\br_{jK}^T\hSig_{BKj}^{-1}\br_{jK})}\\
& &+\cdots+n^{-1}\sum_{{\bxz j=1}}^n\frac{\br_{j,K-1}^T\hSig_{B,K-1,j}^{-1}\br_{j,K-1}\cdot\br_{jK}^T\hSig_{BKj}^{-1}\bSig_{KK}\br_{jK}}
{(1+\br_{j,K-1}^T\hSig_{B,K-1,j}^{-1}\br_{j,K-1})\cdot(1+\br_{jK}^T\hSig_{BKj}^{-1}\br_{jK})}\\
& &+n^{-1}\sum_{i\not=j}\frac{\br_{i2}^T\hSig_{B2i}^{-1}\br_{j2}\cdot\br_{j1}^T\hSig_{B1j}^{-1}\bSig_{11}\br_{i1}}
{(1+\br_{i2}^T\hSig_{B2i}^{-1}\br_{i2})\cdot(1+\br_{j1}^T\hSig_{B1j}^{-1}\br_{j1})}\\
& &+\cdots+n^{-1}\sum_{i\not=j}\frac{\br_{iK}^T\hSig_{BKi}^{-1}\br_{jK}\cdot\br_{j1}^T\hSig_{B1j}^{-1}\bSig_{11}\br_{i1}}
{(1+\br_{iK}^T\hSig_{BKi}^{-1}\br_{iK})\cdot(1+\br_{j1}^T\hSig_{B1j}^{-1}\br_{j1})}+\cdots\\
& &+n^{-1}\sum_{i\not=j}\frac{\br_{i1}^T\hSig_{B1i}^{-1}\br_{j1}\cdot\br_{jK}^T\hSig_{BKj}^{-1}\bSig_{KK}\br_{iK}}
{(1+\br_{i1}^T\hSig_{B1i}^{-1}\br_{i1})\cdot(1+\br_{jK}^T\hSig_{BKj}^{-1}\br_{jK})}\\
& &+\cdots+n^{-1}\sum_{i\not=j}\frac{\br_{i,K-1}^T\hSig_{B,K-1,i}^{-1}\br_{j,K-1}\cdot\br_{jK}^T\hSig_{BKj}^{-1}\bSig_{KK}\br_{iK}}
{(1+\br_{i,K-1}^T\hSig_{B,K-1,i}^{-1}\br_{i,K-1})\cdot(1+\br_{jK}^T\hSig_{BKj}^{-1}\br_{jK})}.
\end{eqnarray*}
Thus, $\tilde{T}_3(\lambda)$ has $2(K-1)K$ summation terms. We will work on the first $(K-1)K$ summation terms and the second $(K-1)K$ summation terms separately in Steps~1 and 2 below. First, for $j=1,\cdots,n$, we have  
\begin{eqnarray}
& &\left|\frac{1}{1+\br_{j2}^T\hSig_{B2j}^{-1}\br_{j2}}-\frac{1}{1+n^{-1}\tr\hSig_{B2j}^{-1}}\right|\nonumber\\
&=&\frac{|\br_{j2}^T\hSig_{B2j}^{-1}\br_{j2}-n^{-1}\tr\hSig_{B2j}^{-1}|}{(1+\br_{j2}^T\hSig_{B2j}^{-1}\br_{j2})(1+n^{-1}\tr\hSig_{B2j}^{-1})}\nonumber\\
&\leq&|\br_{j2}^T\hSig_{B2j}^{-1}\br_{j2}-n^{-1}\tr\hSig_{B2j}^{-1}|=|\epsilon_{n2j}|\nonumber\\
&=&|\epsilon_{n2j}|(I(|\epsilon_{n2j}|\leq\delta))+|\epsilon_{n2j}|(I(|\epsilon_{n2j}|>\delta))\nonumber\\
&\leq&\delta+|\epsilon_{n2j}|(I(|\epsilon_{n2j}|>\delta))\nonumber\\
&\leq&\delta+{\bxz p_2}n^{-1}(\eta_n^2+\lambda^{-1})(I(|\epsilon_{n2j}|>\delta))\label{E1},
\end{eqnarray}
where
$$
\br_{j2}^T\hSig_{B2j}^{-1}\br_{j2}\leq\|\bSig_{22}^{-1/2}\br_{j2}\|^2=n^{-1}\|{\bf w}_{j2}\|^2\leq {\bxz p_2}\eta_n^2
$$
by (\ref{E0}) and 
$$
n^{-1}\tr\hSig_{B2j}^{-1}\leq {\bxz p_2}\lambda^{-1}.
$$
Similarly, we have
\begin{eqnarray}
& &\left|\frac{1}{1+\br_{j1}^T\hSig_{B1j}^{-1}\bSig_{11}\br_{j1}}-\frac{1}{1+n^{-1}\tr\hSig_{B1j}^{-1}\bSig_{11}}\right|\nonumber\\
&\leq&\delta+{\bxz p_1}n^{-1}(\eta_n^2+\lambda^{-1})(I(|{\bxz \epsilon_{n1j}}|>\delta)),\label{E2}
\end{eqnarray}
{\bxz where $|\epsilon_{n1j}|=|\br_{j1}^T\hSig_{B1j}^{-1}\bSig_{11}\br_{j1}-n^{-1}\tr\hSig_{B1j}^{-1}\bSig_{11}|$.}
We have
$$
\rE I(|\epsilon_{n1j}|>\delta)=P(|\epsilon_{n1j}|>\delta)\leq o(n^{-t}),~~\forall t,
$$
and
$$
\rE I(|\epsilon_{n2j}|>\delta)=P(|\epsilon_{n2j}|>\delta)\leq o(n^{-t}),~~\forall t.
$$
It follows that 
\begin{eqnarray}
& &{\bxz 
\rE\left|\frac{1}{1+\br_{j2}^T\hSig_{B2j}^{-1}\br_{j2}}-\frac{1}{1+n^{-1}\tr\hSig_{B2j}^{-1}}\right|\nonumber}\\
&\leq&\rE[\delta+
{\bxz p_2n^{-1}}(\eta_n^2+\lambda^{-1})(I(|\epsilon_{n2j}|>\delta))]\nonumber\\
&=&\delta+{\bxz p_2n^{-1}}(\eta_n^2+\lambda^{-1})P(|\epsilon_{n2j}|>\delta)\nonumber\\
&=&\delta+{\bxz p_2n^{-1}}(\eta_n^2+\lambda^{-1})o(n^{-t}),~~\forall t.\label{E3}
\end{eqnarray}
Similarly,
\begin{eqnarray}
& &\rE\left|\frac{1}{1+\br_{j1}^T\hSig_{B1j}^{-1}\bSig_{11}\br_{j1}}-\frac{1}{1+n^{-1}\tr\hSig_{B1j}^{-1}\bSig_{11}}\right|\nonumber\\
&\leq&\delta+{\bxz p_1}n^{-1}(\eta_n^2+\lambda^{-1})o(n^{-t}),~~\forall t.\label{E4}
\end{eqnarray}
As $\delta\rightarrow0$ and $n\rightarrow \infty$, we have 
\begin{eqnarray*}
& &\rE\left|n^{-1}\sum_{j=1}^n\left[\frac{\br_{j2}^T\hSig_{B2j}^{-1}\br_{j2}\cdot\br_{j1}^T\hSig_{B1j}^{-1}\bSig_{11}\br_{j1}}
{(1+\br_{j2}^T\hSig_{B2j}^{-1}\br_{j2})\cdot(1+\br_{j1}^T\hSig_{B1j}^{-1}\br_{j1})}
-\left(1-\frac{1}{1+n^{-1}\tr\hSig_{B1j}^{-1}}\right)\left(1-\frac{1}{1+n^{-1}\tr\hSig_{B2j}^{-1}}\right)\right]\right|\\
&\leq&\rE\left|n^{-1}\sum_{j=1}^n\left(\frac{1}{1+\br_{j2}^T\hSig_{B2j}^{-1}\br_{j2}}-\frac{1}{1+n^{-1}\tr\hSig_{B2j}^{-1}}\right)
\frac{\br_{j1}^T\hSig_{B1j}^{-1}\bSig_{11}\br_{j1}}{1+\br_{j1}^T\hSig_{B1j}^{-1}\bSig_{11}\br_{j1}}\right|\\
& &+\rE\left|n^{-1}\sum_{j=1}^n\left(1-\frac{1}{1+n^{-1}\tr\hSig_{B1j}^{-1}}\right)
\left(\frac{1}{1+\br_{j1}^T\hSig_{B1j}^{-1}\bSig_{11}\br_{j1}}-\frac{1}{1+n^{-1}\tr\hSig_{B1j}^{-1}\bSig_{11}}\right)\right|\\
&\leq&\rE n^{-1}\sum_{j=1}^n\left|\frac{1}{1+\br_{j2}^T\hSig_{B2j}^{-1}\br_{j2}}-\frac{1}{1+n^{-1}\tr\hSig_{B2j}^{-1}}\right|\\
& &+{\bxz \rE}n^{-1}\sum_{j=1}^n\left|\frac{1}{1+\br_{j1}^T\hSig_{B1j}^{-1}\bSig_{11}\br_{j1}}-\frac{1}{1+n^{-1}\tr\hSig_{B1j}^{-1}\bSig_{11}}\right|\\
&\leq&[\delta+pn^{-1}(\eta_n^2+\lambda^{-1})o(n^{-t})]^2\rightarrow 0,
\end{eqnarray*}
where the last inequality is from (\ref{E3}) and (\ref{E4}) {\bxz and the fact that $\max(p_1, p_2)<p$}.
Therefore, we have 
\begin{eqnarray*}
& &n^{-1}\sum_{j=1}^n\frac{\br_{j2}^T\hSig_{B2j}^{-1}\br_{j2}\cdot\br_{j1}^T\hSig_{B1j}^{-1}\bSig_{11}\br_{j1}}
{(1+\br_{j2}^T\hSig_{B2j}^{-1}\br_{j2})\cdot(1+\br_{j1}^T\hSig_{B1j}^{-1}\br_{j1})}\\
&=&n^{-1}\sum_{j=1}^n\left(1-\frac{1}{1+n^{-1}\tr\hSig_{B1j}^{-1}}\right)\left(1-\frac{1}{1+n^{-1}\tr\hSig_{B2j}^{-1}}\right)+o_p(1).
\end{eqnarray*}
Moreover, we have
\begin{eqnarray}
& &\left|\frac{1}{1+n^{-1}\tr\hSig_{B1j}^{-1}}-\frac{1}{1+n^{-1}\tr\hSig_{B1}^{-1}}\right|\nonumber\\
&=&\left|\frac{n^{-1}\tr\hSig_{B1}^{-1}-n^{-1}\tr\hSig_{B1j}^{-1}}{(1+n^{-1}\tr\hSig_{B1j}^{-1})(1+n^{-1}\tr\hSig_{B1}^{-1})}\right|\nonumber\\
&=&\left|\frac{n^{-1}\br_{j1}^T\hSig_{B1j}^{-2}\br_{j1}}{(1+\br_{j1}^T\hSig_{B1j}^{-1}\br_{j1})(1+n^{-1}\tr\hSig_{B1j}^{-1})(1+n^{-1}\tr\hSig_{B1}^{-1})}\right|
\nonumber\\
&\leq&n^{-1}p\eta_n^2.\label{E5}
\end{eqnarray}
Similarly, \begin{equation}\left|\frac{1}{1+n^{-1}\tr\hSig_{B2j}^{-1}}-\frac{1}{1+n^{-1}\tr\hSig_{B2}^{-1}}\right|\leq n^{-1}p\eta_n^2\label{E6}\end{equation}
{\bf Step 1:}
Then by (\ref{E5}) and (\ref{E6}), we have
\begin{eqnarray*}
& &n^{-1}\sum_{j=1}^n\frac{\br_{j2}^T\hSig_{B2j}^{-1}\br_{j2}\cdot\br_{j1}^T\hSig_{B1j}^{-1}\bSig_{11}\br_{j1}}
{(1+\br_{j2}^T\hSig_{B2j}^{-1}\br_{j2})\cdot(1+\br_{j1}^T\hSig_{B1j}^{-1}\br_{j1})}\\
&=&n^{-1}\sum_{j=1}^n\left(1-\frac{1}{1+n^{-1}\tr\hSig_{B1j}^{-1}}\right)\left(1-\frac{1}{1+n^{-1}\tr\hSig_{B2j}^{-1}}\right)+o_p(1)\\
&=&\left(1-\frac{1}{1+n^{-1}\tr\hSig_{B1}^{-1}}\right)\left(1-\frac{1}{1+n^{-1}\tr\hSig_{B2}^{-1}}\right)+n^{-1}p\eta_n^2+o_p(1)\\
&=&\left(1-\frac{1}{1+n^{-1}\rE\tr\hSig_{B1}^{-1}}\right)\left(1-\frac{1}{1+n^{-1}\rE\tr\hSig_{B2}^{-1}}\right)+n^{-1}p\eta_n^2+o_p(1).
\end{eqnarray*}

Let $a_{n\ell}(\lambda)=p_{\ell}^{-1}\rE\tr(\bSig_{\ell\ell}\hSig_{B\ell}^{-1})$ and $c_{n\ell}=p_{\ell}/n\rightarrow c_{\ell}$, then
\begin{eqnarray*}
& &n^{-1}\sum_{j=1}^n\frac{\br_{j\ell}^T\hSig_{B\ell j}^{-1}\br_{j\ell}\cdot\br_{jh}^T\hSig_{Bhj}^{-1}\bSig_{hh}\br_{jh}}
{(1+\br_{j\ell}^T\hSig_{B\ell j}^{-1}\br_{j\ell})\cdot(1+\br_{jh}^T\hSig_{Bhj}^{-1}\br_{jh})}\\
&=&\frac{ c_{n\ell}a_{n\ell}(\lambda)\cdot c_{nh}a_{nh}(\lambda)}
{[1+c_{n\ell}a_{n\ell}(\lambda)]\cdot[1+c_{nh}a_{nh}(\lambda)]}+o_p(1).
\end{eqnarray*}
Thus, we have
\begin{eqnarray}
& &\sum_{j=1}^n\frac{\br_{j2}^T\hSig_{B2j}^{-1}\br_{j2}\cdot\br_{j1}^T\hSig_{B1j}^{-1}\bSig_{11}\br_{j1}}
{(1+\br_{j2}^T\hSig_{B2j}^{-1}\br_{j2})\cdot(1+\br_{j1}^T\hSig_{B1j}^{-1}\br_{j1})}\nonumber\\
&&+\cdots+\sum_{i=1}^n\frac{\br_{jK}^T\hSig_{BKj}^{-1}\br_{jK}\cdot\br_{j1}^T\hSig_{B1j}^{-1}\bSig_{11}\br_{j1}}
{(1+\br_{jK}^T\hSig_{BKj}^{-1}\br_{jK})\cdot(1+\br_{j1}^T\hSig_{B1j}^{-1}\br_{j1})}+\cdots\nonumber\\
& &+\sum_{j=1}^n\frac{\br_{j1}^T\hSig_{B1j}^{-1}\br_{j1}\cdot\br_{jK}^T\hSig_{BKj}^{-1}\bSig_{KK}\br_{jK}}
{(1+\br_{j1}^T\hSig_{B1j}^{-1}\br_{j1})\cdot(1+\br_{jK}^T\hSig_{BKj}^{-1}\br_{jK})}\nonumber\\
&&+\cdots+\sum_{i=1}^n\frac{\br_{j,K-1}^T\hSig_{B,K-1,j}^{-1}\br_{j,K-1}\cdot\br_{jK}^T\hSig_{BKj}^{-1}\bSig_{KK}\br_{jK}}
{(1+\br_{j,K-1}^T\hSig_{B,K-1,j}^{-1}\br_{j,K-1})\cdot(1+\br_{jK}^T\hSig_{BKj}^{-1}\br_{jK})}\nonumber\\
&=&\sum\limits_{\ell, h=1,...K;~{\bxz l\neq h}}\frac{c_{n\ell}a_{n\ell}(\lambda)\cdot c_{nh}a_{nh}(\lambda)}
{[1+c_{n\ell}a_{n\ell}(\lambda)]\cdot[1+c_{nh}a_{nh}(\lambda)]}+o_p(1)\nonumber\\
&=&\sum\limits_{\ell, h=1,...,K;~{\bxz l\neq h}}\left[1-\frac{1}{1+c_{n\ell}a_{n\ell}(\lambda)}\right]\cdot\left[1-\frac{1}{1+c_{nh}a_{nh}(\lambda)}\right]+o_p(1)\nonumber\\
&=&\sum\limits_{\ell, h=1,...,K;~{\bxz l\neq h}}c_{\ell}c_h[1-\lambda m_{\ell}(-\lambda)][1-\lambda m_h(-\lambda)]+o_p(1),\label{E7}
\end{eqnarray}
where the last equation is from  \cite{bai2008large} and
$$
m_{\ell}(z)=\int\frac{1}{t[1-c_{\ell}-c_{\ell}{\bxz z}m_{\ell}(z)]-z}dH_{\ell}(t)
$$
with $H_{\ell}(t)$ being the limiting spectral distribution of $\bSig_{\ell\ell}$ and $m_{\ell}(z)$ being the Stieltjes transform
of the limiting spectral distribution of the $\ell$th diagonal block $n^{-1}\sum_{i=1}^n\bx_{i\ell}\bx_{i\ell}^T$ of $\hSig_B$.\\
{\bf Step 2:}
Similarly, we have
$$
\sum_{i\not=j}\frac{\br_{i2}^T\hSig_{B2i}^{-1}\br_{j2}\cdot\br_{j1}^T\hSig_{B1j}^{-1}\bSig_{11}\br_{i1}}
{(1+\br_{i2}^T\hSig_{B2i}^{-1}\br_{i2})\cdot(1+\br_{j1}^T\hSig_{B1j}^{-1}\br_{j1})}=
\sum_{i\not=j}\frac{\br_{i2}^T\hSig_{B2i}^{-1}\br_{j2}\cdot\br_{j1}^T\hSig_{B1j}^{-1}\bSig_{11}\br_{i1}}
{(1+n^{-1}\tr\hSig_{B2i}^{-1})\cdot(1+n^{-1}\tr\hSig_{B1j}^{-1})}+o_p(1).
$$
It follows that 
\begin{eqnarray*}
& &\left[\sum_{i\not=j}\frac{\br_{i2}^T\hSig_{B2i}^{-1}\br_{j2}\cdot\br_{j1}^T\hSig_{B1j}^{-1}\bSig_{11}\br_{i1}}
{(1+n^{-1}\tr\hSig_{B2i}^{-1})\cdot(1+n^{-1}\tr\hSig_{B1j}^{-1})}\right]^2\\
&=&\sum_{i_1\not=j_1\not=i_2\not=j_2}\frac{\br_{i_12}^T\hSig_{B2i_1}^{-1}\br_{j_12}\cdot\br_{j_11}^T\hSig_{B1j_1}^{-1}\bSig_{11}\br_{i_11}}
{(1+n^{-1}\tr\hSig_{B2i_1}^{-1})\cdot(1+n^{-1}\tr\hSig_{B1j_1}^{-1})}
\frac{\br_{i_22}^T\hSig_{B2i_2}^{-1}\br_{j_22}\cdot\br_{j_21}^T\hSig_{B1j_2}^{-1}\bSig_{11}\br_{i_21}}
{(1+n^{-1}\tr\hSig_{B2i_2}^{-1})\cdot(1+n^{-1}\tr\hSig_{B1j_2}^{-1})}\\
& &+\sum_{i_1=i_2\not=j_1\not=j_2}\frac{\br_{i_12}^T\hSig_{B2i_1}^{-1}\br_{j_12}\cdot\br_{j_11}^T\hSig_{B1j_1}^{-1}\bSig_{11}\br_{i_11}}
{(1+n^{-1}\tr\hSig_{B2i_1}^{-1})\cdot(1+n^{-1}\tr\hSig_{B1j_1}^{-1})}
\frac{\br_{i_12}^T\hSig_{B2i_1}^{-1}\br_{j_22}\cdot\br_{j_21}^T\hSig_{B1j_2}^{-1}\bSig_{11}\br_{i_11}}
{(1+n^{-1}\tr\hSig_{B2i_1}^{-1})\cdot(1+n^{-1}\tr\hSig_{B1j_2}^{-1})}\\
& &+\sum_{i_1\not=i_2\not=j_1=j_2}\frac{\br_{i_12}^T\hSig_{B2i_1}^{-1}\br_{j_12}\cdot\br_{j_11}^T\hSig_{B1j_1}^{-1}\bSig_{11}\br_{i_11}}
{(1+n^{-1}\tr\hSig_{B2i_1}^{-1})\cdot(1+n^{-1}\tr\hSig_{B1j_1}^{-1})}
\frac{\br_{i_22}^T\hSig_{B2i_2}^{-1}\br_{j_12}\cdot\br_{j_11}^T\hSig_{B1j_1}^{-1}\bSig_{11}\br_{i_21}}
{(1+n^{-1}\tr\hSig_{B2i_2}^{-1})\cdot(1+n^{-1}\tr\hSig_{B1j_1}^{-1})}\\
& &+\sum_{i_1=j_2\not=i_2\not=j_2}\frac{\br_{i_12}^T\hSig_{B2i_1}^{-1}\br_{j_12}\cdot\br_{j_11}^T\hSig_{B1j_1}^{-1}\bSig_{11}\br_{i_11}}
{(1+n^{-1}\tr\hSig_{B2i_1}^{-1})\cdot(1+n^{-1}\tr\hSig_{B1j_1}^{-1})}
\frac{\br_{i_22}^T\hSig_{B2i_2}^{-1}\br_{i_12}\cdot\br_{i_11}^T\hSig_{B1i_1}^{-1}\bSig_{11}\br_{i_21}}
{(1+n^{-1}\tr\hSig_{B2i_2}^{-1})\cdot(1+n^{-1}\tr\hSig_{B1i_1}^{-1})}\\
& &+\sum_{i_1\not=j_2\not=i_2=j_1}\frac{\br_{i_12}^T\hSig_{B2i_1}^{-1}\br_{i_22}\cdot\br_{i_21}^T\hSig_{B1i_2}^{-1}\bSig_{11}\br_{i_11}}
{(1+n^{-1}\tr\hSig_{B2i_1}^{-1})\cdot(1+n^{-1}\tr\hSig_{B1i_2}^{-1})}
\frac{\br_{i_22}^T\hSig_{B2i_2}^{-1}\br_{j_22}\cdot\br_{j_21}^T\hSig_{B1j_2}^{-1}\bSig_{11}\br_{i_21}}
{(1+n^{-1}\tr\hSig_{B2i_2}^{-1})\cdot(1+n^{-1}\tr\hSig_{B1j_2}^{-1})}\\
& &+\sum_{i_1=i_2\not=j_1=j_2}\frac{\br_{i_12}^T\hSig_{B2i_1}^{-1}\br_{j_12}\cdot\br_{j_11}^T\hSig_{B1j_1}^{-1}\bSig_{11}\br_{i_11}}
{(1+n^{-1}\tr\hSig_{B2i_1}^{-1})\cdot(1+n^{-1}\tr\hSig_{B1j_1}^{-1})}
\frac{\br_{i_12}^T\hSig_{B2i_1}^{-1}\br_{j_12}\cdot\br_{j_11}^T\hSig_{B1j_1}^{-1}\bSig_{11}\br_{i_11}}
{(1+n^{-1}\tr\hSig_{B2i_1}^{-1})\cdot(1+n^{-1}\tr\hSig_{B1j_1}^{-1})}\\
& &+\sum_{i_1=j_2\not=i_2=j_1}\frac{\br_{i_12}^T\hSig_{B2i_1}^{-1}\br_{j_12}\cdot\br_{j_11}^T\hSig_{B1j_1}^{-1}\bSig_{11}\br_{i_11}}
{(1+n^{-1}\tr\hSig_{B2i_1}^{-1})\cdot(1+n^{-1}\tr\hSig_{B1j_1}^{-1})}
\frac{\br_{j_12}^T\hSig_{B2j_1}^{-1}\br_{i_12}\cdot\br_{i_11}^T\hSig_{B1i_1}^{-1}\bSig_{11}\br_{j_11}}
{(1+n^{-1}\tr\hSig_{B2j_1}^{-1})\cdot(1+n^{-1}\tr\hSig_{B1i_1}^{-1})}\\
&=&\sum_{i_1\not=j_1\not=i_2\not=j_2}\frac{\br_{i_12}^T\hSig_{B2i_1}^{-1}\br_{j_12}\cdot\br_{j_11}^T\hSig_{B1j_1}^{-1}\bSig_{11}\br_{i_11}
\cdot\br_{i_22}^T\hSig_{B2i_2}^{-1}\br_{j_22}\cdot\br_{j_21}^T\hSig_{B1j_2}^{-1}\bSig_{11}\br_{i_21}}
{(1+n^{-1}\tr\hSig_{B2i_1i_2j_1j_2}^{-1})^4}\\
& &+\sum_{i_1=i_2\not=j_1\not=j_2}\frac{\br_{i_12}^T\hSig_{B2i_1}^{-1}\br_{j_12}\cdot\br_{j_11}^T\hSig_{B1j_1}^{-1}\bSig_{11}\br_{i_11}
\cdot\br_{i_12}^T\hSig_{B2i_1}^{-1}\br_{j_22}\cdot\br_{j_21}^T\hSig_{B1j_2}^{-1}\bSig_{11}\br_{i_11}}
{(1+n^{-1}\tr\hSig_{B2i_1i_2j_1j_2}^{-1})^4}\\
& &+\sum_{i_1\not=i_2\not=j_1=j_2}\frac{\br_{i_12}^T\hSig_{B2i_1}^{-1}\br_{j_12}\cdot\br_{j_11}^T\hSig_{B1j_1}^{-1}\bSig_{11}\br_{i_11}
\cdot\br_{i_22}^T\hSig_{B2i_2}^{-1}\br_{j_12}\cdot\br_{j_11}^T\hSig_{B1j_1}^{-1}\bSig_{11}\br_{i_21}}
{(1+n^{-1}\tr\hSig_{B2i_1i_2j_1j_2}^{-1})^4}\\
& &+\sum_{i_1=j_2\not=i_2\not=j_2}\frac{\br_{i_12}^T\hSig_{B2i_1}^{-1}\br_{j_12}\cdot\br_{j_11}^T\hSig_{B1j_1}^{-1}\bSig_{11}\br_{i_11}\cdot
\br_{i_22}^T\hSig_{B2i_2}^{-1}\br_{i_12}\cdot\br_{i_11}^T\hSig_{B1i_1}^{-1}\bSig_{11}\br_{i_21}}
{(1+n^{-1}\tr\hSig_{B2i_1i_2j_1j_2}^{-1})^4}\\
& &+\sum_{i_1\not=j_2\not=i_2=j_1}\frac{\br_{i_12}^T\hSig_{B2i_1}^{-1}\br_{i_22}\cdot\br_{i_21}^T\hSig_{B1i_2}^{-1}\bSig_{11}\br_{i_11}\cdot
\br_{i_22}^T\hSig_{B2i_2}^{-1}\br_{j_22}\cdot\br_{j_21}^T\hSig_{B1j_2}^{-1}\bSig_{11}\br_{i_21}}
{(1+n^{-1}\tr\hSig_{B2i_1i_2j_1j_2}^{-1})^4}\\
& &+\sum_{i_1=i_2\not=j_1=j_2}\frac{\br_{i_12}^T\hSig_{B2i_1}^{-1}\br_{j_12}\cdot\br_{j_11}^T\hSig_{B1j_1}^{-1}\bSig_{11}\br_{i_11}\cdot
\br_{i_12}^T\hSig_{B2i_1}^{-1}\br_{j_12}\cdot\br_{j_11}^T\hSig_{B1j_1}^{-1}\bSig_{11}\br_{i_11}}
{(1+n^{-1}\tr\hSig_{B2i_1j_1i_2j_2}^{-1})^4}\\
& &+\sum_{i_1=j_2\not=i_2=j_1}\frac{\br_{i_12}^T\hSig_{B2i_1}^{-1}\br_{j_12}\cdot\br_{j_11}^T\hSig_{B1j_1}^{-1}\bSig_{11}\br_{i_11}
\cdot\br_{j_12}^T\hSig_{B2j_1}^{-1}\br_{i_12}\cdot\br_{i_11}^T\hSig_{B1i_1}^{-1}\bSig_{11}\br_{j_11}}
{(1+n^{-1}\tr\hSig_{B2i_1j_1i_2j_2}^{-1})^4}+o_p(1).
\end{eqnarray*}
Note that 
\begin{eqnarray*}
& &\br_{i_12}^T\hSig_{B2i_1}^{-1}\br_{j_12}\\
&=&\br_{i_12}^T\hSig_{B2i_1j_2}^{-1}\br_{j_12}-\frac{\br_{i_12}^T\hSig_{B2i_1j_1}^{-1}
\br_{j_12}\cdot\br_{j_12}^T\hSig_{B2i_1j_1}^{-1}\br_{j_12}}{1+\br_{j_12}^T\hSig_{B2i_1j_1}^{-1}\br_{j_12}}\\
&=&\frac{\br_{i_12}^T\hSig_{B2i_1j_1}^{-1}\br_{j_12}}{1+\br_{j_12}^T\hSig_{B2i_1j_1}^{-1}\br_{j_12}}\\
&=&\frac{\br_{i_12}^T\hSig_{B2i_1j_1i_2}^{-1}\br_{j_12}}{1+\br_{j_12}^T\hSig_{B2i_1j_1}^{-1}\br_{j_12}}
-\frac{\br_{i_12}^T\hSig_{B2i_1j_1i_2}^{-1}\br_{i_22}\cdot\br_{i_22}^T\hSig_{B2i_1j_1i_2}^{-1}\br_{j_12}}
{(1+\br_{i_22}^T\hSig_{B2i_1j_1i_2}^{-1}\br_{i_22})(1+\br_{j_12}^T\hSig_{B2i_1j_1}^{-1}\br_{j_12})}\\
&=&\frac{\br_{i_12}^T\hSig_{B2i_1j_1i_2j_2}^{-1}\br_{j_12}}{1+\br_{j_12}^T\hSig_{B2i_1j_1}^{-1}\br_{j_12}}
-\frac{\br_{i_12}^T\hSig_{B2i_1j_1i_2j_2}^{-1}\br_{j_22}\cdot\br_{j_22}^T\hSig_{B2i_1j_1i_2j_2}^{-1}\br_{j_12}}
{(1+\br_{j_22}^T\hSig_{B2i_1j_1i_2j_2}^{-1}\br_{j_22})(1+\br_{j_12}^T\hSig_{B2i_1j_1}^{-1}\br_{j_12})}\\
&&-\frac{\br_{i_12}^T\hSig_{B2i_1j_1i_2}^{-1}\br_{i_22}\cdot\br_{i_22}^T\hSig_{B2i_1j_1i_2}^{-1}\br_{j_12}}
{(1+\br_{j_12}^T\hSig_{B2i_1j_1}^{-1}\br_{j_12})(1+\br_{i_22}^T\hSig_{B2i_1j_1i_2}^{-1}\br_{i_22})}\\
&=&\frac{\br_{i_12}^T\hSig_{B2i_1j_1i_2j_2}^{-1}\br_{j_12}}{1+\br_{j_12}^T\hSig_{B2i_1j_1}^{-1}\br_{j_12}}\\
&&-\frac{\br_{i_12}^T\hSig_{B2i_1j_1i_2j_2}^{-1}\br_{j_22}\cdot\br_{j_22}^T\hSig_{B2i_1j_1i_2j_2}^{-1}\br_{j_12}}
{(1+\br_{j_22}^T\hSig_{B2i_1j_1i_2j_2}^{-1}\br_{j_22})(1+\br_{j_12}^T\hSig_{B2i_1j_1}^{-1}\br_{j_12})}\\
&&-\frac{\br_{i_12}^T\hSig_{B2i_1j_1i_2j_2}^{-1}\br_{i_22}\cdot\br_{i_22}^T\hSig_{B2i_1j_1i_2j_2}^{-1}\br_{j_12}}
{(1+\br_{j_12}^T\hSig_{B2i_1j_1}^{-1}\br_{j_12})(1+\br_{i_22}^T\hSig_{B2i_1j_1i_2}^{-1}\br_{i_22})}\\
&&+\frac{\br_{i_12}^T\hSig_{B2i_1j_1i_2j_2}^{-1}\br_{j_22}
\cdot\br_{j_22}^T\hSig_{B2i_1j_1i_2j_2}^{-1}\br_{i_22}\cdot\br_{i_22}^T\hSig_{B2i_1j_1i_2j_2}^{-1}\br_{j_12}}
{(1+\br_{j_22}^T\hSig_{B2i_1j_1i_2j_2}^{-1}\br_{j_22})(1+\br_{j_12}^T\hSig_{B2i_1j_1}^{-1}\br_{j_12})(1+\br_{i_22}^T\hSig_{B2i_1j_1i_2}^{-1}\br_{i_22})}\\
&&+\frac{\br_{i_12}^T\hSig_{B2i_1j_1i_2j_2}^{-1}\br_{i_22}\cdot\br_{i_22}^T\hSig_{B2i_1j_1i_2j_2}^{-1}\br_{j_22}
\cdot\br_{j_22}^T\hSig_{B2i_1j_1i_2j_2}^{-1}\br_{j_12}}
{(1+\br_{j_22}^T\hSig_{B2i_1j_1i_2j_2}^{-1}\br_{j_22})(1+\br_{j_12}^T\hSig_{B2i_1j_1}^{-1}\br_{j_12})(1+\br_{i_22}^T\hSig_{B2i_1j_1i_2}^{-1}\br_{i_22})}.
\end{eqnarray*}
We have
$$
\sum_{i_1\not=j_1\not=i_2\not=j_2}\rE\br_{i_12}^T\hSig_{B2i_1j_1i_2j_2}^{-1}\br_{j_12}\br_{j_11}^T\bSig_{11}\hSig_{B2i_1j_1i_2j_2}^{-1}\br_{i_11}
\br_{i_22}^T\hSig_{B2i_1j_1i_2j_2}^{-1}\br_{j_22}\br_{j_21}^T\bSig_{11}\hSig_{B2i_1j_1i_2j_2}^{-1}\br_{i_21}=0.
$$
Similar proofs can be made for other terms. It follows that
$$
\rE\left[\sum_{i\not=j}\frac{\br_{i2}^T\hSig_{B2i}^{-1}\br_{j2}\cdot\br_{j1}^T\hSig_{B1j}^{-1}\bSig_{11}\br_{i1}}
{(1+n^{-1}\tr\hSig_{B2i}^{-1})\cdot(1+n^{-1}\tr\hSig_{B1j}^{-1})}\right]^2=o(1)
$$
and
$$
\rE\sum_{i\not=j}\frac{\br_{i2}^T\hSig_{B2i}^{-1}\br_{j2}\cdot\br_{j1}^T\hSig_{B1j}^{-1}\bSig_{11}\br_{i1}}
{(1+n^{-1}\tr\hSig_{B2i}^{-1})\cdot(1+n^{-1}\tr\hSig_{B1j}^{-1})}=o(1).
$$
Then we can obtain that
\begin{equation}\label{E8}
\sum_{i\not=j}\frac{\br_{i2}^T\hSig_{B2i}^{-1}\br_{j2}\cdot\br_{j1}^T\hSig_{B1j}^{-1}\bSig_{11}\br_{i1}}
{(1+n^{-1}\tr\hSig_{B2i}^{-1})\cdot(1+n^{-1}\tr\hSig_{B1j}^{-1})}=o_p(1).
\end{equation}
By (\ref{E7}) and (\ref{E8}), we have
$$
\tilde{T}_3(\lambda)=\sum\limits_{\ell, h=1,...,K;~{\bxz l\neq h}}c_{\ell}c_h[1-\lambda m_{\ell}(-\lambda)][1-\lambda m_h(-\lambda)]+o_p(1),
$$
where the last equation is from Bai and Zhou (2008) and
$$
m_{\ell}(z)=\int\frac{1}{t[1-c_{\ell}-c_{\ell}m_{\ell}(z)]-z}dH_{\ell}(t)
$$
with $H_{\ell}(t)$ being the limiting spectral distribution of $\bSig_{\ell\ell}$ and $m_{\ell}(z)$ being the Stieltjes transform
of the limiting spectral distribution of the $\ell$ diagonal block $n^{-1}\sum_{i=1}^n\bx_{i\ell}\bx_{i\ell}^T$ of $\hSig_B$. Then Theorem~1 is proved. 

Next, we prove Theorem~2. Under the conditions listed in this paper, we need the limits of

$$
{\rm tr}\{\hat{\bSig}_{Z}(\hat{\bSig}_{B}+\lambda \bbI_p)^{-1}\hat{\bSig} \},
$$
$$
{\rm tr}\{\hat{\bSig}(\hat{\bSig}_{B}+\lambda \bbI_p)^{-1}\hat{\bSig}_{Z}
(\hat{\bSig}_{B}+\lambda \bbI_p)^{-1}\hat{\bSig} \},
$$
and
$$
{\rm tr}\{(\hat{\bSig}_{B}+\lambda \bbI_p)^{-1}\hat{\bSig}_{Z}
(\hat{\bSig}_{B}+\lambda \bbI_p)^{-1}\hat{\bSig} \}.
$$
We have 
\begin{eqnarray*}
&&{\rm tr}\{\hat{\bSig}_{Z}(\hat{\bSig}_{B}+\lambda \bbI_p)^{-1}\hat{\bSig} \}\\
&&={\rm tr}\{\hat{\bSig}_{Z}(\hat{\bSig}_{B}+\lambda \bbI_p)^{-1}(\hat{\bSig}_{B}+\lambda \bbI_p+\hat{\bSig}-\hat{\bSig}_{B}-\lambda \bbI_p) \} \\
&&={\rm tr}[\hat{\bSig}_{Z}\{\bbI_p+ (\hat{\bSig}_{B}+\lambda \bbI_p)^{-1}(\hat{\bSig}-\hat{\bSig}_{B}) -\lambda(\hat{\bSig}_{B}+\lambda \bbI_p)^{-1} \} ] \\
&&={\rm tr}(\hat{\bSig}_{Z})+
{\rm tr}\{\hat{\bSig}_{Z}(\hat{\bSig}_{B}+\lambda \bbI_p)^{-1}(\hat{\bSig}-\hat{\bSig}_{B})\}
-\lambda {\rm tr}\{(\hat{\bSig}_{B}+\lambda \bbI_p)^{-1}\hat{\bSig}_{Z}\}\\
&&={\rm tr}(\hat{\bSig}_{Z})-\lambda {\rm tr}\{(\hat{\bSig}_{B}+\lambda \bbI_p)^{-1}\hat{\bSig}_{Z}\}\\ 
&&=\big[{\rm tr}({\bSig})-\lambda {\rm tr}\{(\hat{\bSig}_{B}+\lambda \bbI_p)^{-1}\bSig\}\big]\cdot(1+o_p(1))\\
&&=\big[p-\lambda \sum^{K}_{\ell=1}{\rm tr}\{ (\hat{\bSig}_{B_\ell}+\lambda \bbI_{p_l})^{-1}\bSig_\ell\}\big]\cdot(1+o_p(1))\\
&&=\big[p-\lambda \sum^{K}_{\ell=1}{\rm tr}\{ (a_\ell\bSig_\ell+\lambda \bbI_{p_l})^{-1}\bSig_\ell\}\big]\cdot(1+o_p(1)). 
\end{eqnarray*}
The last equation follows from \cite{dobriban2019one} and the 
 $a_\ell$ is the unique positive solution of 
$$1-a_\ell
=c_{\ell} \cdot \Big\{1- \lambda \int_{o}^{\infty} (a_\ell t+\lambda)^{-1} dH_\ell(t)  \Big\}
=c_{\ell} \cdot \Big\{1-  \tE_{H_\ell (t)}\big(\frac{\lambda}{a_\ell t+\lambda}\big)\Big\}. $$
Next, we have 
\begin{eqnarray*}
&&{\rm tr}\{\hat{\bSig}(\hat{\bSig}_{B}+\lambda \bbI_p)^{-1}\hat{\bSig}_{Z}
(\hat{\bSig}_{B}+\lambda \bbI_p)^{-1}\hat{\bSig} \}\\
&&={\rm tr}[\hat{\bSig}_{Z}\{(\hat{\bSig}_{B}+\lambda \bbI_p)^{-1}(\hat{\bSig}_{B}+\lambda \bbI_p+\hat{\bSig}-\hat{\bSig}_{B}-\lambda \bbI_p)\}^2 ] \\
&&={\rm tr}[{\bSig}\{\bbI_p+ (\hat{\bSig}_{B}+\lambda \bbI_p)^{-1}(\hat{\bSig}-\hat{\bSig}_{B}) -\lambda(\hat{\bSig}_{B}+\lambda \bbI_p)^{-1} \}^2 ]\cdot(1+o_p(1)) \\
&&=\Big(
{\rm tr}[\bSig+{\bSig}\{(\hat{\bSig}_{B}+\lambda \bbI_p)^{-1}(\hat{\bSig}-\hat{\bSig}_{B})\}^2
+\lambda^2(\hat{\bSig}_{B}+\lambda \bbI_p)^{-2}{\bSig}
-2\lambda(\hat{\bSig}_{B}+\lambda \bbI_p)^{-1}{\bSig}]\Big)\cdot(1+o_p(1))
 \\
&&=\Big(p+\lambda^2\sum_{\ell=1}^{K}\tr\{(a_\ell\bmSigma_\ell+\lambda\I_{p_\ell})^{-2}(\I_{p_\ell}-\dot{a}_\ell\bmSigma_\ell)\bmSigma_\ell\}-2\lambda\sum^{K}_{\ell=1}{\rm tr}\{ (a_\ell\bSig_\ell+\lambda \bbI_{p_l})^{-1}\bSig_\ell\}\\
&&+
\sum_{\ell,h=1,\cdots,K;~{\bxz \ell\neq h}} \omega^{-1} c_\ell c_h \{1-\lambda m_\ell(-\lambda)\}\{1-\lambda m_h(-\lambda)\}\Big)\cdot(1+o_p(1))
\end{eqnarray*}
The last equation is based on the above Theorem~1 and \cite{dobriban2019one} and
$$\dot{a}_\ell
=\frac{c_\ell \cdot \tE_{H_\ell(t)}\big\{\frac{a_\ell t}{(a_\ell t+\lambda)^2}\big\} }{
-1-c_\ell\lambda\cdot \tE_{H_\ell(t)}\big\{\frac{t}{(a_\ell t+\lambda)^2}\big\}}. $$
Similarly, we have 
\begin{eqnarray*}
&&{\rm tr}\{(\hat{\bSig}_{B}+\lambda \bbI_p)^{-1}\hat{\bSig}_{Z}
(\hat{\bSig}_{B}+\lambda \bbI_p)^{-1}\hat{\bSig} \}\\
&&={\rm tr}[(\hat{\bSig}_{B}+\lambda \bbI_p)^{-1}\hat{\bSig}_{Z}\{(\hat{\bSig}_{B}+\lambda \bbI_p)^{-1}(\hat{\bSig}_{B}+\lambda \bbI_p+\hat{\bSig}-\hat{\bSig}_{B}-\lambda \bbI_p)\} ] \\
&&={\rm tr}[(\hat{\bSig}_{B}+\lambda \bbI_p)^{-1}{\bSig}\{\bbI_p+ (\hat{\bSig}_{B}+\lambda \bbI_p)^{-1}(\hat{\bSig}-\hat{\bSig}_{B}) -\lambda(\hat{\bSig}_{B}+\lambda \bbI_p)^{-1} \} ]\cdot(1+o_p(1)) \\
&&={\rm tr}[(\hat{\bSig}_{B}+\lambda \bbI_p)^{-1}{\bSig}-\lambda(\hat{\bSig}_{B}+\lambda \bbI_p)^{-1}{\bSig}(\hat{\bSig}_{B}+\lambda \bbI_p)^{-1} ]\cdot(1+o_p(1)) \\
&&=[\sum^{K}_{\ell=1}{\rm tr}\{ (a_\ell\bSig_\ell+\lambda \bbI_{p_l})^{-1}\bSig_\ell\}-\lambda\sum_{\ell=1}^{K}\tr\{(a_\ell\bmSigma_\ell+\lambda\I_{p_\ell})^{-2}(\I_{p_\ell}-\dot{a}_\ell\bmSigma_\ell)\bmSigma_\ell\} ]\cdot(1+o_p(1)) 
\end{eqnarray*}
Then following  \cite{zhao2019cross} and \cite{dobriban2018high}, Theorem~2 is proved by the Lemma B.26 in \cite{bai2010spectral} and continuous mapping theory.

\subsection{Theorems~3~and~4 (reference panel)}
Under the conditions listed in this paper, we first define some additional notations
\begin{itemize}
\item $c_{n\ell}=p_{\ell}/n\rightarrow c_{\ell}\in(0, \infty)$, $\ell=1,...,K$;
\item $b_{p\ell}=p_{\ell}/p\rightarrow b_{\ell}\in(0, \infty)$, $\ell=1,...,K$;
\item $\bx_i=\bSig^{1/2}{\bf w}_i$, $i=1,...,n$; and 
\item $\bSig_{\ell\ell}$ is the $l$th diagonal block of $\bSig$, which 
has the limiting spectral distribution $H_{\ell}(t)$, $\ell=1,...,K$.
\end{itemize}
The first goal is to obtain the limits of 
$$
{T}_4(\lambda)=p^{-1}{\rm tr}\{(\hat{\bSig}_{W_\ell}+\lambda\bI_{p_\ell})^{-1}\bSig_\ell(\hat{\bSig}_{W_\ell}+\lambda\bI_{p_\ell})^{-1}\bSig^2_\ell\}
$$
and
$$
{T}_5(\lambda)=p^{-1}{\rm tr}\{(\hat{\bSig}_{W_\ell}+\lambda\bI_{p_\ell})^{-1}\bSig_\ell(\hat{\bSig}_{W_\ell}+\lambda\bI_{p_\ell})^{-1}\bSig_\ell\}, 
$$
for $\ell=1,\cdots,K$. 
Let
\begin{eqnarray*}
&&\hat{\bSig}_{W_{\ell}}=\sum\limits_{i=1}^n\br_{i\ell}\br_{i\ell}^T,~~~~\hat{\bSig}_{W_{\ell},-k}=\sum\limits_{i\neq k}\br_{i\ell}\br_{i\ell}^T,\\
&&\bB_{\ell}(\lambda)=\hat{\bSig}_{W_{\ell}}+\lambda\bI_{p_{\ell}},~~~~\bB_{\ell,-k}(\lambda)=\hat{\bSig}_{W_{\ell},-k}+\lambda\bI_{p_{\ell}}, \quad \mbox{and} \\
&&\bK_{\ell}=\{1+n^{-1}\rE\tr(\bB_{\ell}(\lambda)^{-1}\bSig_{\ell})\}^{-1}\bSig_{\ell}.
\end{eqnarray*}
For simplicity, below we write $\bB_{\ell}(\lambda)$ as $\bB_{\ell}$ and $\bB_{\ell,-k}(\lambda)$ and $\bB_{\ell,-k}$. 
Then we have
\begin{equation}\label{eq1}
(\bK_{\ell}+\lambda\bI_{p_{\ell}})^{-1}-(\hat{\bSig}_{W_{\ell}}+\lambda\bI_{p_{\ell}})^{-1}
=\sum_{k=1}^n\frac{(\bK_{\ell}+\lambda\bI_{p_{\ell}})^{-1}\br_{k\ell}\br_{k\ell}^T\bB_{\ell,-k}^{-1}}{1+\br_{k\ell}^T{\bxz \bB_{\ell,-k}^{-1}}\br_{k\ell}}
-(\bK_{\ell}+\lambda\bI_{p_{\ell}})^{-1}\bK_{\ell}\bB_{\ell}^{-1}.
\end{equation}
Note that
\begin{eqnarray*}
& &p^{-1}\tr\left((\hat{\bSig}_{\ell}+\lambda\bI_{p_{\ell}})^{-1}\bSig_{\ell}
(\hat{\bSig}_{\ell}+\lambda\bI_{p_{\ell}})^{-1}\bSig_{\ell}^2\right)-p^{-1}\tr\left((\bK_{\ell}+\lambda\bI_{p_{\ell}})^{-1}
\bSig_{\ell}(\bK_{\ell}+\lambda\bI_{p_{\ell}})^{-1}\bSig_{\ell}^2\right)\\
&=&p^{-1}\tr\left((\hat{\bSig}_{\ell}+\lambda\bI_{p_{\ell}})^{-1}\bSig_{\ell}(\hat{\bSig}_{\ell}+\lambda\bI_{p_{\ell}})^{-1}\bSig_{\ell}^2\right)
-p^{-1}\tr\left((\bK_{\ell}+\lambda\bI_{p_{\ell}})^{-1}\bSig_{\ell}(\hat{\bSig}_{\ell}+\lambda\bI_{p_{\ell}})^{-1}\bSig_{\ell}^2\right)\\
& &+p^{-1}\tr\left((\bK_{\ell}+\lambda\bI_{p_{\ell}})^{-1}\bSig_{\ell}(\hat{\bSig}_{\ell}+\lambda\bI_{\ell})^{-1}\bSig_{\ell}^2\right)
-p^{-1}\tr\left((\bK_{\ell}+\lambda\bI_{p_{\ell}})^{-1}\bSig_{\ell}(\bK_{\ell}+\lambda\bI_{p_{\ell}})^{-1}\bSig_{\ell}^2\right)\\
&=&p^{-1}\tr\left[\left((\hat{\bSig}_{\ell}+\lambda\bI_{p_{\ell}})^{-1}-(\bK_{\ell}+\lambda\bI_{p_{\ell}})^{-1}\right)\bSig_{\ell}
(\hat{\bSig}_{\ell}+\lambda\bI_{p_{\ell}})^{-1}\bSig_{\ell}^2\right]\\
& &+p^{-1}\tr\left[(\bK_{\ell}+\lambda\bI_{p_{\ell}})^{-1}\bSig_{\ell}\left((\hat{\bSig}_{\ell}+\lambda\bI_{p_{\ell}})^{-1}
-(\bK_{\ell}+\lambda\bI_{p_{\ell}})^{-1}\right)\bSig_{\ell}^2\right].
\end{eqnarray*}
{\bf Goal}: We want to prove 
\begin{eqnarray}\label{eq2}
& &p^{-1}\tr\left[\left((\hat{\bSig}_{\ell}+\lambda\bI_{p_{\ell}})^{-1}-(\bK_{\ell}+\lambda\bI_{p_{\ell}})^{-1}\right)
\bSig_{\ell}(\hat{\bSig}_{\ell}+\lambda\bI_{p_{\ell}})^{-1}\bSig_{\ell}^2\right]\nonumber\\
&=&\frac{p^{-1}\tr(\bB_{\ell}^{-1}\bSig_{\ell}\bB_{\ell}^{-1}\bSig_{\ell})
n^{-1}\tr(\bB_{\ell}^{-1}\bSig_{\ell}^2(\bK_{\ell}+\lambda\bI_{p_{\ell}})^{-1}\bSig_{\ell})}{\big(1+n^{-1}\tr(\bB_{\ell}^{-1}\bSig_{\ell}))^2}+o_p(1)
\end{eqnarray}
and
\begin{equation}\label{eq3}
p^{-1}\tr\left[(\bK_{\ell}+\lambda\bI_{p_{\ell}})^{-1}\bSig_{\ell}\left((\hat{\bSig}_{\ell}+\lambda\bI_{p_{\ell}})^{-1}
-(\bK_{\ell}+\lambda\bI_{p_{\ell}})^{-1}\right)\bSig_{\ell}^2\right]=o_p(1),
\end{equation}
which lead to 
\begin{eqnarray}\label{eq4}
&&p^{-1}\tr\left((\hat{\bSig}_{\ell}+\lambda\bI_{p_{\ell}})^{-1}\bSig_{\ell}(\hat{\bSig}_{\ell}+\lambda\bI_{p_{\ell}})^{-1}\bSig_{\ell}^2\right)
-p^{-1}\tr\left((\bK_{\ell}+\lambda\bI_{p_{\ell}})^{-1}\bSig_{\ell}(\bK_{\ell}+\lambda\bI_{p_{\ell}})^{-1}\bSig_{\ell}^2\right)\nonumber\\
&=&\frac{p^{-1}\tr(\bB_{\ell}^{-1}\bSig_{\ell}\bB_{\ell}^{-1}\bSig_{\ell})n^{-1}\tr(\bB_{\ell}^{-1}\bSig_{\ell}^2(\bK_{\ell}+\lambda\bI_{p_{
\ell}})^{-1}\bSig_{\ell})}{(1+n^{-1}\tr(\bB_{\ell}^{-1}\bSig_{\ell}))^2}+o_p(1).
\end{eqnarray}
To achieve this goal, we use the following two steps. \\
{\bf Step 1}: First, we consider the term in (\ref{eq3}). Based on Equation~(\ref{eq1}), we have
\begin{eqnarray*}
& &p^{-1}\tr\left[(\bK_{\ell}+\lambda\bI_{p_{\ell}})^{-1}\bSig_{\ell}\left((\hat{\bSig}_{\ell}+\lambda\bI_{p_{\ell}})^{-1}
-(\bK_{\ell}+\lambda\bI_{p_{\ell}})^{-1}\right)\bSig_{\ell}^2\right]\\
&=&-p^{-1}\tr\left[(\bK_{\ell}+\lambda\bI_{p_{\ell}})^{-1}\bSig_{\ell}\left(\sum_{k=1}^n\frac{(\bK_{\ell}+\lambda\bI_{\ell})^{-1}
\br_{k\ell}\br_{k\ell}^T\bB_{\ell,-k}^{-1}}{1+\br_{k\ell}^T\bB_{\ell,-k}^{-1}\br_{k\ell}}
   -(\bK_{\ell}+\lambda\bI_{p_{\ell}})^{-1}\bK_{\ell}\bB_{\ell}^{-1}\right)\bSig_{\ell}^2\right]\\
&=&-p^{-1}\sum_{k=1}^n\frac{\br_{k\ell}^T\bB_{\ell,-k}^{-1}\bSig_{\ell}^2(\bK_{\ell}+\lambda\bI_{p_{\ell}})^{-1}
\bSig_{\ell}(\bK_{\ell}+\lambda\bI_{p_{\ell}})^{-1}\br_{k\ell}}{1+\br_{k\ell}^T\bB_{\ell,-k}^{-1}\br_{k\ell}}\\
& &+p^{-1}\tr\left((\bK_{\ell}+\lambda\bI_{p_{\ell}})^{-1}\bSig_{\ell}(\bK_{\ell}+\lambda\bI_{p_{\ell}})^{-1}\bK_{\ell}\bB_{\ell}^{-1}\bSig_{\ell}^2\right)\\
&=&-p^{-1}\sum_{k=1}^n\frac{d_{2k}}{1+\br_{k\ell}^T\bB_{\ell,-k}^{-1}\br_{k\ell}},
\end{eqnarray*}
where
$$d_{2k}=\br_{k\ell}^T\bB_{\ell,-k}^{-1}\bA_2\br_{k\ell}-(1+\br_{k\ell}^T\bB_{\ell,-k}^{-1}\br_{k\ell})n^{-1}\tr\left(\bB_{\ell}^{-1}\bA_2\bK_{\ell}\right)$$
with $$\bA_2=\bSig_{\ell}^2(\bK_{\ell}+\lambda\bI_{p_{\ell}})^{-1}\bSig_{\ell}(\bK_{\ell}+\lambda\bI_{p_{\ell}})^{-1}.$$
Write $d_{2k}=d_{2k1}+d_{2k2}+d_{2k3}$ with
\begin{eqnarray*}
d_{2k1}&=&\br_{k\ell}^T\bB_{\ell,-k}^{-1}\bA_2\br_{k\ell}-n^{-1}\tr\big(\bB_{k\ell}^{-1}\bA_2\bSig_{\ell}\big),\\
d_{2k2}&=&n^{-1}\tr\big(\bB_{\ell,-k}^{-1}\bA_2\bSig_{\ell}\big)-n^{-1}\tr\big(\bB_{\ell}^{-1}\bA_2\bSig_{\ell}\big),\\
d_{2k3}&=&\left(1-\frac{1+\br_{k\ell}^T\bB_{\ell,-k}^{-1}\br_{k\ell}}{1+n^{-1}\rE\tr(\bB_{\ell}^{-1}\bSig_{\ell})}\right)
\cdot n^{-1}\tr\big(\bB_{\ell}^{-1}\bA_2\bSig_{\ell}\big).
\end{eqnarray*}
Following the discussions in \cite{bai2008large}, we obtain
$$p^{-1}\tr\left[(\bK_{\ell}+\lambda\bI_{p_{\ell}})^{-1}\bSig_{\ell}
\left((\hat{\bSig}_{\ell}+\lambda\bI_{p_{\ell}})^{-1}-(\bK_{\ell}+\lambda\bI_{p_{\ell}})^{-1}\right)\bSig_{\ell}^2\right]=o_p(1).$$
{\bf Step 2}: Next, we consider the term in (\ref{eq2}). Again, based on Equation~(\ref{eq1}), we have
\begin{eqnarray*}
& &p^{-1}\tr\left[\left((\hat{\bSig}_{\ell}+\lambda\bI_{p_{\ell}})^{-1}-(\bK_{\ell}+\lambda\bI_{p_{\ell}})^{-1}\right)
\bSig_{\ell}(\hat{\bSig}_{\ell}+\lambda\bI_{p_{\ell}})^{-1}\bSig_{\ell}^2\right]\\
&=&-p^{-1}\tr\left[\left(\sum_{k=1}^n\frac{(\bK_{\ell}+\lambda\bI_{p_{\ell}})^{-1}\br_{k\ell}\br_{k\ell}^T\bB_{\ell,-k}^{-1}}
{1+\br_{k\ell}^T\bB_{\ell,-k}^{-1}\br_{k\ell}}
   -(\bK_{\ell}+\lambda\bI_{p_{\ell}})^{-1}\bK_{\ell}\bB_{\ell}^{-1}\right)\bSig_{\ell}(\hat{\bSig}_{\ell}+\lambda\bI_{p_{\ell}})^{-1}\bSig_{\ell}^2\right]\\
&=&-p^{-1}\sum_{k=1}^n\frac{\br_{k\ell}^T\bB_{\ell,-k}^{-1}\bSig_{\ell}\bB_{\ell}^{-1}\bSig_{\ell}^2(\bK_{\ell}+\lambda\bI_{p_{\ell}})^{-1}\br_{k\ell}}
{1+\br_{k\ell}^T\bB_{\ell,-k}^{-1}\br_{k\ell}}
   +p^{-1}\tr\left((\bK_{\ell}+\lambda\bI_{p_{\ell}})^{-1}\bK_{\ell}\bB_{\ell}^{-1}\bSig_{\ell}\bB_{\ell}^{-1}\bSig_{\ell}^2\right)\\
&=&-p^{-1}\sum_{k=1}^n\frac{d_{1k}}{1+\br_{k\ell}^T\bB_{\ell,-k}^{-1}\br_{k\ell}},
\end{eqnarray*}
where
$$d_{1k}=\br_{k\ell}^T\bB_{\ell,-k}^{-1}\bSig_{\ell}\bB_{\ell}^{-1}\bA_1\br_{k\ell}
-(1+\br_{k\ell}^T\bB_{\ell,-k}^{-1}\br_{k\ell})n^{-1}\tr\left(\bB_{\ell}^{-1}\bSig_{\ell}\bB_{\ell}^{-1}\bA_1\bK_{\ell}\right)$$
with $$\bA_1=\bSig_{\ell}^2(\bK_{\ell}+\lambda\bI_{p_{\ell}})^{-1}.$$
Since
\begin{eqnarray*}
& &\br_{k\ell}^T\bB_{\ell,-k}^{-1}\bSig_{\ell}\bB_{\ell}^{-1}\bA_1\br_{k\ell}\\
&=&\br_{k\ell}^T\bB_{\ell,-k}^{-1}\bSig_{\ell}\bB_{\ell,-k}^{-1}\bA_1\br_{k\ell}
-\frac{\br_{k\ell}^T\bB_{\ell,-k}^{-1}\bSig_{\ell}\bB_{\ell,-k}^{-1}\br_{k\ell}\br_{k\ell}^T\bB_{\ell,-k}^{-1}\bA_1\br_{k\ell}}
{1+\br_{k\ell}^T\bB_{\ell,-k}^{-1}\br_{k\ell}}\\
&=&n^{-1}\tr(\bB_{\ell,-k}^{-1}\bSig_{\ell}\bB_{\ell,-k}^{-1}\bA_1\bSig_{\ell})
-\frac{n^{-1}\tr(\bB_{\ell,-k}^{-1}\bSig_{\ell}\bB_{\ell,-k}^{-1}\bSig_{\ell})n^{-1}\tr(\bB_{\ell,-k}^{-1}\bA_1\bSig_{\ell})}
{1+n^{-1}\tr(\bB_{\ell,-k}^{-1}\bSig_{\ell})}+o_p(1),
\end{eqnarray*}
and
\begin{eqnarray*}
&&n^{-1}\tr({\bxz \bB_{\ell,-k}^{-1}}\bSig_{\ell}{\bxz \bB_{\ell,-k}^{-1}}\bA_1\bSig_{\ell})-n^{-1}\tr(\bB_{\ell}^{-1}\bSig_{\ell}\bB_{\ell}^{-1}\bA_1\bSig_{\ell})=o_p(1),\\
&&n^{-1}\tr(\bB_{\ell}^{-1}\bSig_{\ell}\bB_{\ell}^{-1}\bA_1\bSig_{\ell})-(1+\br_{k\ell}^T\bB_{\ell,-k}^{-1}\br_{k\ell})
n^{-1}\tr\left(\bB_{\ell}^{-1}\bSig_{\ell}\bB_{\ell}^{-1}\bA_1\bK_{\ell}\right)=o_p(1),
\end{eqnarray*}
then we have
\begin{eqnarray*}
& &p^{-1}\tr\left[\left((\hat{\bSig}_{\ell}+\lambda\bI_{p_{\ell}})^{-1}-(\bK_{\ell}+\lambda\bI_{p_{\ell}})^{-1}\right)
\bSig_{\ell}(\hat{\bSig}_{\ell}+\lambda\bI_{p_{\ell}})^{-1}\bSig_{\ell}^2\right]\\
&=&p^{-1}\sum_{k=1}^n\frac{n^{-1}\tr(\bB_{\ell,-k}^{-1}\bSig_{\ell}\bB_{\ell,-k}^{-1}\bSig_{\ell})
n^{-1}\tr(\bB_{\ell,-k}^{-1}\bA_1\bSig_{\ell})}{\big(1+n^{-1}\tr(\bB_{\ell,-k}^{-1}\bSig_{\ell})\big)^2}+o_p(1)\\
&=&\frac{p^{-1}\tr(\bB_{\ell}^{-1}\bSig_{\ell}\bB_{\ell}^{-1}\bSig_{\ell})n^{-1}
\tr(\bB_{\ell}^{-1}\bA_1\bSig_{\ell})}{\big(1+n^{-1}\tr(\bB_{\ell}^{-1}\bSig_{\ell})\big)^2}+o_p(1).
\end{eqnarray*}
Combining Steps~1 and 2, we have proved Equation~($\ref{eq4}$). 
Now we consider the limit of ${T}_5(\lambda)=p^{-1}\tr(\bB_{\ell}^{-1}\bSig_{\ell}\bB_{\ell}^{-1}\bSig_{\ell})$. 
Similar to Equation~($\ref{eq4}$), we can show that 
\begin{eqnarray*}
& &p^{-1}\tr(\bB_{\ell}^{-1}\bSig_{\ell}\bB_{\ell}^{-1}\bSig_{\ell})
-p^{-1}\tr\left((\bK_{\ell}+\lambda\bI_{p_{\ell}})^{-1}\bSig_{\ell}(\bK_{\ell}+\lambda\bI_{p_{\ell}})^{-1}\bSig_{\ell}\right)\\
&=&\frac{p^{-1}\tr(\bB_{\ell}^{-1}\bSig_{\ell}\bB_{\ell}^{-1}\bSig_{\ell})
n^{-1}\tr(\bB_{\ell}^{-1}\bSig_{\ell}(\bK_{\ell}+\lambda\bI_{p_{\ell}})^{-1}\bSig_{\ell})}
{\big(1+n^{-1}\tr(\bB_{\ell}^{-1}\bSig_{\ell})\big)^2}+o_p(1).
\end{eqnarray*}
It follows that
\begin{eqnarray*}
            & &\left(1-\frac{n^{-1}\tr(\bB_{\ell}^{-1}\bSig_{\ell}(\bK_{\ell}+\lambda\bI_{p_{\ell}})^{-1}\bSig_{\ell})}
               {\big(1+n^{-1}\tr(\bB_{\ell}^{-1}\bSig_{\ell})\big)^2}\right)p^{-1}\tr(\bB_{\ell}^{-1}\bSig_{\ell}\bB_{\ell}^{-1}\bSig)\\
            &=&p^{-1}\tr\left((\bK_{\ell}+\lambda\bI_{p_{\ell}})^{-1}\bSig_{\ell}(\bK_{\ell}+\lambda\bI_{p_{\ell}})^{-1}\bSig_{\ell}\right)+o_p(1)  
\end{eqnarray*}
Therefore, we have
\begin{eqnarray}
            & &{T}_5(\lambda)\nonumber\\
            &=&p^{-1}\tr(\bB_{\ell}^{-1}\bSig_{\ell}\bB_{\ell}^{-1}\bSig_{\ell})\label{eq5}\\
            &=&\left(1-\frac{n^{-1}\tr(\bB_{\ell}^{-1}\bSig_{\ell}(\bK_{\ell}+\lambda\bI_{p_{\ell}})^{-1}\bSig_{\ell})}
               {\big(1+n^{-1}\tr(\bB_{\ell}^{-1}\bSig_{\ell})\big)^2}\right)^{-1}\nonumber\\
           & &\cdot p^{-1}\tr\left((\bK_{\ell}+\lambda\bI_{p_{\ell}})^{-1}\bSig_{\ell}(\bK_{\ell}+\lambda\bI_{p_{\ell}})^{-1}\bSig_{\ell}\right)+o_p(1)\nonumber\\
            &=&\left(1-\frac{n^{-1}\tr((\bK_{\ell}+\lambda\bI_{p_{\ell}})^{-1}\bSig_{\ell}(\bK_{\ell}+\lambda\bI_{p_{\ell}})^{-1}\bSig_{\ell})}
               {\big(1+n^{-1}\tr((\bK_{\ell}+\lambda\bI_{p_{\ell}})^{-1}\bSig_{\ell})\big)^2}\right)^{-1}\nonumber\\
               & & \cdot p^{-1}\tr\left((\bK_{\ell}+\lambda\bI_{p_{\ell}})^{-1}\bSig_{\ell}(\bK_{\ell}+\lambda\bI_{p_{\ell}})^{-1}\bSig_{\ell}\right)+o_p(1).\nonumber
\end{eqnarray}
Now we consider the limit of ${T}_4(\lambda)=p^{-1}\tr(\bB_{\ell}^{-1}\bSig_{\ell}\bB_{\ell}^{-1}\bSig_{\ell}^2)$. 
Based on (\ref{eq4}) and (\ref{eq5}), we have
\begin{eqnarray*}
& &p^{-1}\tr\left((\hat{\bSig}_{\ell}+\lambda\bI_{p_{\ell}})^{-1}\bSig_{\ell}
(\hat{\bSig}_{\ell}+\lambda\bI_{p_{\ell}})^{-1}\bSig_{\ell}^2\right)
-p^{-1}\tr\left((\bK_{\ell}+\lambda\bI_{p_{\ell}})^{-1}\bSig_{\ell}(\bK_{\ell}+\lambda\bI_{p_{\ell}})^{-1}\bSig_{\ell}^2\right)\\
&=&\frac{p^{-1}\tr\left((\bK_{\ell}+\lambda\bI_{p_{\ell}})^{-1}\bSig_{\ell}(\bK_{\ell}+\lambda\bI_{p_{\ell}})^{-1}
\bSig_{\ell}\right)n^{-1}\tr(\bB_{\ell}^{-1}\bSig_{\ell}^2(\bK_{\ell}+\lambda\bI_{p_{\ell}})^{-1}\bSig_{\ell})}
   {\big(1+n^{-1}\tr(\bB_{\ell}^{-1}\bSig_{\ell})\big)^2-n^{-1}\tr(\bB_{\ell}^{-1}\bSig_{\ell}(\bK_{\ell}+\lambda\bI_{p_{\ell}})^{-1}\bSig_{\ell})}+o_p(1).
\end{eqnarray*}
That is,
\begin{eqnarray*}
{T}_4(\lambda) &=&p^{-1}\tr\left((\hat{\bSig}_{\ell}+\lambda\bI_{p_{\ell}})^{-1}\bSig_{\ell}(\hat{\bSig}_{\ell}+\lambda\bI_{p_{\ell}})^{-1}\bSig_{\ell}^2\right)\\
&=&\frac{p^{-1}\tr\left((\bK_{\ell}+\lambda\bI_{p_{\ell}})^{-1}\bSig_{\ell}(\bK_{\ell}+\lambda\bI_{p_{\ell}})^{-1}
\bSig_{\ell}\right)n^{-1}\tr(\bB_{\ell}^{-1}\bSig_{\ell}^2(\bK_{\ell}+\lambda\bI_{p_{\ell}})^{-1}\bSig_{\ell})}
   {\big(1+n^{-1}\tr(\bB_{\ell}^{-1}\bSig_{\ell})\big)^2-n^{-1}\tr(\bB_{\ell}^{-1}\bSig_{\ell}(\bK_{\ell}+\lambda\bI_{p_{\ell}})^{-1}\bSig_{\ell})}\\
& &+p^{-1}\tr\left((\bK_{\ell}+\lambda\bI_{p_{\ell}})^{-1}\bSig_{\ell}(\bK_{\ell}+\lambda\bI_{p_{\ell}})^{-1}\bSig_{\ell}^2\right)+o_p(1).
\end{eqnarray*}
From \cite{bai2008large}, we have
\begin{eqnarray*}
&&n^{-1}\tr(\bB_{\ell}^{-1}\bSig_{\ell}^2(\bK_{\ell}+\lambda\bI_{\ell})^{-1}\bSig_{\ell})
=n^{-1}\tr((\bK_{\ell}+\lambda\bI_{p_{\ell}})^{-1}\bSig_{\ell}^2(\bK_{\ell}+\lambda\bI_{p_{\ell}})^{-1}\bSig_{\ell})+o_p(1),\\
&&n^{-1}\tr(\bB_{\ell}^{-1}\bSig_{\ell}(\bK_{\ell}+\lambda\bI_{p_{\ell}})^{-1}\bSig_{\ell})=n^{-1}\tr((\bK_{\ell}+\lambda\bI_{p_{\ell}})^{-1}
\bSig_{\ell}(\bK_{\ell}+\lambda\bI_{p_{\ell}})^{-1}\bSig_{\ell})+o_p(1),
\quad\mbox{and}\\
&&n^{-1}\tr(\bB_{\ell}^{-1}\bSig_{\ell})=n^{-1}\tr((\bK_{\ell}+\lambda\bI_{p_{\ell}})^{-1}\bSig_{\ell})+o_p(1).
\end{eqnarray*}
It follows that 
\begin{eqnarray*}
{T}_4(\lambda) &=&p^{-1}\tr\left((\hat{\bSig}_{\ell}+\lambda\bI_{p_{\ell}})^{-1}\bSig_{\ell}(\hat{\bSig}_{\ell}+\lambda\bI_{p_{\ell}})^{-1}\bSig_{\ell}^2\right)\\
            &=&\frac{\big(1+n^{-1}\tr((\bK_{\ell}+\lambda\bI_{p_{\ell}})^{-1}\bSig_{\ell})\big)^2p^{-1}\tr\left((\bK_{\ell}+\lambda\bI_{p_{\ell}})^{-1}
               \bSig_{\ell}(\bK_{\ell}+\lambda\bI_{p_{\ell}})^{-1}\bSig_{\ell}^2\right)}
               {\big(1+n^{-1}\tr((\bK_{\ell}+\lambda\bI_{p_{\ell}})^{-1}\bSig_{\ell})\big)^2
               -n^{-1}\tr((\bK_{\ell}+\lambda\bI_{p_{\ell}})^{-1}\bSig_{\ell}(\bK_{\ell}+\lambda\bI_{p_{\ell}})^{-1}\bSig_{\ell})}+o_p(1).
\end{eqnarray*}
Then Theorem~3 is proved. 
Next, we prove Theorem~4. Under the conditions listed in this paper, we need the limits of

$$
{\rm tr}\{\hat{\bSig}_{Z}(\hat{\bSig}_{BW}+\lambda \bbI_p)^{-1}\hat{\bSig} \},
$$
$$
{\rm tr}\{\hat{\bSig}(\hat{\bSig}_{BW}+\lambda \bbI_p)^{-1}\hat{\bSig}_{Z}
(\hat{\bSig}_{BW}+\lambda \bbI_p)^{-1}\hat{\bSig} \},
$$
$$
{\rm tr}\{(\hat{\bSig}_{BW}+\lambda \bbI_p)^{-1}\hat{\bSig}_{Z}
(\hat{\bSig}_{BW}+\lambda \bbI_p)^{-1}\hat{\bSig} \},
$$
$$
{\rm tr}\{\hat{\bSig}_{Z}(\hat{\bSig}_{BZ}+\lambda \bbI_p)^{-1}\hat{\bSig} \},
$$
$$
{\rm tr}\{\hat{\bSig}(\hat{\bSig}_{BZ}+\lambda \bbI_p)^{-1}\hat{\bSig}_{Z}
(\hat{\bSig}_{BZ}+\lambda \bbI_p)^{-1}\hat{\bSig} \},
$$
$$
{\rm tr}\{(\hat{\bSig}_{BZ}+\lambda \bbI_p)^{-1}\hat{\bSig}_{Z}
(\hat{\bSig}_{BZ}+\lambda \bbI_p)^{-1}\hat{\bSig} \},
$$
We have 
\begin{eqnarray*}
&&{\rm tr}\{\hat{\bSig}_{Z}(\hat{\bSig}_{BW}+\lambda \bbI_p)^{-1}\hat{\bSig} \}\\
&&={\rm tr}\{(\hat{\bSig}_{BW}+\lambda \bbI_p)^{-1}\bSig^2 \}\cdot(1+o_p(1))\\
&&=\sum^{K}_{\ell=1}{\rm tr}\{ (\hat{\bSig}_{BW_\ell}+\lambda \bbI_{p_\ell})^{-1}\bSig_\ell^2\}\cdot(1+o_p(1))=\sum^{K}_{\ell=1}{\rm tr}\{ (a_{w_\ell}\bSig_\ell+\lambda \bbI_{p_\ell})^{-1}\bSig_\ell^2\}\cdot(1+o_p(1)), 
\end{eqnarray*}
where $a_{w_\ell}$ is the unique positive solution of 
$$1-a_{w_\ell}
=c_{w_\ell} \cdot \Big\{1- \lambda \int_{o}^{\infty} (a_{w_\ell} t+\lambda)^{-1} dH_\ell(t)  \Big\}
=c_{w_\ell} \cdot \Big\{1-  \tE_{H_\ell (t)}\big(\frac{\lambda}{a_{w_\ell} t+\lambda}\big)\Big\} $$
with $c_{w_{\ell_n}}=p_\ell/n_w \to c_{w_{\ell}} \in (0,\infty)$. 
We also have 
\begin{eqnarray*}
&&{\rm tr}\{\hat{\bSig}(\hat{\bSig}_{BW}+\lambda \bbI_p)^{-1}\hat{\bSig}_{Z}
(\hat{\bSig}_{BW}+\lambda \bbI_p)^{-1}\hat{\bSig} \}\\
&&={\rm tr}\{(\hat{\bSig}_{BW}+\lambda \bbI_p)^{-1}\bSig
(\hat{\bSig}_{BW}+\lambda \bbI_p)^{-1}\hat{\bSig}^2 \} \cdot(1+o_p(1)) \\
&&=\Big(
{\rm tr}\{(\hat{\bSig}_{BW}+\lambda \bbI_p)^{-1}\bSig
(\hat{\bSig}_{BW}+\lambda \bbI_p)^{-1}\bSig^2 \} \\
&&\qquad +\omega \cdot
{\rm tr}\{(\hat{\bSig}_{BW}+\lambda \bbI_p)^{-1}\bSig
(\hat{\bSig}_{BW}+\lambda \bbI_p)^{-1}\bSig \}\Big)\cdot(1+o_p(1))
 \\
&&=\Big(\sum_{\ell=1}^{K} \tr\{(\widehat{\bmSigma}_{W_\ell}+\lambda \I_{p_\ell})^{-1}\bmSigma_\ell(\widehat{\bmSigma}_{W_\ell}+\lambda\I_{p_\ell})^{-1}\bmSigma_\ell^2\}\\
&&\qquad + \omega \cdot \sum_{\ell=1}^{K} \tr\{(\widehat{\bmSigma}_{W_\ell}+\lambda \I_{p_\ell})^{-1}\bmSigma_\ell(\widehat{\bmSigma}_{W_\ell}+\lambda\I_{p_\ell})^{-1}\bmSigma_\ell\}
\Big)\cdot(1+o_p(1))
\end{eqnarray*}
Then we have the limit of ${\rm tr}\{\hat{\bSig}(\hat{\bSig}_{BW}+\lambda \bbI_p)^{-1}\hat{\bSig}_{Z}
(\hat{\bSig}_{BW}+\lambda \bbI_p)^{-1}\hat{\bSig} \}$ by the above Theorem~3. 
Similarly, the limit of ${\rm tr}\{(\hat{\bSig}_{BW}+\lambda \bbI_p)^{-1}\hat{\bSig}_{Z}
(\hat{\bSig}_{BW}+\lambda \bbI_p)^{-1}\hat{\bSig} \}$ can also be obtained from Theorem~3. Next, we have 
\begin{eqnarray*}
&&{\rm tr}\{\hat{\bSig}_{Z}(\hat{\bSig}_{BZ}+\lambda \bbI_p)^{-1}\hat{\bSig} \}\\
&&={\rm tr}\{(\hat{\bSig}_{BZ}+\lambda \bbI_p+\hat{\bSig}_Z-\hat{\bSig}_{BZ}-\lambda \bbI_p)(\hat{\bSig}_{BZ}+\lambda \bbI_p)^{-1}\hat{\bSig} \} \\
&&={\rm tr}[\{\bbI_p+ (\hat{\bSig}_Z-\hat{\bSig}_{BZ})(\hat{\bSig}_{BZ}+\lambda \bbI_p)^{-1} -\lambda(\hat{\bSig}_{BZ}+\lambda \bbI_p)^{-1} \}\hat{\bSig} ] \\
&&={\rm tr}(\hat{\bSig})+
{\rm tr}\{(\hat{\bSig}_Z-\hat{\bSig}_{BZ})(\hat{\bSig}_{BZ}+\lambda \bbI_p)^{-1}\hat{\bSig}\}
-\lambda {\rm tr}\{(\hat{\bSig}_{BZ}+\lambda \bbI_p)^{-1}\hat{\bSig}\}\\
&&={\rm tr}(\hat{\bSig})-\lambda {\rm tr}\{(\hat{\bSig}_{BZ}+\lambda \bbI_p)^{-1}\hat{\bSig}\}\\ 
&&=\big[{\rm tr}({\bSig})-\lambda {\rm tr}\{(\hat{\bSig}_{BZ}+\lambda \bbI_p)^{-1}\bSig\}\big]\cdot(1+o_p(1))\\
&&=\big[p-\lambda \sum^{K}_{\ell=1}{\rm tr}\{ (\hat{\bSig}_{BZ_\ell}+\lambda \bbI_{p_\ell})^{-1}\bSig_\ell\}\big]\cdot(1+o_p(1))\\
&&=\big[p-\lambda \sum^{K}_{\ell=1}{\rm tr}\{ (a_{z_\ell}\bSig_\ell+\lambda \bbI_{p_\ell})^{-1}\bSig_\ell\}\big]\cdot(1+o_p(1)), 
\end{eqnarray*}
where $a_{z_\ell}$ is the unique positive solution of 
$$1-a_{z_\ell}
=c_{z_\ell} \cdot \Big\{1- \lambda \int_{o}^{\infty} (a_{z_\ell} t+\lambda)^{-1} dH_\ell(t)  \Big\}
=c_{z_\ell} \cdot \Big\{1-  \tE_{H_\ell (t)}\big(\frac{\lambda}{a_{z_\ell} t+\lambda}\big)\Big\} $$
with $c_{z_{\ell_n}}=p_\ell/n_z \to c_{z_{\ell}} \in (0,\infty)$. 
In addition, 
\begin{eqnarray*}
&&{\rm tr}\{\hat{\bSig}(\hat{\bSig}_{BZ}+\lambda \bbI_p)^{-1}\hat{\bSig}_{Z}
(\hat{\bSig}_{BZ}+\lambda \bbI_p)^{-1}\hat{\bSig} \},\\
&&={\rm tr}[\{(\hat{\bSig}_{BZ}+\lambda \bbI_p)^{-1}-\lambda(\hat{\bSig}_{BZ}+\lambda \bbI_p)^{-2} \} \hat{\bSig}^2 ] \\
&&={\rm tr}\{(\hat{\bSig}_{BZ}+\lambda \bbI_p)^{-1}\hat{\bSig}^2\}-\lambda{\rm tr}\{(\hat{\bSig}_{BZ}+\lambda \bbI_p)^{-2}\hat{\bSig}^2 \} \\
&&=\Big(
{\rm tr}\{(\hat{\bSig}_{BZ}+\lambda \bbI_p)^{-1}{\bSig}^2\}+\omega {\rm tr}\{(\hat{\bSig}_{BZ}+\lambda \bbI_p)^{-1}{\bSig}\} \\
&&\qquad -\lambda{\rm tr}\{(\hat{\bSig}_{BZ}+\lambda \bbI_p)^{-2}{\bSig}^2\}-\lambda\omega{\rm tr}\{(\hat{\bSig}_{BZ}+\lambda \bbI_p)^{-2}{\bSig} \}
\Big)\cdot(1+o_p(1))
 \\
&&=\Big(\sum^{K}_{\ell=1}{\rm tr}\{ (a_{z_\ell}\bSig_\ell+\lambda \bbI_{p_\ell})^{-1}\bSig_\ell^2\}+\omega\sum^{K}_{\ell=1}{\rm tr}\{ (a_{z_\ell}\bSig_\ell+\lambda \bbI_{p_\ell})^{-1}\bSig_\ell\}\\
&&\qquad-\lambda
\sum_{\ell=1}^{K}\tr\{(a_{z_\ell}\bmSigma_\ell+\lambda\I_{p_\ell})^{-2}(\I_{p_\ell}-\dot{a}_{z_\ell}\bmSigma_\ell)\bmSigma_\ell^2\}\\
&&\qquad-\lambda\omega
\sum_{\ell=1}^{K}\tr\{(a_{z_\ell}\bmSigma_\ell+\lambda\I_{p_\ell})^{-2}(\I_{p_\ell}-\dot{a}_{z_\ell}\bmSigma_\ell)\bmSigma_\ell\}\Big)\cdot(1+o_p(1)),
\end{eqnarray*}
where
$$\dot{a}_{z_\ell}
=\frac{c_{z_\ell} \cdot \tE_{H_\ell(t)}\big\{\frac{a_{z_\ell} t}{(a_{z_\ell} t+\lambda)^2}\big\} }{
-1-c_{z_\ell}\lambda\cdot \tE_{H_\ell(t)}\big\{\frac{t}{(a_{z_\ell} t+\lambda)^2}\big\}}. $$
Similarly, we have 
\begin{eqnarray*}
&&{\rm tr}\{(\hat{\bSig}_{BZ}+\lambda \bbI_p)^{-1}\hat{\bSig}_{Z}
(\hat{\bSig}_{BZ}+\lambda \bbI_p)^{-1}\hat{\bSig} \},\\
&&={\rm tr}[\{(\hat{\bSig}_{BZ}+\lambda \bbI_p)^{-1}-\lambda(\hat{\bSig}_{BZ}+\lambda \bbI_p)^{-2} \} \hat{\bSig} ] \\
&&={\rm tr}\{(\hat{\bSig}_{BZ}+\lambda \bbI_p)^{-1}\hat{\bSig}\}-\lambda{\rm tr}\{(\hat{\bSig}_{BZ}+\lambda \bbI_p)^{-2}\hat{\bSig} \} \\
&&=\Big(
{\rm tr}\{(\hat{\bSig}_{BZ}+\lambda \bbI_p)^{-1}{\bSig}\}
 -\lambda{\rm tr}\{(\hat{\bSig}_{BZ}+\lambda \bbI_p)^{-2}{\bSig}\}
\Big)\cdot(1+o_p(1))
 \\
&&=\Big(\sum^{K}_{\ell=1}{\rm tr}\{ (a_{z_\ell}\bSig_\ell+\lambda \bbI_{p_\ell})^{-1}\bSig_\ell\}-\lambda
\sum_{\ell=1}^{K}\tr\{(a_{z_\ell}\bmSigma_\ell+\lambda\I_{p_\ell})^{-2}(\I_{p_\ell}-\dot{a}_{z_\ell}\bmSigma_\ell)\bmSigma_\ell\}\Big)\cdot(1+o_p(1)).
\end{eqnarray*}
Then Theorem~4 is proved by the Lemma B.26 in \cite{bai2010spectral} and continuous mapping theory.  
\subsection{Independent random effects assumption}\label{sec1.3}
In Condition~3 of the main paper, we assume 
\begin{flalign*}
\begin{matrix} 
\bmbeta_{(1)}
\end{matrix}
\sim F 
\left (
\begin{matrix} 
\bm{0_m}
\end{matrix},
\begin{matrix} 
p^{-1} \cdot\bmSigma_{\beta} 
\end{matrix}
\right ),
\end{flalign*} 
where $\bmSigma_{\beta}=\sigma_{\beta}^2 \cdot \bbI_m$. Based on this condition, to obtain the asymptotic results in our Theorem~2,~4,~5, we additionally need some mild conditions on a few tract functions, detailed in Conditions~4,~5,~6, respectively. 

An alternative condition for $\bmbeta_{(1)}$ that allows variation of genetic effects could be the independent random effect assumption, in which $\bmSigma_{\beta}=\diagg(\sigma_{\beta_1}^2,\cdots,\sigma_{\beta_{m}}^2)$, with $\sigma_{\beta_i}^2$ being some constants, $i=1,\cdots,m$. 
Let $\bmSigma_{\beta}^*$ be a $p \times p$ diagonal matrix such that there are $m$ nonzero diagonal entries corresponding to the $m$ causal genetic variants with effects in $\bmSigma_{\beta}$ and other $p-m$  diagonal entries are zeros. Let $\tr(\bmSigma_{\beta}^*)/p=\sum_{i=1}^{m}\sigma_{\beta_i}^2=\sigma^2_{\beta}$ be the averaged per-variant genetic effect. Our results in the Theorem~2,~4,~5 still hold under the independent random effect assumption, if we use the following new versions of Conditions~4,~5,~6, respectively. 

\paragraph{Condition~$4^*$.}
Let $\A$ be a generic $p\times p$ matrix and $\A_m$ be an $m\times m$ sub-matrix of $\A$ corresponding to the $m$ causal variants with nonzero effects. As $\mbox{min}(n,p)\to \infty$,   we assume
$\tr\{\bmSigma(\widehat{\bmSigma}_B+\lambda \I_p)^{-1}\widehat{\bmSigma}\bmSigma_{\beta}^*\}=$ $\sigma^2_{\beta}\cdot \tr\{\bmSigma(\widehat{\bmSigma}_B+\lambda \I_p)^{-1}\widehat{\bmSigma}\}\cdot(1+o_p(1))$ and 
$\tr\{\widehat{\bmSigma}(\widehat{\bmSigma}_B+\lambda \I_p)^{-1}\bmSigma(\widehat{\bmSigma}_B+\lambda \I_p)^{-1}\widehat{\bmSigma}\bmSigma_{\beta}^*\}=$ $\sigma^2_{\beta}\cdot\tr\{\widehat{\bmSigma}(\widehat{\bmSigma}_B+\lambda \I_p)^{-1}\bmSigma(\widehat{\bmSigma}_B+\lambda \I_p)^{-1}\widehat{\bmSigma}\}\cdot(1+o_p(1))$. 

\paragraph{Condition~$5^*$.}
As $\mbox{min}(n,n_w,n_z,p)\to \infty$,    we assume  \\
$\tr\{\widehat{\bmSigma}(\widehat{\bmSigma}_{BZ}+\lambda \I_p)^{-1}\widehat{\bmSigma}_Z(\widehat{\bmSigma}_{BZ}+\lambda \I_p)^{-1}\widehat{\bmSigma}\bmSigma_{\beta}^*\}=$ $\sigma^2_{\beta}\cdot\tr\{\widehat{\bmSigma}(\widehat{\bmSigma}_{BZ}+\lambda \I_p)^{-1}\widehat{\bmSigma}_Z(\widehat{\bmSigma}_{BZ}+\lambda \I_p)^{-1}\widehat{\bmSigma}\}\cdot(1+o_p(1))$, 
$\tr\{\widehat{\bmSigma}(\widehat{\bmSigma}_{BZ}+\lambda \I_p)^{-1}\bmSigma\bmSigma_{\beta}^*\}$= 
$\sigma^2_{\beta}\cdot\tr\{\widehat{\bmSigma}(\widehat{\bmSigma}_{BZ}+\lambda \I_p)^{-1}\bmSigma\}\cdot(1+o_p(1))$, $\tr\{\bmSigma(\widehat{\bmSigma}_{BW}+\lambda \I_p)^{-1}\bmSigma\bmSigma_{\beta}^*\}=$
$\sigma^2_{\beta}\cdot\tr\{\bmSigma(\widehat{\bmSigma}_{BW}+\lambda \I_p)^{-1}\bmSigma\}\cdot(1+o_p(1))$, and 
$\tr\{\widehat{\bmSigma}(\widehat{\bmSigma}_{BW}+\lambda \I_p)^{-1}\bmSigma(\widehat{\bmSigma}_{BW}+\lambda \I_p)^{-1}\widehat{\bmSigma}\bmSigma_{\beta}^*\}=$
$\sigma^2_{\beta}\cdot\tr\{\widehat{\bmSigma}(\widehat{\bmSigma}_{BW}+\lambda \I_p)^{-1}\bmSigma(\widehat{\bmSigma}_{BW}+\lambda \I_p)^{-1}\widehat{\bmSigma}\}\cdot(1+o_p(1))$,
where $\widehat{\bmSigma}_Z=n_z^{-1}\Z^T\Z$.

\paragraph{Condition~$6^*$.}
As $\mbox{min}(n,n_z,p)\to \infty$, we assume \\
$\tr\{\bmSigma\V\R(\lambda)\V^T\widehat{\bmSigma}\bmSigma_{\beta}^*\}=$
$\sigma^2_{\beta}\cdot \tr\{\bmSigma\V\R(\lambda)\V^T\widehat{\bmSigma}\}\cdot(1+o_p(1))$, 
$\tr\{\widehat{\bmSigma}\V\R(\lambda)\V^T\bmSigma\V\R(\lambda)\V^T\widehat{\bmSigma}\bmSigma_{\beta}^*\}=$
$\sigma^2_{\beta}\cdot\tr\{\widehat{\bmSigma}\V\R(\lambda)\V^T\bmSigma\V\R(\lambda)\V^T\widehat{\bmSigma}\cdot(1+o_p(1))\}$, and 
$\tr\{\widehat{\bmSigma}\V\V^T\bmSigma\V\V^T\widehat{\bmSigma}\bmSigma_{\beta}^*\}=$
$\sigma^2_{\beta}\cdot\tr\{\widehat{\bmSigma}\V\V^T\bmSigma\V\V^T\widehat{\bmSigma}\}\cdot(1+o_p(1))$.

\section{Supplementary figures and tables}\label{sec3}
\begin{suppfigure}
\centering
\includegraphics[page=2,width=1\linewidth]{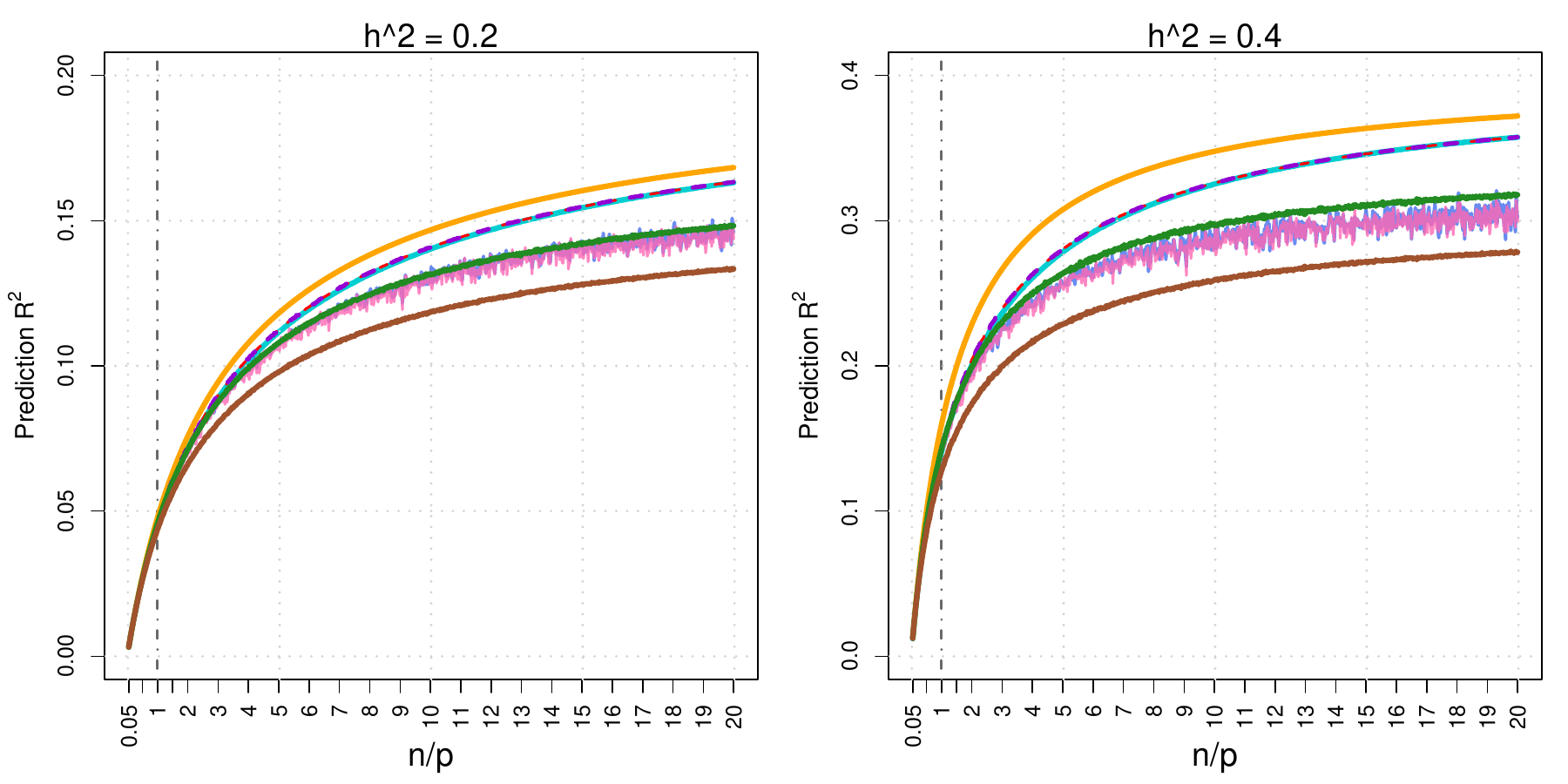}
\includegraphics[page=3,width=1\linewidth]{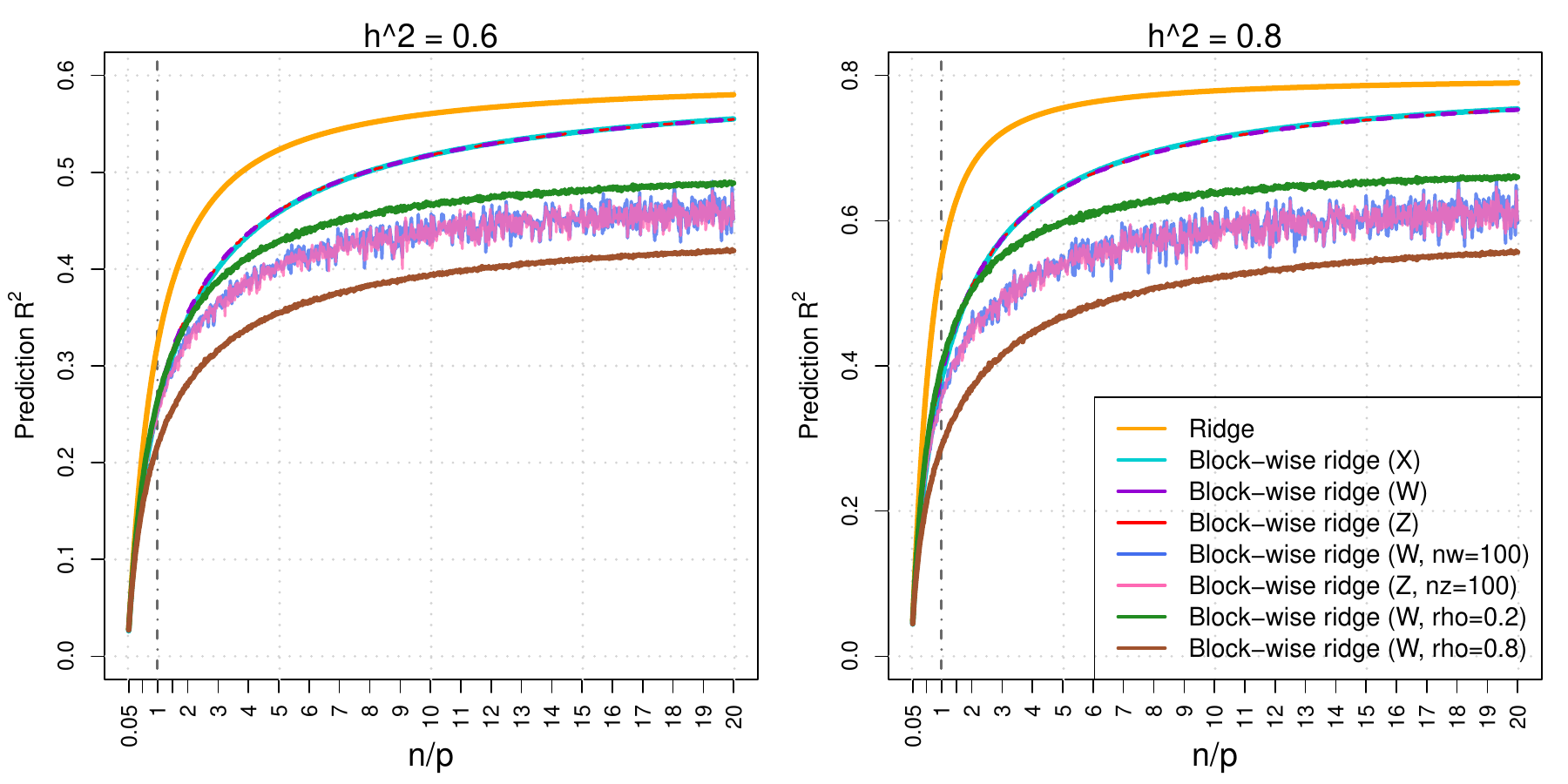}
  \caption{Comparing the prediction accuracy of block-wise ridge estimators $\widehat{\bmbeta}_{B}(\lambda^{*})$, $\widehat{\bmbeta}_{BW}(\lambda^{*})$, $\widehat{\bmbeta}_{BZ}(\lambda^{*})$,
  and ridge estimator $\widehat{\bmbeta}_{R}(\lambda^{*})$ at different $n/p$ ratios with $\h_\beta^2$ ranging from $0.2$ to $0.8$.
We simulate the data with $20$ independent blocks, each of which has a block size $p_\ell=50$ ($p=1,000$). An auto-correlation structure is given within each block and the auto-correlation coefficient is $\rho_b=0.5$. We also illustrate the 
performance of $\widehat{\bmbeta}_{BW}(\lambda^{*})$ and $\widehat{\bmbeta}_{BZ}(\lambda^{*})$ when $n_w=n_z=100$, and  the
performance of $\widehat{\bmbeta}_{BW}(\lambda^{*})$ when the $\rho_b$ in $\W$ is $0.8$ or $0.2$. The vertical line represents $n/p=1$. 
}
\label{figs1}
\end{suppfigure}
\begin{suppfigure}
\centering
\includegraphics[page=1,width=1\linewidth]{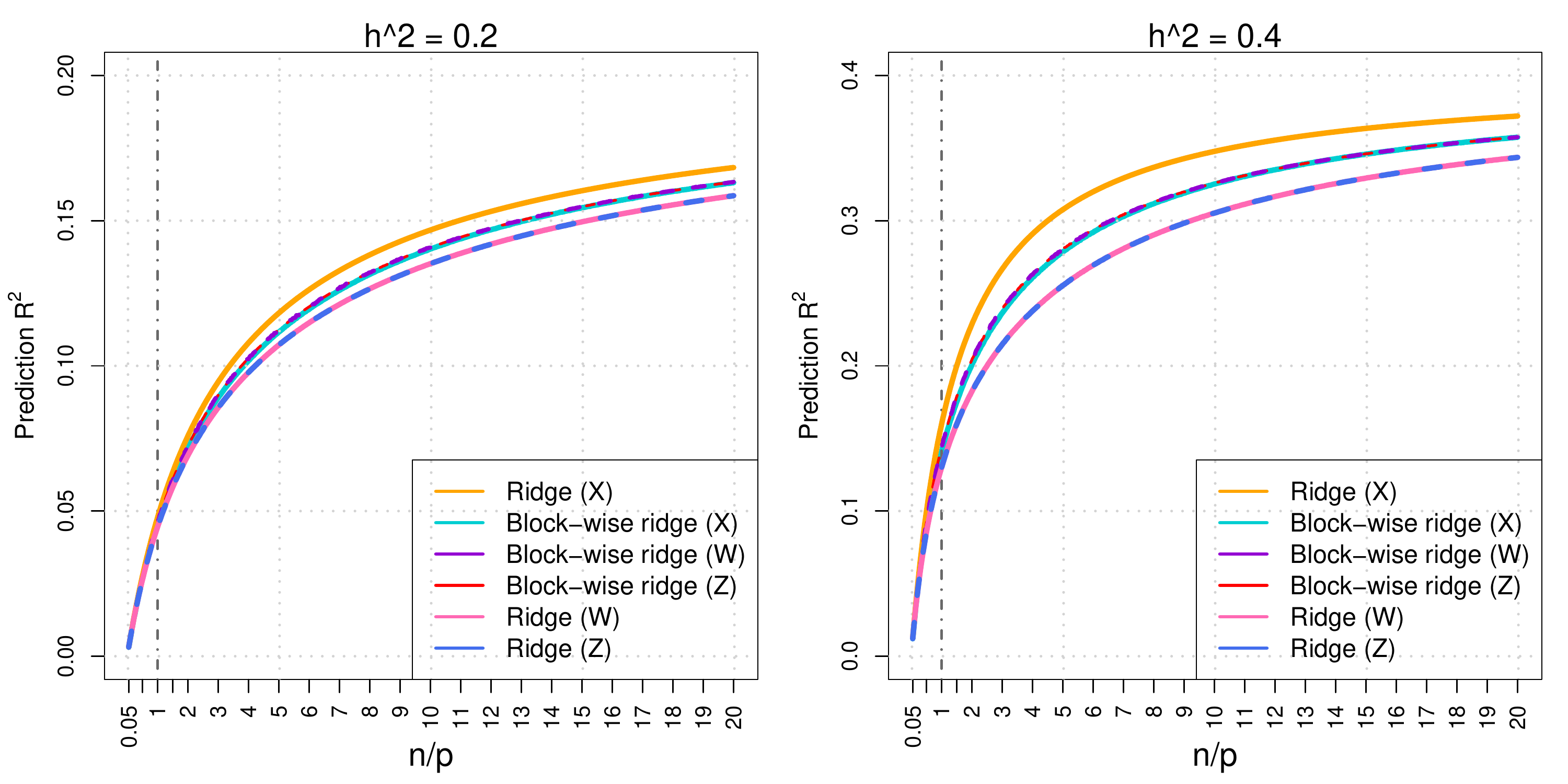}
\includegraphics[page=2,width=1\linewidth]{Results-Upper-Oct22-2021-4.pdf}
  \caption{Comparing the prediction accuracy of block-wise ridge estimators $\widehat{\bmbeta}_{B}(\lambda^{*})$, $\widehat{\bmbeta}_{BW}(\lambda^{*})$, $\widehat{\bmbeta}_{BZ}(\lambda^{*})$,
 ridge estimator $\widehat{\bmbeta}_{R}(\lambda^{*})$, and reference panel-based ridge estimators 
 $\widehat{\bmbeta}_{RW}(\lambda^{*})$ and $\widehat{\bmbeta}_{RZ}(\lambda^{*})$
at different $n/p$ ratios with $\h_\beta^2$ ranging from $0.2$ to $0.8$.
We simulate the data with $20$ independent blocks, each of which has a block size $p_\ell=50$ ($p=1,000$). An auto-correlation structure is given within each block and the auto-correlation coefficient is $\rho_b=0.5$. 
The vertical line represents $n/p=1$. 
}
\label{figs2}
\end{suppfigure}
\begin{suppfigure}
\centering
\includegraphics[page=1,width=1\linewidth]{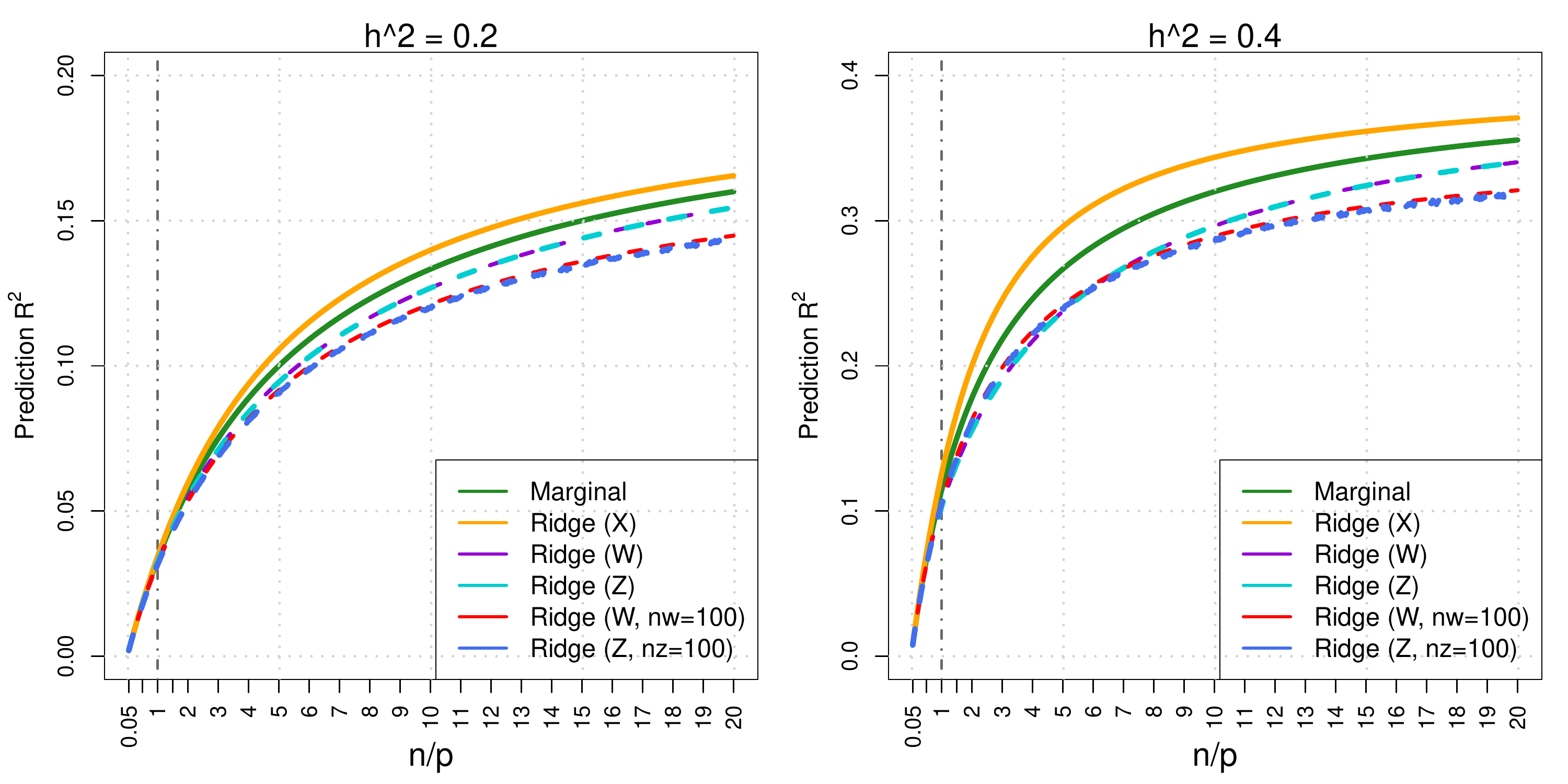}
\includegraphics[page=2,width=1\linewidth]{Results-Upper-Oct22-2021-5.pdf}
  \caption{Comparing the prediction accuracy of ridge estimator $\widehat{\bmbeta}_{R}(\lambda^{*})$, marginal estimator $\widehat{\bmbeta}_{S}$,  and reference panel-based ridge estimators 
 $\widehat{\bmbeta}_{RW}(\lambda^{*})$ and $\widehat{\bmbeta}_{RZ}(\lambda^{*})$
at different $n/p$ ratios with $\h_\beta^2$ ranging from $0.2$ to $0.8$.
We simulate the data with $1,000$ independent features ($p=1,000$). 
We also illustrate the 
performance of $\widehat{\bmbeta}_{RW}(\lambda^{*})$ and $\widehat{\bmbeta}_{RZ}(\lambda^{*})$ when $n_w=n_z=100$. The vertical line represents $n/p=1$. 
}
\label{figs3}
\end{suppfigure}
\begin{suppfigure}
\centering
\includegraphics[page=1,width=1\linewidth]{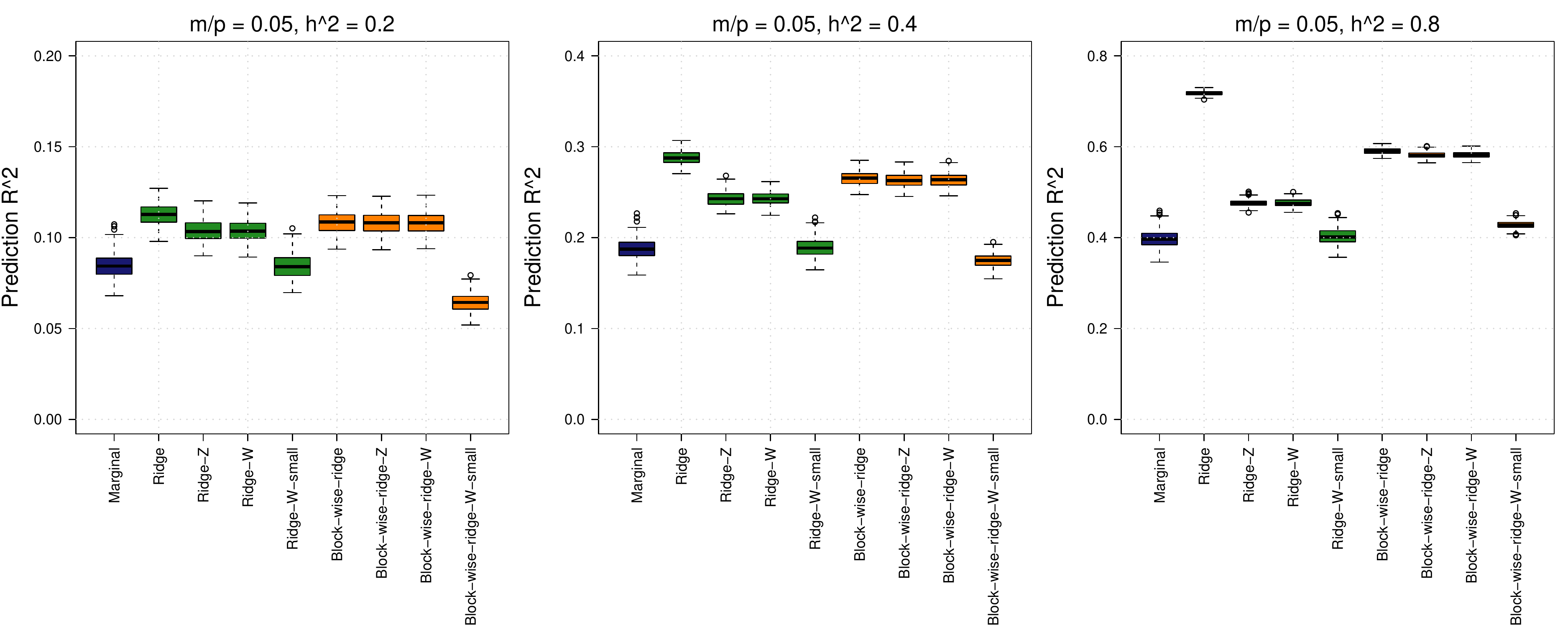}
\includegraphics[page=3,width=1\linewidth]{Summary-Simu-block-Oct-22-2021-1-sigma.pdf}
  \caption{Comparing the prediction accuracy of ridge estimator $\widehat{\bmbeta}_{R}(\lambda^{*})$, marginal estimator $\widehat{\bmbeta}_{S}$,  and reference panel-based ridge estimators 
 $\widehat{\bmbeta}_{RW}(\lambda^{*})$ and $\widehat{\bmbeta}_{RZ}(\lambda^{*})$
at different $n/p$ ratios with $\h_\beta^2$ ranging from $0.2$ to $0.8$.
We simulate the data with $20$ independent blocks, each of which has a block size $p_\ell=50$ ($p=1,000$). An auto-correlation structure is given within each block and the auto-correlation coefficient is $\rho_b=0.5$. We also illustrate the 
performance of $\widehat{\bmbeta}_{RW}(\lambda^{*})$ and $\widehat{\bmbeta}_{RZ}(\lambda^{*})$ when $n_w=n_z=100$. The vertical line represents $n/p=1$. 
}
\label{figs4}
\end{suppfigure}
\begin{suppfigure}
\centering
\includegraphics[page=4,width=1\linewidth]{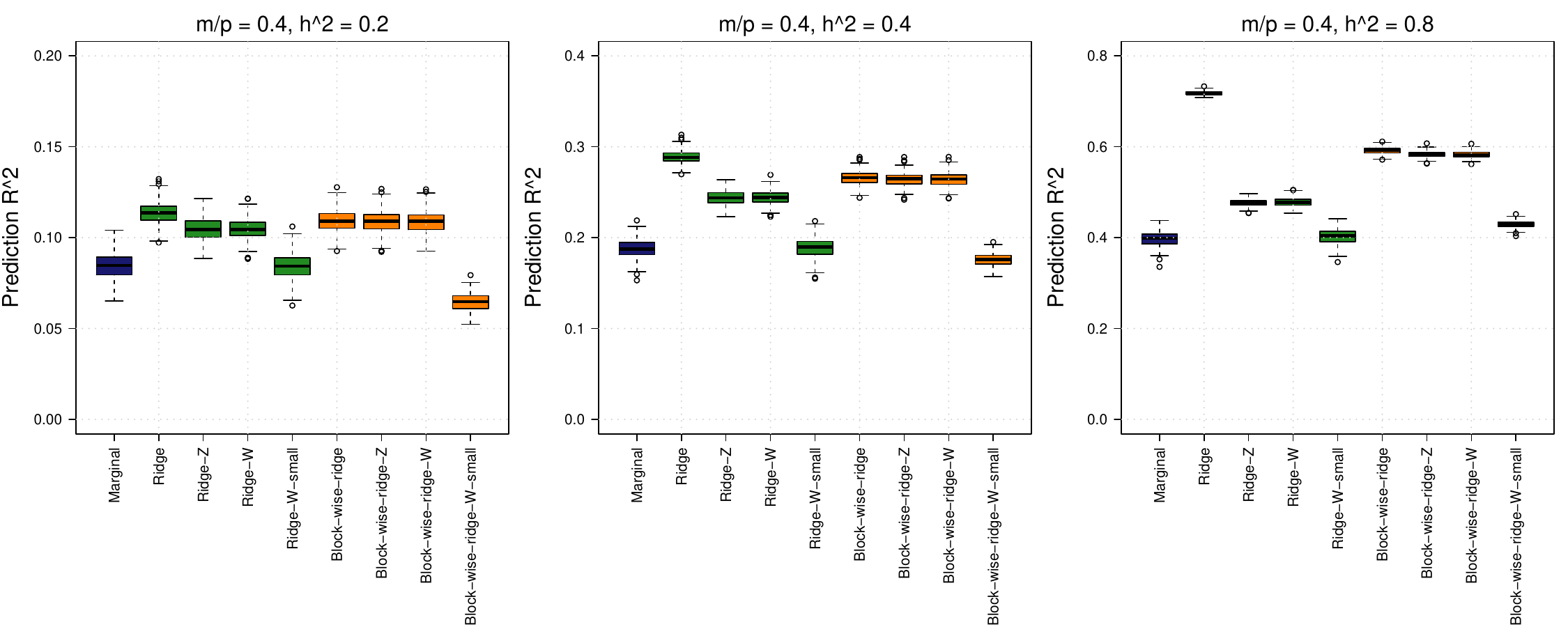}
\includegraphics[page=5,width=1\linewidth]{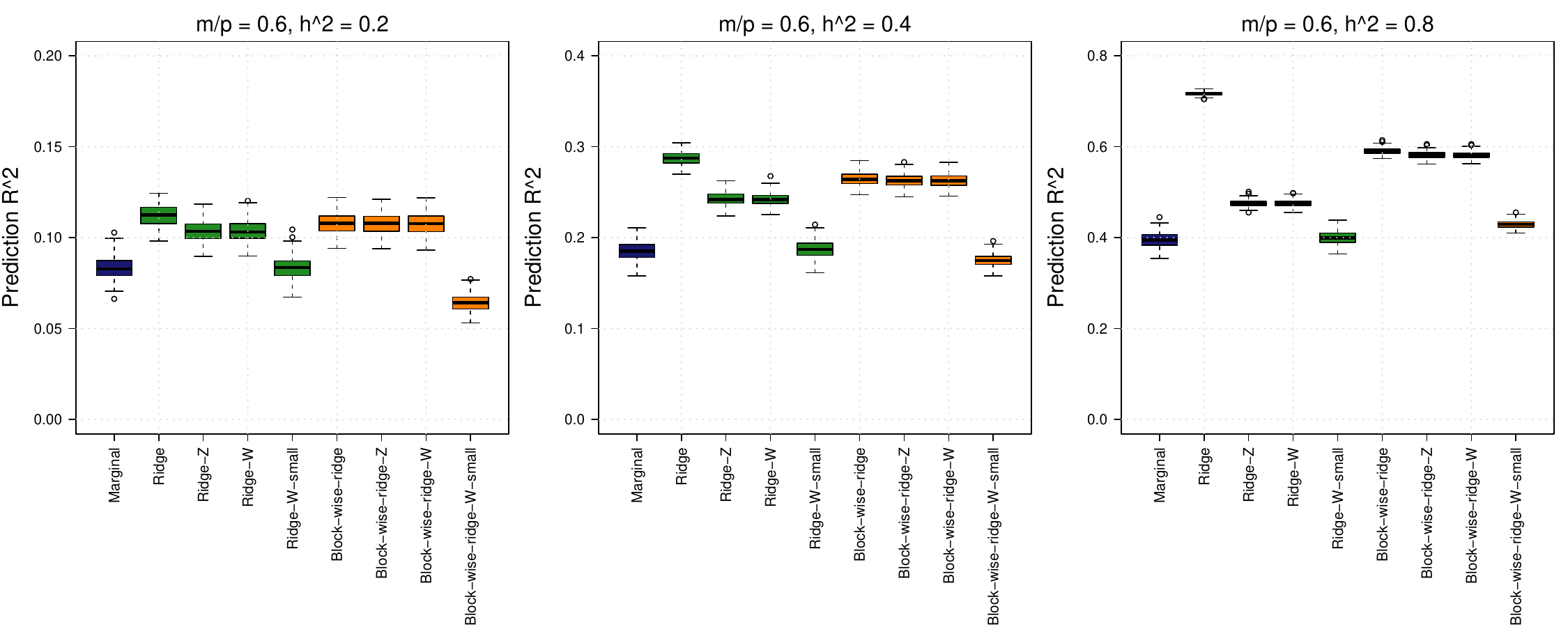}
  \caption{Comparing the prediction accuracy of ridge estimator $\widehat{\bmbeta}_{R}(\lambda^{*})$, marginal estimator $\widehat{\bmbeta}_{S}$,  and reference panel-based ridge estimators 
 $\widehat{\bmbeta}_{RW}(\lambda^{*})$ and $\widehat{\bmbeta}_{RZ}(\lambda^{*})$
at different $n/p$ ratios with $\h_\beta^2$ ranging from $0.2$ to $0.8$.
We simulate the data with $20$ independent blocks, each of which has a block size $p_\ell=50$ ($p=1,000$). An auto-correlation structure is given within each block and the auto-correlation coefficient is $\rho_b=0.5$. We also illustrate the 
performance of $\widehat{\bmbeta}_{RW}(\lambda^{*})$ and $\widehat{\bmbeta}_{RZ}(\lambda^{*})$ when $n_w=n_z=100$. The vertical line represents $n/p=1$. 
}
\label{figs5}
\end{suppfigure}
\begin{suppfigure}
\centering
\includegraphics[page=1,width=1\linewidth]{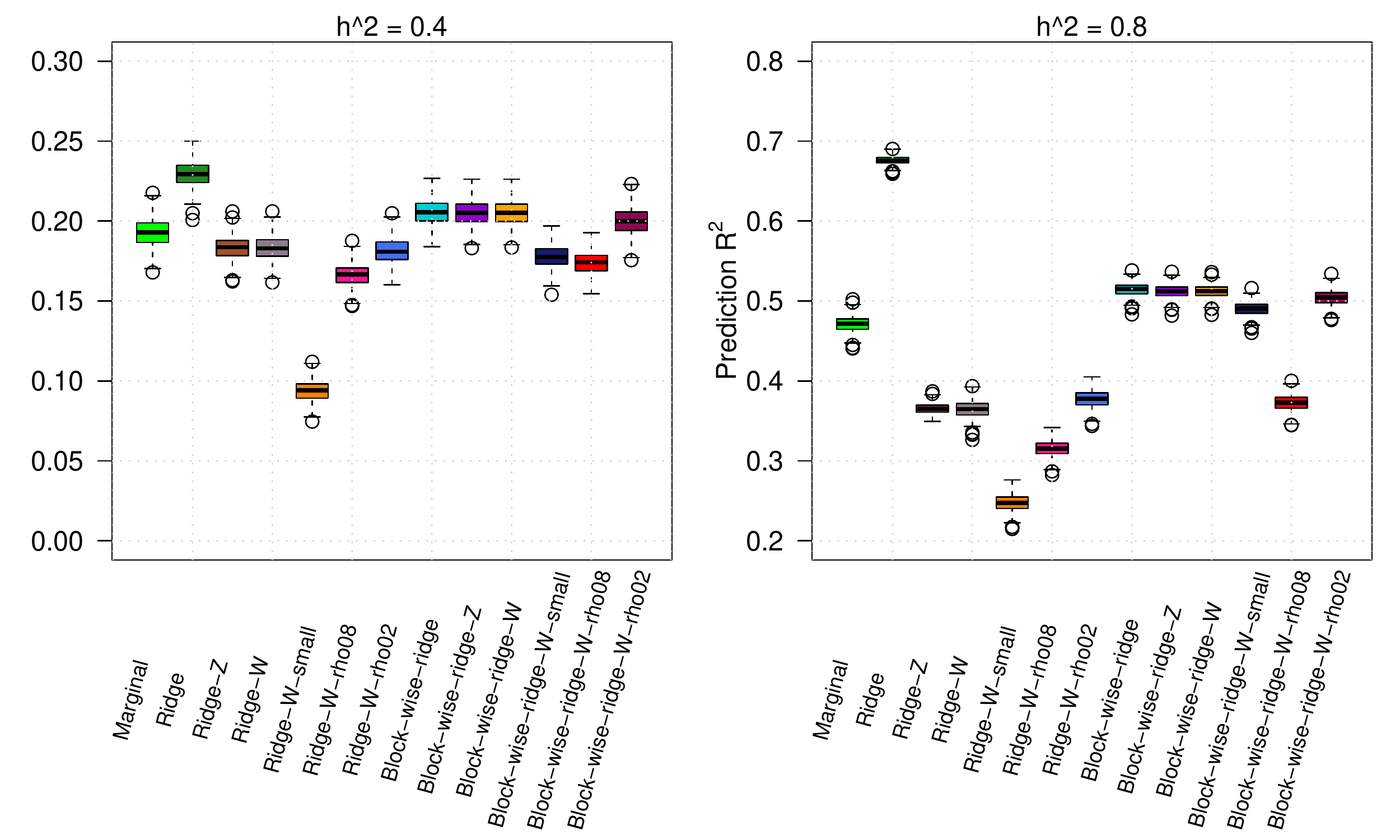}
  \caption{
  Prediction accuracy of different estimators. 
    We set $n=n_z=n_w=10,000$, $p=5,000$, and $\h_\beta^2=0.4$ (left panel) and $0.8$ (right panel).
  Marginal: $\widehat{\bmbeta}_S$; 
  Ridge: $\widehat{\bmbeta}_R(\lambda^*)$; 
  Ridge-Z: $\widehat{\bmbeta}_{RZ}(\lambda^*)$; 
  Ridge-W: $\widehat{\bmbeta}_{RW}(\lambda^*)$; 
Ridge-W-small: $\widehat{\bmbeta}_{RW}(\lambda^*)$ with $n_w=100$; 
Ridge-W-rho08: $\widehat{\bmbeta}_{RW}(\lambda^*)$ with $\rho_b=0.8$; 
Ridge-W-rho02: $\widehat{\bmbeta}_{RW}(\lambda^*)$ with $\rho_b=0.2$;
Block-wise-ridge: $\widehat{\bmbeta}_{B}(\lambda^*)$; 
  Block-wise-ridge-Z: $\widehat{\bmbeta}_{BZ}(\lambda^*)$;
  Block-wise-ridge-W: $\widehat{\bmbeta}_{BW}(\lambda^*)$;
    Block-wise-ridge-W-small: $\widehat{\bmbeta}_{BW}(\lambda^*)$ with $n_w=100$; 
    Block-wise-ridge-W-rho08: $\widehat{\bmbeta}_{BW}(\lambda^*)$ with $\rho_b=0.8$; and 
        Block-wise-ridge-W-rho02: $\widehat{\bmbeta}_{BW}(\lambda^*)$ with $\rho_b=0.2$. 
}
\label{figs6}
\end{suppfigure}
\begin{suppfigure}
\centering
\includegraphics[page=2,width=1\linewidth]{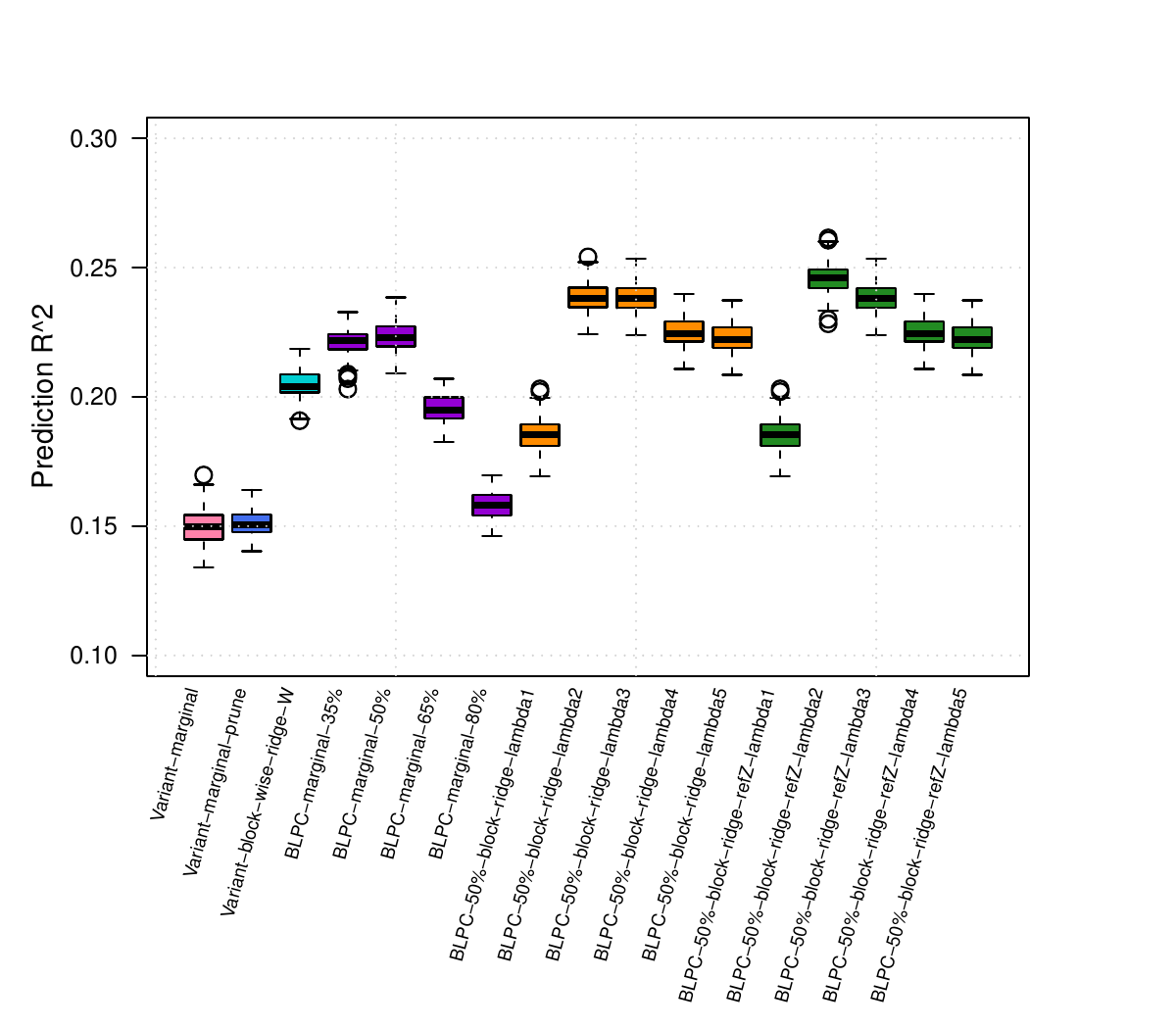}
  \caption{
  Prediction accuracy of different estimators in the UK Biobank data simulation. 
    We set $n=100,000$, $n_z=10,000$, $p=653,122$, and $\h_\beta^2=0.6$.
  Variant-marginal: marginal estimator with genetic variants; 
  Variant-marginal-prune: marginal estimator with genetic variants after LD-based pruning; 
  Variant-block-wise-ridge-W: block-wise reference panel-based ridge estimator with genetic variants;
 BLPC-marginal-35\%/-50\%/-65\%/-80\%: marginal estimator with top BLPCs explaining $35\%$, $50\%$, $65\%$, and $80\%$  genetic variations, respectively; 
 BLPC-50\%-block-ridge-lambda1/-lambda2/-lambda3/-lambda4/-lambda5: block-wise ridge estimator with BLPCs explaining $50\%$ genetic variations and different tuning parameters lambda. 
 BLPC-50\%-block-ridge-refZ-lambda1/-lambda2/-lambda3/-lambda4/-lambda5: block-wise reference panel-based ridge estimator with BLPCs explaining $50\%$ genetic variations and different tuning parameters lambda. 
}
\label{figs7}
\end{suppfigure}
\begin{suppfigure}
\centering
\includegraphics[page=2,width=1\linewidth]{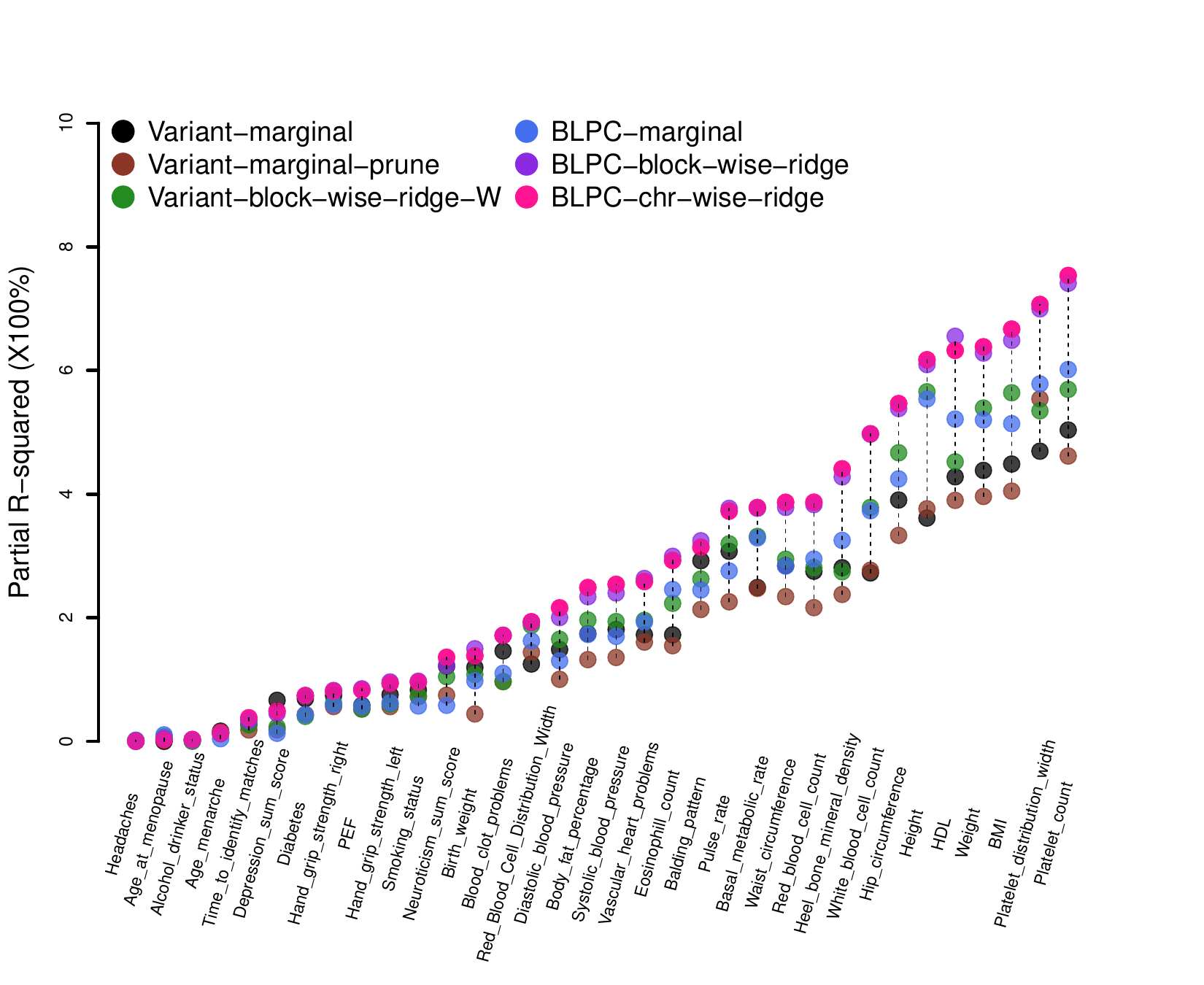}
  \caption{
 Prediction accuracy of different estimators in the UK Biobank data analysis for Asian testing samples.
  Variant-marginal: marginal screening with genetic variants; 
  Variant-marginal-prune: marginal screening with pruned genetic variants; 
  Variant-block-wise-ridge-W: block-wise ridge estimator with LDs being estimated from the 1000 Genomes reference panel; 
  BLPC-marginal: marginal screening with BLPCs;
  BLPC-block-wise-ridge: block-wise ridge estimator with BLPCs; and 
  BLPC-chr-wise-ridge: chromosome-wise ridge estimator with BLPCs. 
}
\label{figs8}
\end{suppfigure}

\begin{supptable}[ht]
\centering
\scalebox{0.8}{
\begin{tabular}{rrrrrrrrr}
  \hline
Trait & V\_1 & V\_2 & V\_3 & PC\_1 & PC\_2 & PC\_3 & Diff & Prop \\ 
  \hline
Heel\_bone\_mineral\_density & 5.86 & 6.24 & 7.73 & 7.07 & 8.74 & 8.47 & 1.01 & 0.13 \\ 
  Hand\_grip\_strength\_left & 1.33 & 1.19 & 1.46 & 1.36 & 1.74 & 1.80 & 0.34 & 0.24 \\ 
  Hand\_grip\_strength\_right & 1.37 & 1.23 & 1.41 & 1.34 & 1.79 & 1.78 & 0.38 & 0.27 \\ 
  Waist\_circumference & 4.73 & 4.03 & 5.24 & 5.01 & 6.17 & 6.27 & 1.03 & 0.20 \\ 
  Hip\_circumference & 5.82 & 5.87 & 7.68 & 7.52 & 8.95 & 9.20 & 1.53 & 0.20 \\ 
  Height & 7.76 & 8.98 & 12.03 & 11.69 & 12.50 & 12.97 & 0.94 & 0.08 \\ 
  BMI & 6.50 & 6.46 & 8.89 & 8.65 & 10.18 & 10.36 & 1.47 & 0.17 \\ 
  Basal\_metabolic\_rate & 2.98 & 3.26 & 4.27 & 4.06 & 4.69 & 4.86 & 0.59 & 0.14 \\ 
  Weight & 5.53 & 5.76 & 7.66 & 7.40 & 8.64 & 8.97 & 1.31 & 0.17 \\ 
  Body\_fat\_percentage & 4.07 & 3.61 & 4.76 & 4.63 & 5.49 & 5.58 & 0.82 & 0.17 \\ 
  Balding\_pattern & 8.38 & 6.29 & 8.12 & 7.48 & 9.67 & 9.65 & 1.30 & 0.15 \\ 
  HDL & 5.65 & 6.63 & 8.03 & 8.44 & 9.64 & 10.00 & 1.96 & 0.24 \\ 
  Eosinophill\_count & 2.40 & 3.48 & 4.30 & 4.49 & 5.17 & 5.36 & 1.07 & 0.25 \\ 
  Platelet\_count & 8.03 & 9.54 & 11.36 & 11.15 & 12.68 & 12.92 & 1.55 & 0.14 \\ 
  Platelet\_distribution\_width & 5.56 & 6.77 & 9.10 & 9.16 & 9.93 & 10.33 & 1.23 & 0.13 \\ 
  Red\_blood\_cell\_count & 4.24 & 5.32 & 6.12 & 6.19 & 7.17 & 7.43 & 1.31 & 0.21 \\ 
  Red\_blood\_cell\_distribution\_width & 2.07 & 2.55 & 3.63 & 3.46 & 4.19 & 4.44 & 0.82 & 0.23 \\ 
  White\_blood\_cell\_count & 2.43 & 3.52 & 5.11 & 4.74 & 5.83 & 6.63 & 1.52 & 0.30 \\ 
  Diastolic\_blood\_pressure & 2.81 & 2.92 & 3.93 & 3.58 & 4.49 & 4.56 & 0.62 & 0.16 \\ 
  Systolic\_blood\_pressure & 3.24 & 3.07 & 3.78 & 3.47 & 4.36 & 4.47 & 0.70 & 0.18 \\ 
  Pulse\_rate & 3.50 & 3.33 & 3.89 & 3.80 & 4.75 & 4.88 & 0.99 & 0.25 \\ 
  FEV & 3.73 & 4.85 & 6.33 & 6.13 & 7.59 & 7.84 & 1.51 & 0.24 \\ 
  FVC & 4.66 & 4.68 & 5.75 & 5.29 & 6.85 & 6.86 & 1.11 & 0.19 \\ 
  PEF & 1.10 & 1.16 & 1.41 & 1.21 & 1.72 & 1.81 & 0.40 & 0.28 \\ 
  Vascular\_heart\_problems & 2.53 & 2.59 & 2.78 & 2.78 & 3.58 & 3.53 & 0.80 & 0.29 \\ 
  Blood\_clot\_problems & 1.87 & 1.64 & 1.91 & 1.83 & 2.31 & 2.37 & 0.46 & 0.24 \\ 
  Diabetes & 0.57 & 0.51 & 0.47 & 0.43 & 0.74 & 0.73 & 0.16 & 0.28 \\ 
  Headaches & 0.15 & 0.21 & 0.18 & 0.17 & 0.23 & 0.23 & 0.02 & 0.09 \\ 
  Age\_at\_menopause & 0.22 & 0.16 & 0.25 & 0.28 & 0.30 & 0.30 & 0.05 & 0.21 \\ 
  Age\_menarche & 1.21 & 0.71 & 0.90 & 0.60 & 1.28 & 1.19 & 0.07 & 0.06 \\ 
  Birth\_weight & 1.72 & 1.58 & 1.68 & 1.39 & 2.00 & 1.99 & 0.28 & 0.16 \\ 
  Time\_to\_identify\_matches & 1.12 & 0.78 & 1.07 & 0.95 & 1.37 & 1.39 & 0.26 & 0.23 \\ 
  Depression\_sum\_score & 0.83 & 0.79 & 0.79 & 0.64 & 0.88 & 0.94 & 0.11 & 0.14 \\ 
  Neuroticism\_sum\_score & 1.86 & 1.60 & 1.85 & 1.69 & 2.38 & 2.31 & 0.51 & 0.28 \\ 
  Smoking\_status & 2.12 & 1.99 & 2.24 & 2.09 & 2.88 & 2.97 & 0.73 & 0.33 \\ 
  Alcohol\_drinker\_status & 0.24 & 0.18 & 0.15 & 0.15 & 0.27 & 0.27 & 0.03 & 0.10 \\ 
   \hline
\end{tabular}
}
\caption{Partial R-squared ($\times 100\%$) of different estimators in the UK Biobank prediction analysis on $36$ complex traits.
The first column is the ID of complex traits. 
V\_1, marginal screening with genetic variants (Variant-marginal); 
V\_2, marginal screening with pruned genetic variants; (Variant-marginal-prune); 
V\_3, block-wise ridge estimator with LDs being estimated from the 1000 Genomes reference panel;
PC\_1, marginal screening with BLPCs (BLPC-marginal);
PC\_2, block-wise ridge estimator with BLPCs (BLPC-block-wise-ridge); and 
PC\_3, chromosome-wise ridge estimator with BLPCs (BLPC-chr-wise-ridge). 
The Diff = max\{PC\_1, PC\_2, PC\_3\} - max\{V\_1, V\_2, V\_3\}, showing the difference between the best performance among the three BLPC-based methods and the best performance among the three variant-based methods. The Prop = Diff$/$max\{V\_1, V\_2, V\_3\}
}
\end{supptable}

\clearpage
\bibliographystyle{rss}
\bibliography{sample.bib}
